\documentclass[reprint,aps,longbibliography,superscriptaddress]{revtex4-1}  
\usepackage{amsmath,amssymb,amsfonts} 
\usepackage{bm} 
\usepackage{braket}          
\usepackage[pdftex]{graphicx}
\usepackage[dvipsnames,usenames]{color} 
\usepackage{textcomp} 
\usepackage{dsfont}

\def\rme{\mathrm{e}}
\def\rmd{\mathrm{d}}
\def\rmi{\mathrm{i}}

\def\bfr{\textbf{\em r}}
\def\bfk{\textbf{\em k}}
\def\Ip{I_\mathrm{p}}

\def\mathbi#1{\textbf{\em #1}}

\def\abs#1{\left|#1\right|^2}

\def\EXUV{\mathbi{E}_{\mathrm{\tiny{XUV}}}}
\def\epsXUV{\bm{\epsilon}}

\newcommand{\myfigure}[3]{
            \begin{figure}
                      \includegraphics[width=#1\columnwidth]{Figures/#2}
                    \caption{#3}
          \label{fig:#2}
        \end{figure}
    }

\def\stef#1{#1}

\begin{document}

\title{{\large PhD TUTORIAL} \\ Self-probing of Molecules with High Harmonic Generation}

\author{S Haessler}\email{stefan.haessler@tuwien.ac.at}
\affiliation{Photonics Institute, Vienna  University of Technology, Gu\ss hausstra\ss e 27/387, A-1040 Vienna, Austria}

\author{J Caillat}
\affiliation{UPMC, Universit\'e Paris 06, CNRS, UMR 7614, LCPMR, 11 rue Pierre et Marie Curie, 75231 Paris Cedex 05, France}

\author{P Sali\`eres}
\affiliation{CEA-Saclay, IRAMIS, Service des Photons, Atomes et Mol\'ecules, 91191 Gif-sur-Yvette, France}

\begin{abstract}
This tutorial presents the most important aspects of the molecular self-probing paradigm, which views the process of high harmonic generation as ``a molecule being probed by one of its own electrons''. Since the properties of the electron wavepacket acting as a probe allow a combination of attosecond and \AA ngstr\"om resolutions in measurements, this idea bears great potential for the observation, and possibly control, of ultrafast quantum dynamics in molecules at the electronic level.
Theoretical as well as experimental methods and concepts at the basis of self-probing measurements are introduced. Many of these are discussed on the example of molecular orbital tomography.
\end{abstract}

\maketitle

%
\section{Introduction}
\subsubsection{Combining attosecond and \AA ngstr\"om resolutions}
The measurement of structural changes of matter as it undergoes processes important to physics, chemistry and biology has always been one of the prime goals of experimentalists. While in the 19th century questions like ``How do cats manage to always land on their feet?'' were answered \cite{[][ Accessible via the conservatoire nationale des arts et m\'etiers: \url{http://cnum.cnam.fr}]Marey1894kitty}, current technology allows to address \mbox{(sub-)femtosecond} and \AA ngstr\"om scales, i.e. the natural scales of electrons at the atomic/molecular level  (1 atomic unit of time $\approx$ 24 as $=24\times10^{-18}\:$s, 1 atomic unit of length $\approx$ 0.53 \AA $=5.3\times10^{-11}\:$m). 

A very instructive overview of various methods currently developed to achieve such extreme resolutions can be found in \cite{[***][]Altucci2010review}. One of these emerged during the last decade as a by-product of efforts to generate light pulses of attosecond duration \cite{[{*** For an excellent didactic introduction to attosecond science, see }][]Bucksbaum2007review,[{** For the most recent and extensive review, see }][]Krausz2009} and is currently attracting tremendous attention: the self-probing of molecules by their own electrons.

This idea is inherent to the three-step recollision model \cite{Corkum1993Plasma,*Schafer1993Above} ubiquitous in strong-field physics, the field concerned with the highly non-linear response of matter to optical electric fields of similar strength to atomic/molecular binding fields: \emph{(i)}~The strong laser field tears a valence electron away from the molecule through tunnel ionization. \emph{(ii)}~The ``freed'' electron is accelerated in the laser field, which soon reverses its sign to toss the electron back. \emph{(iii)}~It may now happen that the laser driven electron recollides with its parent ion and scatters in different ways, such as inelastic scattering leading to, e.g., core excitation or double ionization; or simply elastic scattering of the electron. The idea of self-probing is now nothing more than considering the electron-ion scattering as a probe of the molecule. An excellent overview of self-probing based on different scattering mechanisms can be found in \cite{[***][]Lein2007Molecular}. Each of these possible ``third steps'' demands its specific theoretical modelling.

For this tutorial, we choose to delve into only one particular outcome of inelastic electron-ion scattering; namely the recombination of the electron with the hole it had left behind. As a result of recombination, the kinetic energy of the recolliding electron plus the binding energy are released through the emission of a photon, typically in the extreme-ultraviolet (XUV, 10--120 eV or 0.37--4.4 atomic units (a.u.)) or even soft x-ray (120--1200 eV or 4.4--44 a.u.) spectral range. This strong-field process is the so called \emph{high harmonic generation} (HHG), discovered simultaneously in Saclay \cite{Ferray1988Multipleharmonic} and Chicago \cite{Mcpherson1987Studies} in 1987.

In the self-probing paradigm, the recolliding electron takes the role of a probe pulse and the emitted photons that of the signal carrying information on the molecule to the detector. What makes this scheme particularly attractive is that this probe pulse turns out to have a set of beautiful properties found nowhere else in this combination:

\myfigure{.9}{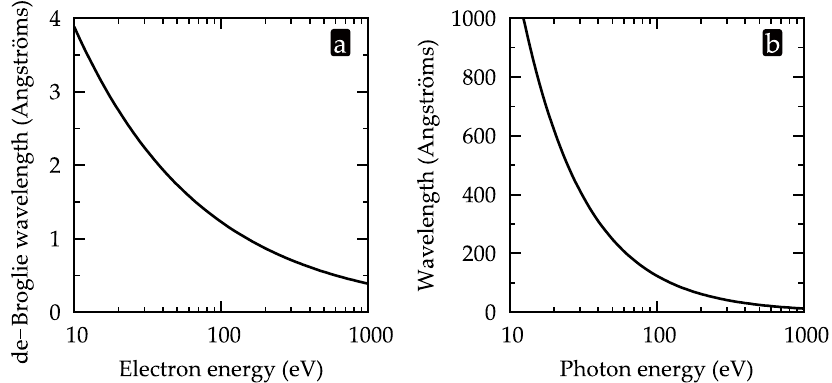}{Wavelength associated with an \stef{electron} (a) and a \stef{photon} (b) as function of the energy. Due to their different dispersion relation, electrons are suitable to probe Angstr\"om structures at much lower energies than photons.}
\emph{(i)}~The electron energies, $E$, correspond to de-Broglie-wavelengths of $\lambda_\mathrm{e}=2\pi/\sqrt{2 E}$  (atomic units \cite{[{*** Atomic units are based on the proposition of }][{. We strongly recommend to unfamiliar readers to study an introduction to atomic units (a.u.), where $\hbar=m_\mathrm{e}=e=4\pi\epsilon_0=1$. This can be found in most atomic/molecular physics text books or, e.g., on wikipedia. It follows, for example, that 1 a.u. of energy = 27.2 eV, 1 a.u. of length = 0.53 \AA, 1 a.u. of time = 24.2 as, 1 a.u. of intensity = $3.54\times10^{16}\:$W cm$^{-2}$ and 1 a.u. of electric field strength = $5.1\times10^{11}\:$V m$^{-1}$.}]Shull1959} will be used throughout the tutorial) in the range of few \AA ngstr\"oms or even sub-\AA ngstr\"om. Obtaining the same wavelengths with photons, for which $\lambda= 2\pi c/E$, where $c$ is the vacuum light velocity, requires several keV energy, i.e. hard x-rays! Such x-rays are not only hard to make and control, but also primarily interact with electronic core-states rather than the valence shell, relevant for dynamics in chemical and biological systems. A comparison of the wavelengths of photons and electrons is shown in figure~\ref{fig:debroglie-and-photon}.

\emph{(ii)}~The whole HHG three-step process of tearing the electron away, accelerating it and finally making it recollide and recombine happens in only a fraction of a driving laser field cycle; e.g. in less than 2.7~fs with the commonly used 800-nm lasers. The total duration of the electron probe pulse is only $\sim1\:$fs long \cite{[***][]Niikura2002Sublasercycle}. Pushing this idea further and splitting the electron into narrow spectral components (quantum-mechanically, it is of course a wave packet), one can make use of its intrinsic chirp and devise the \emph{chirp-encoded recollision} scheme allowing for $\sim0.1\:$fs temporal resolution \cite{[**][]Baker2006Probing,*[***][]Bucksbaum2006pacerNV,[*][]Lein2005,[**][]Smirnova2009co2,*[***][]Vrakking2009newsviews,Niikura2005Mapping,[**][]Mairesse2003Attosecond} (cf. section \ref{sec:decode:chirpencoded}). To date, not even gigantic x-ray free-electron lasers such as LCLS in Stanford can provide x-ray pulses with such short durations.

\emph{(iii)}~The HHG process is driven by the laser field in a fully \emph{coherent} way, i.e. it happens in a macroscopic number of molecules in the laser focus in a perfectly synchronized way, leading to a ``macroscopic'' light signal from a single-molecule effect---a very pleasant situation for experimentalists.

In total, we have a probe electron wave packet (EWP) composed of wavelengths of typical molecular dimensions ($\sim1\:$\AA) and with a duration similar to intra-molecular electron dynamics ($\sim1\:$fs or less) or very rapid nuclear dynamics, e.g. of protons. This clearly holds promise for directly time-resolving such dynamics and probably even for ultrafast \emph{imaging} of electrons in molecules. ``Imaging'' means obtaining structural information of any kind---what kind, depends very strongly on the underlying theoretical model.

Tracking ``electrons at work'' in molecules has been a dream of physicists for a long time because this could tackle fundamental questions such as:  When exactly is the Born-Oppenheimer approximation valid and when does it break down \cite{[*][]Lodi2010tutorial,[*][]Bransdenbook,[*][]Piela2007book,Muskatel2009post-bo}? How does the correlated electron cloud re-arrange after a fast perturbation and how long does it take? How does this influence the subsequent slower nuclear motion? Could we control and drive the electron re-arrangement? When and how do correlations and couplings gain importance in molecular dynamics? Of particular interest here will be the migration of the electron-hole after sudden ionization of a molecule by an attosecond light pulse or via tunnelling in a strong infrared field \cite{[**][]Breidbach2005,Luennemann2010ultrafast,[**][]Remacle2006,[***][]Smirnova2009pnas}, since the first experiments related to such hole-dynamics have recently been demonstrated \cite{[**][]Smirnova2009co2,*[***][]Vrakking2009newsviews,[**][]Haessler2010tomo,*[***][]Smirnova2010NV} using the self-probing scheme.

Despite these encouraging  achievements, we appreciate the great difficulty of the development of reliable models for the interaction of molecules with strong laser fields, upon which to base the extraction of information from self-probing experiments, and the many issues that are still open. We will not fail to mention here the most important ``construction sites'' of the theories behind self-probing; a more involved introduction into this vivid debate is attempted in \cite{[***][]SalieresRPP}. We write this tutorial because we believe that a pedagogical overview of the ideas and concepts behind self-probing can help many researchers with different backgrounds and different levels of experience, who share the common goal of dynamic imaging at the \mbox{(sub-)femtosecond} and \AA ngstr\"om scales, understand both the great potential and challenges of self-probing, and hopefully discover synergies and new opportunities.

\subsubsection{A bit of history}

In this section, instead of narrating the detailed development of the field over the last decade, we would just like to highlight some works we consider to be seminal milestones for the interpretation and exploitation of high harmonics generated in {\em molecules}.

In 2000, the Imperial college group of Jon Marangos initiated the studies of HHG in aligned molecules and announced that it could become a tool for studying molecular dynamics \cite{[*][]Hay2000pulselength,Velotta2001,[*][]Hay2002,deNalda2004}. Shortly after, Manfred Lein achieved a theoretical breakthrough when he described numerical results obtained for H$_2^+$ molecules with a very simple analytical model, demonstrating that destructive interference between the recolliding EWP and the electron bound-state \emph{wavefunction} during the recombination step of HHG leaves a clear signature in the harmonic emission \cite{[*][]Lein2002Role,[***][]Lein2002Interference}.

Simultaneously, the Paul Corkum group at NRC Ottawa was the first to explicitly formulate the self-probing paradigm, in the context of non-sequential double ionization of H$_2$ \cite{[***][]Niikura2002Sublasercycle}. In 2004, based on the picture of the electron-hole recombination in HHG as an interference between a probe EWP and the bound state, they pushed the idea of molecular imaging by self-probing to the extreme, proposing a tomographic analysis to reconstruct the electron bound-state \emph{wavefunction} \cite{Itatani2004Tomographic}. Although in this work, a static wavefunction had been reconstructed, the potential of ultra-fast---possibly attosecond---time resolved imaging of electrons bound in molecules was evident. This certainly constituted a conceptual breakthrough, although based on controversial assumptions, and initiated a great number of studies by both theoreticians and experimentalists.

In the following years, a lot of data was produced around the world: the intensity of harmonics generated in aligned molecules was shown in Italy and Japan to exhibit spectral minima \cite{[*][]Vozzi2005Controlling,[*][]Kanai2005Quantum,[*][]Torres2007Probing}. The connection of these minima with intramolecular interferences was confirmed soon after by harmonic phase measurements in Saclay, Boulder, Stanford and Ottawa \cite{[*][]Boutu2008Coherent,[**][]Smirnova2009co2,*[***][]Vrakking2009newsviews,[**][]Lock2009,[*][]McFarland2009n2phase,[**][]Wagner2007Extracting}. The harmonic polarization was also investigated and revealed surprisingly high ellipticities \cite{[*][]Mairesse2008Polarizationresolved,[*][]Levesque2007Polarization,[*][]Zhou2009,[*][]Mairesse2010Multichannel}. Ingenious experimental schemes were demonstrated to observe molecular dynamics, such as the mentioned \emph{chirp-encoded recollision} (cf. section \ref{sec:decode:chirpencoded}), or \emph{transient grating spectroscopy}: \cite{[**][]Mairesse2008Transient,[*][]Woerner2010} (cf. section \ref{sec:phasemeas:spatial}).

A lot of confusion was lifted around 2008 when Olga Smirnova convinced the community that analyses of experiments require not only to consider the highest occupied molecular orbital (HOMO) (or the probed ion in the electronic ground state), but also lower lying orbitals (or electronically excited ionic states) which may play a significant role as well, giving rise to rich \mbox{(multi-)electronic} dynamics within the ion \cite{[*][]Smirnova2009circular,[**][]Smirnova2009co2,*[***][]Vrakking2009newsviews}. First experimental indications of such multi-channel contributions have been observed almost simultaneously in Stanford \cite{[**][]McFarland2008}. Based on this idea we could interpret our measurements performed in aligned N$_2$ molecules and tomographically reconstruct with \AA ngstr\"om resolution a snapshot of an electron-hole evolving in the N$_2^+$ ion 1.5~fs after ionization with 600~as ``exposure time'' \cite{[**][]Haessler2010tomo,*[***][]Smirnova2010NV}.

Vibrational and dissociative dynamics in unaligned molecules have also been investigated with great success using HHG in Boulder and Ottawa \cite{Wagner2006Inaugural,Li2008,Woerner2010}.

On the theory side, mainly three approaches are being pursued: \emph{(i)}~tests of the applicability of the \emph{strong-field approximation} (SFA)---a well-proven model for HHG in atoms---against numerical solutions of the time-dependent Schr\"odinger equation (TDSE) for single-electron molecules, mainly done in the groups of Manfred Lein in Hannover \cite{Leinwebsite} and Lars Bojer Madsen in Aarhus \cite{Madsenwebsite}; as well as Chii-Dong Lin's proposal of an extension of the SFA with stationary scattering theory, dubbed \emph{quantitative rescattering theory} \cite{Le2009QRS,[**][]Lin2010review};  \emph{(ii)} the solution of the 3D TDSE for various model systems, pursued mainly in the group of Andr\'e Bandrauk in Sherbrooke \cite{Bandraukwebsite}; \emph{(iii)}~a fully time-domain model developed by Olga Smirnova, Misha Ivanov et al. \cite{[*][]Smirnova2009circular,[**][]Smirnova2009co2,*[***][]Vrakking2009newsviews}, composed of building blocks from advanced strong-field theories and proper multi-electron wavefunctions from quantum chemistry codes.

Advances in theory play a particularly important role for self-probing, since they do not only allow to understand specific features of HHG spectra generated in molecules, but also to exploit them, i.e. to retrieve information on the molecule and its dynamics from the measured HHG spectra.

\subsubsection{Content of this tutorial}

We will not be able to cover all contributions and approaches to the field to everyone's satisfaction, but this is not the aim of the tutorial. Instead, we want to introduce newcomers and since we explain best what we know best, we obviously will talk more about our own work than that of colleagues. This does not mean that we have to overly restrict the topics covered, though. This tutorial will focus on experimental aspects, i.e. the reader should understand what should be measured for a certain goal and what ways there are to do so. To this end, we will devote special attention to orbital tomography because we feel that specifics tend to be more colourful than generalities; i.e. even if a meaningful self-probing experiment does not at all necessarily need to aim at tomographic images, this example allows to discuss all observables that may also serve as input for other schemes for the extraction of information from measurements.

A theoretical introduction is attempted in section \ref{sec:theory}. We will treat descriptions of the HHG process with increasing complexity: from a purely classical treatment of the electron continuum dynamics which shapes our probe EWP, to a quantum mechanical description in the strong-field approximation, and also note the crucial step from the single-molecule high harmonic emission to that of a macroscopic medium. Finally, we will briefly introduce more advanced theoretical concepts relevant to a more accurate description of HHG in molecules. Section \ref{sec:decoding} will then present some ways to decode the HHG signal based on the models introduced before. Among others, here we will describe the chirp-encoded recollision concept and give a quite detailed description of molecular orbital tomography and its requirements on the experiment, establishing a ``to-do'' list of quantities to be measured. This list will guide through section \ref{sec:exp}, where experimental techniques will be presented that set the proper conditions for measurements and give access to all necessary observables. An example for some of the experimental and theoretical concepts discussed before will then be given in section \ref{sec:tomoexp}, where we described experiments on molecular orbital tomography. Finally, section \ref{sec:outlook} will conclude. At the end of the text, a list of the abbreviations used in this tutorial---all defined somewhere in the text but not always easy to recognize at first sight---will help readers who are not yet familiar with the slang used in the strong field community. 

\section{Theory of HHG}\label{sec:theory}
In a world with computing power so gigantic that we could solve exactly the equations governing the complete dynamics of the studied systems, we would calculate the outcome of all experiments and the properties of our system at any instant in time: we would conduct numerical experiments. To gain an actual understanding of what is happening in our system, though, we would have to do the same interpreting of the computed results that we have to do with experimental data today. Such ``full simulations'' would thus merely be able to fill up databases and some sort of subsequent data mining could perhaps discover patterns---but very little of (human) understanding. In some cases, turning on/off some interaction terms might allow to extract information on their relative importance, but this is often not possible without affecting the consistency of the simulations. Consequently, the exact equations and theories will have to be simplified---i.e. \emph{models} will have to be developed in a sort of ``top-down'' approach. Simplified models, based on approximations of the exact description, are thus not inferior fragments of the proper exact theories we have to deal with due to limited computing power, but they are at the very heart of any possible understanding of complex phenomena.

Anyhow, computing power available to most of us is not sufficient to solve multi-particle dynamics in 3D-space exactly for more than a few electrons \cite{Jordan2008Corepolarization}. The development of the theory behind self-probing has thus happened more like a ``bottom-up'' approach, starting from rather simple models with very strong approximations---which nonetheless had great success in describing HHG in atoms---and trying to add refinements to them while saving the schemes to extract information from observables. Here, we will follow the same route: we first present the basics of HHG in atoms (sections A and B) and then discuss some important aspects specific to HHG in molecules (section C).

	\subsection{Basic description of high harmonic generation}\label{sec:basichhg}
	\subsubsection{Tunnel ionization} \label{sec:tunnel}
\stef{The way a bound electron makes a transition to the continuum when interacting non-perturbatively with a strong laser field is one of the central elements in modelling self-probing, since it determines how our probe-electron-wavepacket will look. M. Ivanov and coworkers have written an excellent didactic introduction \cite{[***][]Ivanov2005Anatomy}. Here, we only want to briefly grasp some basic aspects of strong-field ionization.}

\myfigure{.6}{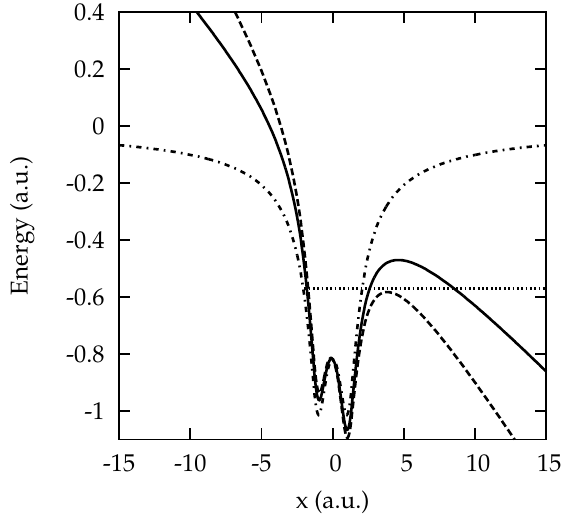}{Cut along the x-direction through the \stef{length-gauge potential,} felt by a single active electron in a diatomic molecule, which is subjected to a strong electric field, $E_0$, pointing along the negative x-direction. The field is $E_0=0$ (dash-dotted line), $E_0=-0.053\:$a.u. (solid line) and $E_0=-0.08\:$a.u. (dashed line). The horizontal dotted line marks $-\Ip=-0.57\:$a.u., the energy of the N$_2$ HOMO.}
For this purpose, it is always helpful to draw a representative image. Figure \ref{fig:tunnelplot} shows a cut along the x-direction through the \stef{length-gauge potential~\footnote{As a reminder: \textit{length gauge} means a specific calibration of the electrodynamic vector potential, $\mathbi{A}(t)=-\int_{-\infty}^t \mathbi{E}(t') \rmd t'$, such that the interaction Hamiltonian writes \stef{$\mathcal{H}_\mathrm{I}=\bfr\cdot \mathbi{E}(t)$}, and the canonical momentum $\mathbi{p}_\mathrm{can}$ is identical to the mechanical momentum $\bfk$. In \textit{velocity gauge}, reached through a unitary transformation corresponding to a gauge transformation of $\mathbi{A}(t)$, the interaction Hamiltonian transforms to \stef{$\mathcal{H}_\mathrm{I}=\mathbi{p}\cdot \mathbi{A}(t)$}, and the canonical momentum is related to the mechanical momentum of the electron by $\mathbi{p}_\mathrm{can}=\bfk-\mathbi{A}(t)$. Although the physics we are describing here is gauge invariant, the length gauge has the advantage to explicitly display the effective potential barrier through which the electron tunnels out.}} 
felt by a single active electron in a diatomic molecule, which is subjected to a strong \emph{static} electric field, $E_0$, pointing along the negative x-direction. One can clearly see how the strong electric field lowers the binding potential on the positive-x side and at some finite distance, the potential falls below the binding energy, $-\Ip$, of the electron. \stef{The electron can thus make a transition to the continuum by \emph{tunnelling}.} Its wavefunction falls exponentially within the classically forbidden region and the tunnelling rate, $\Gamma$, depends on the finite amplitude at the ``exit'' of the \stef{barrier}. Since the 1960's, it is known that this rate is \cite{Keldysh1965,PPT1966,*Bisgaard2004tunneling}:
\begin{equation}
	 \Gamma= \Gamma_0\exp[-2(2\Ip)^{3/2}/(3\vert E_0\vert)],
\label{eq:ADKrate}
\end{equation}
where the pre-factor, $\Gamma_0$, depends on the spatial structure of the bound state. The most general derivation of this rate \stef{for atoms} has been demonstrated by Ammosov, Delone and Krainov \cite{ADK1986}, which is why one often speaks of the \emph{ADK rate}. 

\stef{For \emph{molecules}, $\Gamma_0$ will be a function of the orientation of the laser field with respect to the molecule. Computing this dependence is a particularly difficult task because one needs an accurate representation of the asymptotic form of the molecular wave functions, not available from standard quantum chemistry codes. See \cite{[**][]Murray2011tunnel,[][]Murray2010partial,Tong2002moadk,*Zhao2011effect,[][]Brabec2005tunnel,Gallup2010semiclassical,[**][]Spanner2009oneelectron} for some recent approaches to modelling tunnel ionization of molecules. These not only aim at computing rates, but also the initial momentum distribution of the electron ``born'' in the continuum.}

\stef{In the following, we will only focus on the dependence of the tunnelling rate on the instantaneous electric field, which is exponential in (\ref{eq:ADKrate}).} This result, valid for static fields, can also be used for ``slowly'' oscillating laser fields, such that the potential barrier does not change too much while the electron tunnels. ``Not too much'' is quantified via the adiabaticity parameter,  $\gamma=\omega_0\sqrt{2\Ip}/E_0$, defined by Keldysh \cite{Keldysh1965} as the product of the laser frequency, $\omega_0$, and a ``tunnelling time'', i.e. the time it would take a classical particle to traverse the barrier if the motion were classically allowed. If $\gamma\ll 1$, the tunnelling barrier can be considered as quasi-static and $\vert E_0\vert$ in (\ref{eq:ADKrate}) can simply be replaced by an oscillating $\vert E(t)\vert$. Then, the exponential in (\ref{eq:ADKrate}) leads to the liberation of extremely short electron-bursts around each electric field extremum of the laser.

The other limiting case, $\gamma\gg1$, occurs for weak laser fields and high frequencies, and means that ionization is \stef{better understood as} multi-photon absorption (i.e. it proceeds ``vertically'' in figure \ref{fig:tunnelplot}, as opposed to the ``horizontal'' tunnelling) \stef{and the ionization amplitude will no longer depend on the instantaneous laser field}.

\stef{The Keldysh parameter has to be taken with some care: it loses its sense when $\omega_0\gtrsim\Ip$; or as soon as one reaches a peak laser field, $E_0$, so strong that it completely suppresses the potential barrier, i.e.  \mbox{$V(x)\leq -\Ip$} everywhere on one side. This starts to happen for a field strength \mbox{$\vert E_\mathrm{BS}\vert\approx\Ip^2/4$,} represented in figure \ref{fig:tunnelplot} by the dashed line. It is clear that then, although $\gamma$ might be very small, tunnelling is no longer a relevant concept since the electron can now simply exit the atom/molecule by moving classically over the lowered barrier.}

\stef{This regime is to be avoided in HHG anyway since due to the disappearance of the strong exponential damping associated with tunnelling, the properties of an electron wavepacket ``liberated over the barrier'' will be quite different from those of a wavepacket that has tunnelled out. In particular, the electron will leave so quickly (for the most part already before the laser field cycle maximum) that it cannot be driven back to recollide (see the discussion of trajectories in the next section), thus shutting off HHG. For $\Ip=0.5\:$a.u., the barrier suppression intensity, also called \emph{saturation} intensity, is \mbox{${E_\mathrm{BS}}^2=1.4\times10^{14}\:$W cm$^{-2}$} and, from the perspective of HHG, larger molecules with generally lower $\Ip$ can only withstand lower laser intensities.}

\stef{Typical HHG experiments take place within a parameter region where $\gamma$ does have a sense, though, and it will tell us that we are in a somewhat grey area: for an \mbox{800-nm} laser \mbox{($\stef{\omega_0}=0.057$),} focused to an intensity of \mbox{$1\times10^{14}\:$W cm$^{-2}$} ($\stef{E_0}=0.053$) and small molecules with \mbox{$\Ip\sim0.5\:$a.u.,} we find $\gamma\approx1$. Even with a, say, three times longer driving wavelength, $\gamma\approx0.33$.} In these conditions, tunnelling acquires a non-adiabatic component, i.e. the barrier moves during tunnelling. The effect may be imagined as ``multi-photon absorption facilitating tunnelling'', or the tail of the electron wavefunction being ``vertically heated' in the classically forbidden region \cite{Ivanov2005Anatomy}. \stef{An analytic expression for the non-adiabatic tunnelling rate in these conditions has been derived by Yudin and Ivanov \cite{[ *][]Yudin2001}. It turns out that for $\gamma\approx1$  it remains a highly nonlinear function of the field modulus, only a bit less sharply peaked than the ADK rate (\ref{eq:ADKrate}). The way the electron is born in the continuum in HHG experiments is thus well understood in a tunnelling picture.}

\stef{The laser intensity range where a tunnelling picture works well and where HHG is experimentally feasible thus has an upper bound of $\approx{E_\mathrm{BS}}^2$ and a lower bound of only a few times less (due to the rapidly declining tunnelling rate and thus HHG signal). This may seem absurdly narrow \cite{[][]Reiss2008Limits,*Reiss2010Unsuitability}, especially when comparing to the range of intensities that are available nowadays, covering many orders of magnitude~\cite{[][]Mourou2011more}. It is however the high non-linearity of the tunnel-induced responses obtained within that interval that makes it such a surprisingly rich regime to investigate~\footnote{\stef{Note that the very convenient ``tunnelling'' picture holds because we implicitly express the laser-electron interaction within the dipole approximation, as stressed \emph{e.g.} in \cite{Reiss2008Limits,*Reiss2010Unsuitability}. It is nevertheless a safe approach in the intensity range we are dealing with in the present tutorial, as non-dipole contributions become significant only far beyond the barrier suppression threshold.}}.}

		\subsubsection{Classical} \label{sec:class}
The three-step model for HHG \cite{Corkum1993Plasma,*Schafer1993Above} was mentioned already in the introductory section and is---despite its astonishing simplicity---probably the most often used theoretical tool of strong-field physicists. Many essential features of the HHG physics are captured and a very descriptive framework is set which allows to readily comprehend the self-probing paradigm. 

We call this section ``classical''  although the model described here includes the pure quantum effects tunnelling and recombination. These are, however, described only schematically as ``sudden'' events, while the actual model is a purely classical treatment of the continuum electron dynamics inbetween. This will turn out to be so simple that a reader can easily program his/her own simulations based on it and learn a great deal by simply playing with parameters.

In the \textit{first step}, at some time $t_\rmi$, tunnel ionization leads to an electron being ``born" in the continuum \footnote{The electrons we are concerned with---the ones that eventually recollide with the parent ion---are never really \emph{ionized} because this would mean they remain in the continuum after the process is over. In fact, they must be considered \textit{quasi-bound}. It is thus more precise to speak of ``birth in the continuum''. Nonetheless, within this tutorial, we will often speak of ``ionization'', knowing that this is a bit sloppy.} with initially zero velocity. From this moment on, the electron is treated as a classical point charge and during the \textit{second step}, one considers that its evolution is completely governed by the linearly polarized driving laser field, $\mathbi{E}(t)=\hat{\mathbi{x}}E_0\cos(\omega_0 t)$, with amplitude $E_0$ and angular frequency $\omega_0$. Although our reasoning describes the full 3D-space, it is sufficient to focus on the laser polarization direction $x$ for which the classical equation of motion reads:
	\begin{equation}
		\ddot{x} (t) = -E_0\cos(\omega_0 t)\:.
	\label{eq:theory:classacell}
	\end{equation}
Integration of (\ref{eq:theory:classacell}) with initial conditions $\dot{x}(t_\rmi)=0$ and $x(t_\rmi)=0$ \footnote{The second of these is a simplification neglecting the finite distance from the nucleus (cp. figure \ref{fig:tunnelplot}, where $x_0\approx10\:$a.u.$=0.5\:$ nm) at which the electron is ``born''. Although the electron will follow trajectories leading it only a few nm away from its parent ion, it is not worth worrying about this, since at this birth instant, the electron has so little kinetic energy (actually, in this model none at all), that its de Broglie wavelength is much larger than the `problem' anyhow.} leads to 
	\begin{align}
		\dot{x} (t)& = -\frac{E_0}{\omega_0} \left[ \sin(\omega_0 t) - \sin(\omega_0 t_\rmi)\right]\:, \label{eq:theory:classvelocity} \\
		x(t)& = \frac{E_0}{\omega_0^2} \left[ \cos(\omega_0 t) - \cos(\omega_0 t_\rmi) \right] + \frac{E_0}{\omega_0}\sin(\omega_0 t_\rmi) (t-t_\rmi)\:.  \label{eq:theory:classtraj}
	\end{align}
The last equation shows that not for every electron birth time, $t_\rmi$, does the electron trajectory lead back to the parent ion at $x=0$---the slope of the second member could dominate the trajectory and the electron just drifts away. Within each laser field period \stef{$T_0=2\pi/\omega_0$}, the trajectory does, however, reach $x=0$ again for $0\leq t_\rmi\leq T_0/4$ and $T_0/2\leq t_\rmi\leq 3T_0/4$. By numerically finding these roots, one determines pairs of ionizations times, $t_\rmi$, and recollision times, $t_\mathrm{r}$. Obviously, there may be several roots, i.e. recollisions, in the trajectory corresponding to one $t_\rmi$. We will, however, only consider the first one, because the continuum electron, which quantum mechanically is of course a wavepacket, will spread during propagation in all three dimensions \footnote{As we consider only the $x$-component of the electron movement with one initial condition, this quantum spreading is not included in our description. It can be described by launching for every birth time many trajectories with different small initial velocities in all three dimensions and counting all trajectories that pass through a given finite cross-section around the origin as ``returns''. This was done in \cite{Kitzler2005}.}, thus reducing the efficiency of the subsequent recollisions. Phase matching effects further reduce the contribution of longer trajectories to the macroscopic signal measured in experiments \cite{Antoine1996Attosecond}.

\myfigure{1}{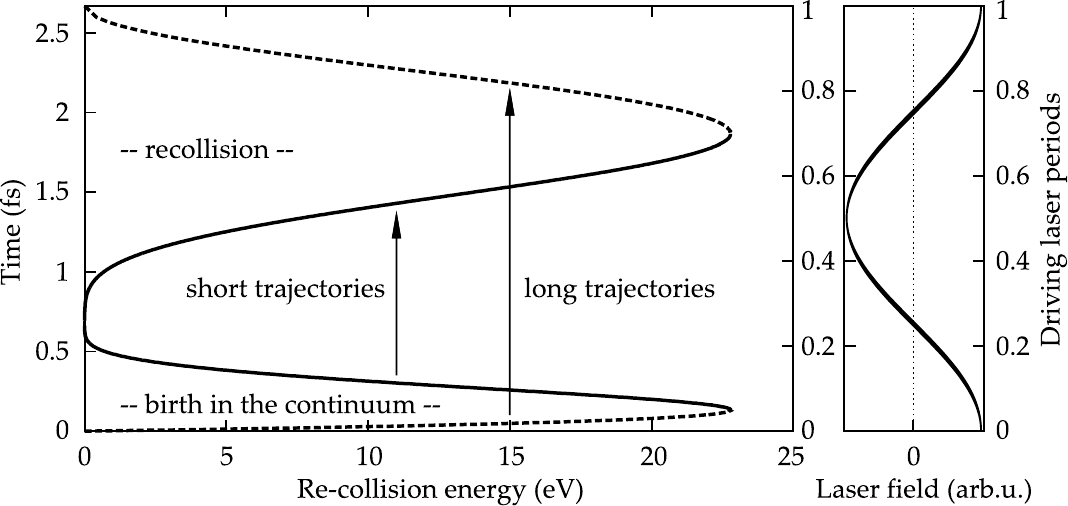}{Left panel: Classical calculation of ionization and recollision times as a function of the electron recollision energy, for an 800 nm laser and an intensity of $I=1.2\times10^{14}\:$W/cm$^2$. The electric field of the driving laser has a cosine time-dependence, i.e. time zero marks the field maximum. Full and dashed lines mark the short and long trajectories, respectively. Right panel: Driving laser field on same time axis.}

For every pair $(t_\rmi,t_\mathrm{r})$, the kinetic energy at recollision, $\dot{x}(t_\mathrm{r})^2/2$ can be determined. Figure \ref{fig:class-ti-tr} shows a plot of the ionization times, $t_\rmi$, and recollision times, $t_\mathrm{r}$, as a function of the associated electron energy at the instant of recollision. With every recollision energy, a long (dashed lines) and a short trajectory (solid lines) are associated, which join for the very highest recollision energy. Electrons are born in the continuum during the first quarter period of the driving laser. The short trajectories then lead to recollision mainly after the subsequent laser field extremum (i.e. after $T_0/2$) and electrons with the highest return energies recollide at $\approx0.7\, T_0$, i.e. close to a zero-crossing of the driving laser field. In the last quarter period, the long trajectories recollide. 

At recollision, the electrons may recombine to the ground state, which is the \textit{third step} of the three-step model. The emitted XUV photon has an energy of $\dot{x}(t_\mathrm{r})^2/2+\Ip$. The highest recollision energy turns out to be $\dot{x}_\mathrm{max}(t_\mathrm{r})^2/2=3.2\, U_\mathrm{P}$, which constitutes the so-called classical cut-off law. Here, $U_\mathrm{P}=E_0^2/4\omega_0^2$ is the ponderomotive potential, i.e. the mean quiver energy of a free electron in a laser field.

\emph{At this point, we can already learn quite a number of things about HHG:}
\begin{itemize}
\item With a multi-cycle driving laser pulse with symmetrical carrier wave, the three-step process is repeated: \emph{(i)} with the same properties every cycle of the driving laser field, and \emph{(ii)} with inversion symmetry (for atoms and symmetric molecules) every half-cycle of the driving laser field, i.e., the recollision direction of the EWP switches its sign. This implies a sign change in the molecular dipole and thus in the XUV emission. In a spectrometer, the contributions of the individual XUV bursts in the attosecond pulse train (APT) \stef{interfere} and one \stef{easily shows}~\footnote{Write, in the time domain, the pulse train as a convolution of one XUV burst with a variant of the dirac comb: $\Delta(t)=\sum_{n=-\infty}^{+\infty} (-1)^n \delta(t-nT_0/2)$. This comb-function is even in $t$ and $T_0$-periodic (mind the sign change!). It can thus be expanded into a Fourier series $\sum_{m=1}^{\infty}a_m\cos[m (2\pi/T_0) t]$, with coefficients  $a_m=2/T_0\int_{t_0}^{t_0+T_0}\Delta(t)\cos[m (2\pi/T_0) t] \rmd t =$ \mbox{$2/T_0 [1-\cos(m\pi)]$}, which turn out to be non-zero only for odd m.} that due to \emph{(i)}, the \stef{measured} spectrum consists of harmonics of the driving laser \stef{frequency}, and due to (\emph{ii}), only the odd harmonic orders are present.  Due to the laser envelope, the XUV emission has a finite duration and this XUV envelope in the time domain leads to a broadening of the harmonic peaks in the spectral domain.

\item From figure \ref{fig:class-ti-tr}, one can infer that recollision \stef{\emph{with energies above a certain threshold value}---say, 10 eV---}takes place only during a fraction of the driving laser cycle, i.e. the recolliding EWP is extremely short and XUV emission takes place in short bursts. Furthermore, different spectral components recollide at different instants: the EWP has a chirp, which is of different sign for the short and long trajectories. 

\item In (\ref{eq:theory:classvelocity}) and (\ref{eq:theory:classtraj}), we can see that the laser field amplitude/intensity acts merely as an amplitude scaling factor on the trajectories, while the time dependence---e.g. the recollision instant for the most energetic (``cut-off'') trajectory---is given by the laser frequency/wavelength only. Thus, as the achievable recollision energies increase linearly with laser intensity, the EWP chirp decreases linearly. Graphically, the curves shown in the left panel of figure \ref{fig:class-ti-tr} are stretched/compressed \emph{horizontally} by an increasing/decreasing laser intensity.

\item The driving laser frequency/wavelength has two main effects: The recollision energy scales as \mbox{$\propto\omega_0^{-2}$}, while the timing of the trajectories scales as  \mbox{$\propto\omega_0^{-1}$}. Thus, increasing the driving laser wavelength leads to quadratically increasing recollision energies and linearly decreasing  EWP chirp. However, it also leads to linearly increasing continuum electron excursion durations $\tau=t_\mathrm{r}-t_\rmi$, which gives the EWP more time to spread, thus decreasing the recollision amplitude \cite{Schiessl2007qpi,Tate2007,Shiner2009}. While using longer wavelength driving lasers is the most promising route to self-probing experiments with much increased EWP bandwidth, this drop in HHG efficiency represents a major difficulty. However, macroscopic effects may help to compensate at least partly for this drop \cite{Yakovlev2007,Colosimo2008Scaling, Popmintchev2009}.

\item There is a clear connection/competition between HHG and ionization. While higher laser intensity leads to an exponentially higher tunnelling rate and linearly increasing recollision energy, medium saturation (i.e. unity ionization probability already during a fraction of the driving laser pulse) sets a strict limit on the maximum usable intensity for HHG. This intensity not only depends on the ionization potential (as suggested by the barrier suppression described section  \ref{sec:tunnel}), but also on the laser pulse duration. The
longer the pulse, the more half-cycles add to the total ionization probability: of all electrons born in the continuum, only a fraction recollide again, and of these only a fraction recombines, while all others eventually drift far away from the ion and thus properly ionize it. This leads to the depletion of the medium ground state as well as the creation of a free-electron gas in the HHG medium causing considerable dispersion and thus degrading macroscopic phase matching conditions (cf. section \ref{sec:phasematch}).
\end{itemize}

Based on what was just introduced, it is now quite easy to extend the model to more amusing laser fields synthesized using several colour-components. Since integration is a linear functional, one can simply sum up the trajectories calculated with (\ref{eq:theory:classtraj}) separately for each colour component, and then search for recollisions. Similarly, one can easily extend this model to two dimensions and treat laser fields with shaped polarization state. One may then allow for a non-zero initial velocity of the electron transverse to the tunnelling direction (which corresponds to the lateral spread of the EWP during its propagation) in order to find closed trajectories that recollide with the core. The probability for such a component does, however, drop quickly and so does the recombination amplitude.

		\subsubsection{Quantum mechanical}\label{sec:sfa}
\paragraph{Strong Field Approximation.\\}
Instead of a classical electron flying along a trajectory and releasing a flash of light as it bounces back on its parent ion, the more appropriate picture is an electron wavefunction, initially bound in an atom or molecule, which is drastically deformed by a strong laser field. Part of the wavefunction is pulled away from the binding potential through the classically forbidden barrier and eventually interferes with the part left in the bound state. The simplest version of a fully quantum mechanical model is the SFA, or Lewenstein model \cite{[*][]Lewenstein1994Theory}. We will give a brief guide through the SFA with the aim of clearly pointing out the approximations made and how the connection to the classical dynamics of the previous section can be recovered.

For a single active electron, the TDSE reads (in length gauge):
\begin{equation}
	\rmi \frac{\partial}{\partial t} \psi(\bfr, t) = \left[ -\frac{1}{2}\nabla^2 + V_0(\bfr) + \bfr\cdot \mathbi{E}(t) \right] \psi(\bfr, t) \,,
\label{eq:theory:sfatdse}
\end{equation}
where $\mathbi{E}(t)$ is the electric field of the laser and $V_0(\bfr)$ represents the interaction of the electron with the nuclei shielded by the remaining bound electrons, which will in the following be referred to as \textit{the core}. Initially, the molecule is supposed to be in its ground state, i.e. $\psi(\bfr, t=0)$ is given by the orbital, $\psi_0(\bfr)$, of our active electron in this ground state. Due to the exponential sensitivity of the tunnelling rate (\ref{eq:ADKrate}) on the binding energy, this will be one of the energetically highest occupied  orbitals. Note that we treat the laser field and its interaction with the electron classically, justified by the high field strengths considered and the associated high photon numbers per unit volume. 
	
Classical electro-dynamics tells us that the radiated XUV spectrum $\tilde{\epsXUV}(\omega)$, where $\omega$ is the XUV light frequency, is given by the Fourier transform $\mathcal{F}_{t\rightarrow\omega}$ of the dipole acceleration, i.e. of the electron acceleration times its charge~\cite{[\stef{Actually, this relation, taken as `obvious' until very recently, is being challenged by }][\stef{, who rather propose that it is the dipole \emph{velocity} that directly relates to the harmonic field. This would only change equations (\ref{eq:theory:speclen})--(\ref{eq:theory:specacc}), but otherwise have no further consequences for the equations and argumentations to follow in this tutorial.}]Baggesen2011Onthe}. As there is in principle no ``acceleration operator'' in quantum mechanics, let us just use the double time derivative of the electron position:
\begin{align}
	\tilde{\epsXUV}(\omega)& = \mathcal{F}_{t\rightarrow\omega}\left[ -\frac{\rmd^2}{\rmd t^2} \braket{\psi \left\vert \hat{\mathbf{r}} \right\vert \psi } \right] 
			= \omega^2  \mathcal{F}_{t\rightarrow\omega}\left[\braket{\psi \left\vert \hat{\mathbf{r}} \right\vert \psi } \right], \label{eq:theory:speclen}
\end{align}
where, for the time derivatives, we have used the differentiation theorem for the Fourier transform. With the Ehrenfest theorem, this can be transformed to:
\begin{align}
	\tilde{\epsXUV}(\omega)&= \mathcal{F}_{t\rightarrow\omega}\left[ -\frac{\rmd}{\rmd t}\braket{ \psi \left\vert \hat{\mathbf{p}} \right\vert \psi } \right] 
				=\left.\left. -\rmi\omega  \mathcal{F}_{t\rightarrow\omega}\right[\braket{\psi \left\vert \hat{\mathbf{p}} \right\vert \psi} \right], \label{eq:theory:specvel} \\  & \nonumber\\
			& = \mathcal{F}_{t\rightarrow\omega}\left[ -\braket{ \psi \left\vert -\nabla V(\hat{\bfr}) \right\vert \psi } \right] 
				= \left.\left.-\mathcal{F}_{t\rightarrow\omega}\right[\braket{\psi \left\vert \hat{\mathbf{a}} \right\vert \psi }\right],
	\label{eq:theory:specacc}
\end{align}
where 
we can recognize {\em a posteriori} an acceleration operator $\hat{\mathbf{a}}=-\nabla V(\hat{\bfr})$, with $V(\hat{\bfr})=V_0(\hat{\bfr})+ \hat{\bfr}\cdot \mathbi{E}(t)$.
We thus find three different ways to compute the complex XUV spectrum, commonly referred to as `length', `velocity' or `acceleration' form for (\ref{eq:theory:speclen}), (\ref{eq:theory:specvel}) and (\ref{eq:theory:specacc}), respectively. At this point, all three forms are equivalent and give the same result, irrespective of the basis on which $\ket{\psi}$ is represented.

The direct numerical solution of (\ref{eq:theory:sfatdse}) is nowadays possible, at least for atoms and simple molecules, using e.g. a pseudo-potential for $V_0(\bfr)$. Suitable approximations can, however, make possible a fully analytical solution which will make it easier to shed light on the physics involved than an interpretation of a numerical solution could. Such an approximative solution has been demonstrated by Maciej Lewenstein and coworkers \cite{Lewenstein1994Theory} shortly after the classical three-step model was proposed. The derivation is based on the strong-field approximation of the TDSE \cite{Keldysh1965,[][]Faisal1973sfa,*Reiss1980Effect,*Reiss1990Complete}, which, additionally to the single active electron, makes the following assumptions:
\begin{enumerate}
	\item Of the bound states, only the field-free ground state of the atom/molecule is considered, all other excited bound states are neglected.
	\item The influence of the core-potential $V_0(\bfr)$ on the electron in the continuum is neglected, i.e. $V_0(\bfr) \ll \bfr\cdot \mathbi{E}(t)$ for the continuum electron.
\end{enumerate}
The laser field has to be sufficiently strong for the second assumption to hold, and of sufficiently low frequency for the first. These conditions overlap with those defining the tunnelling regime (cp. section \ref{sec:tunnel}). Assuming for simplicity the ground state depletion to be negligible, we can now make the ansatz
\begin{equation}
	\ket{\psi(t)} = \rme^{\rmi \Ip t} \left( \ket{\psi_0} + \int \frac{\rmd^3 \bfk}{(2\pi)^3} a(\bfk,t) \ket{\bfk} \right)\,,
\label{eq:theory:sfaansatz}
\end{equation}
i.e. the electron is in a superposition of states: mainly in its bound state $\ket{\psi_0}$ with energy $-\Ip$, but with the small time-dependent amplitudes $a(\bfk,t)$ also in continuum states $\ket{\bfk}$ (designated by their asymptotic momenta $\bfk$). The amplitudes $a(\bfk,t)$ are complex valued and their phases are defined relative to that of the stationary initial state, $\ket{\psi_0}\rme^{\rmi \Ip t}$. Introducing this ansatz into the TDSE (\ref{eq:theory:sfatdse}) and projecting onto
$\ket{\bfk}$ transforms the TDSE into an equation for $a(\bfk,t)$. The latter can be solved analytically when we choose the continuum states to be free-particle states, $\braket{\bfr\vert\bfk}=\rme^{\rmi\bfk\cdot\bfr}$, i.e. assume them to be eigenstates of the truncated Hamiltonian with $V_0(\bfr)$ omitted~\footnote{\stef{These time-dependent eigenstates are also called Volkov waves. In (\ref{eq:theory:sfaansatz}), their time-dependence is implicitly included in the amplitudes  $a(\bfk,t)$.}} (cf. assumption 2). Having found $a(\bfk,t)$, we can write the time-dependent electron wavefunction $\psi(\bfr,t)$ which contains the complete information about the system. The detailed derivation can be found in the original paper \cite{[*][]{Lewenstein1994Theory}} treating atoms, or with a focus on molecules, e.g., in \cite{[** Many rigorous and detailed derivations concerning self-probing and HHG can be found in: ][]Elmarthesis}. 

\vspace{6pt}
\paragraph{From the wavefunction to the harmonics: the radiating dipole.\\} \label{sec:dipoleform}
We can now choose to either calculate the dipole moment, $\braket{\psi\vert \hat{\mathbf{r}}\vert \psi}$, the dipole momentum, $\braket{\psi\left\vert \hat{\mathbf{p}} \right\vert \psi}$, or the dipole acceleration, $\braket{\psi \left\vert \hat{\mathbf{a}} \right\vert \psi}$, and then use (\ref{eq:theory:speclen}), (\ref{eq:theory:specvel}) or (\ref{eq:theory:specacc}) to calculate the complex XUV spectrum $\epsXUV(\omega)$ radiated by a single molecule. Let a general ``dipole operator'', $\hat{\mathbf{d}}$, stand for any of the three we can choose from. The time dependent dipole expectation value, $\mathbi{d}(t)=\braket{\psi(t)\vert\hat{\mathbf{d}}\vert\psi(t)}$, then writes:
\begin{multline}
	\mathbi{d}(t)=-\rmi\int_0^t \rmd t_\rmi \int\rmd^3\mathbi{p}\: \mathbi{d}_\mathrm{rec}[\mathbi{p}+\mathbi{A}(t)] \: \exp[\rmi S(\mathbi{p},t_\rmi,t)] \\
\times d^\mathrm{L}_\mathrm{ion}[\mathbi{p} + \mathbi{A}(t_\rmi),t_\rmi]\, + \mathrm{c.c.},
\label{eq:theory:sfadipole}
\end{multline}
where $\mathbi{p}=\bfk - \mathbi{A}(t)$ is a ``drift momentum'' of the continuum electron (a conserved quantity during propagation in the continuum because $V_0(\bfr)$ is neglected) and  \mbox{$\mathbi{A}(t)=-\int_{-\infty}^t \mathbi{E}(t') \rmd t'$} is the vector potential of the laser field. Note that in (\ref{eq:theory:sfadipole}), only bound-continuum cross-terms are considered. Continuum-continuum transitions, which should be very weak since $a(\bfk,t)$ is very small, are omitted, as is the time-independent $\braket{\psi_0\vert\hat{\mathbf{d}}\vert\psi_0}$. The latter anyway vanishes for bound states with defined parity as is the case for atoms and symmetric molecules. We recover the three steps of the classical approach as three factors in the integrand as follows:

\emph{(i)}~At time $t_\rmi$, part of the electron wavefunction makes a transition to a continuum state with momentum $\bfk=\mathbi{p} + \mathbi{A}(t_\rmi)$, the transition amplitude for which is 
\begin{equation}
d^\mathrm{L}_\mathrm{ion}(\bfk,t_\rmi) = \mathbi{E}(t_\rmi)\cdot \braket{\bfk\vert\hat{\bfr}\vert\psi_0(\bfr)}.
\label{eq:dion}
\end{equation}
The ionization amplitude, $d^\mathrm{L}_\mathrm{ion}(\bfk,t_\rmi)$, already shows up in the expression of the amplitudes $a(\mathbi{k},t)$ after the approximate resolution of the TDSE. It contains a dipole matrix element (DME) in \textit{length form} since the operator here is not selected by our choice of $\hat{\mathbf{d}}$, but comes from the length gauge interaction Hamiltonian $\bfr\cdot \mathbi{E}(t)$ in the TDSE (\ref{eq:theory:sfatdse}).

\emph{(ii)}~In the continuum, the electron propagates under the influence of the laser field only, acquiring a phase relative to the ground state, which equals the semiclassical action
\begin{equation}
	S(\mathbi{p},t_\rmi,t) = -\int_{t_\rmi}^t \rmd t'' \left[ \frac{[\mathbi{p} + \mathbi{A}(t'')]^2}{2} + \Ip \right].
\label{eq:theory:sfaaction}	
\end{equation}

\emph{(iii)}~At time $t$, the electron has a mechanical momentum $\bfk=\mathbi{p} + \mathbi{A}(t)$ and recombines with the core, the amplitude of which is given by the DME
\begin{equation}
	\mathbi{d}_\mathrm{rec}(\bfk) = \braket{\psi_0\vert\hat{\mathbf{d}}\vert\bfk},
\label{eq:drec}
\end{equation}
the only term in (\ref{eq:theory:sfadipole}) depending on the choice for $\hat{\mathbf{d}}$. 

Since the time propagation within the SFA is not rigorously consistent with the system's hamiltonian,
the length (\ref{eq:theory:speclen}), velocity (\ref{eq:theory:specvel}), and acceleration forms (\ref{eq:theory:specacc}) of the radiating dipole are no longer equivalent and one cannot say which of the three is \emph{a priori} the best choice. For a long time, it was the length form, i.e. (\ref{eq:theory:speclen}), that was used almost exclusively, including Lewenstein's seminal paper \cite{Lewenstein1994Theory}. While one can find arguments to prefer one form or the other---Gordon et al. \cite{Gordon2005Quantitative} argue in favour of the acceleration form whereas Chiril\u{a} et al. \cite{[*][]Chirila2007Assessing} showed the velocity form to give reliable results---this issue is at heart a result of the approximations made in the SFA, of which the most severe one is certainly the plane-wave approximation.

According to (\ref{eq:theory:speclen}) to (\ref{eq:theory:specacc}), the complex XUV spectrum, $\epsXUV(\omega)$ is in any case proportional to the Fourier transform of (\ref{eq:theory:sfadipole}):
\begin{align}
\epsXUV(\omega)&\propto \int \mathbi{d}(t)\: \rme^{\rmi\omega t} \rmd t  \nonumber\\
	&= -\rmi \int \rmd t \int_0^t \rmd t_\rmi \int \rmd^3\mathbi{p}\:  \mathbi{d}_\mathrm{rec}[\mathbi{p}+\mathbi{A}(t)] \nonumber \\ 
	&\quad\times   \exp[ \rmi\omega t +\rmi S(\mathbi{p},t_\rmi,t)] d^\mathrm{L}_\mathrm{ion}[\mathbi{p} + \mathbi{A}(t_\rmi),t_\rmi] \,, 
\label{eq:theory:dipolespec} 
\end{align}
where integration over $t$ runs over the duration of the driving laser pulse, or, for a monochromatic driving laser, from $-\infty$ to $+\infty$. Note that here, we dropped the ``c.c.'' from (\ref{eq:theory:sfadipole}), i.e. we Fourier transformed a complex valued dipole and in order to obtain the radiated electric field,  on has to take twice the real part of $\mathcal{F}_{\omega\rightarrow t}[\epsilon(\omega)]$ \footnote{In discarding the ``c.c.'' from (\ref{eq:theory:sfadipole}), we discarded the negative frequency components of the spectrum of $\mathbi{d}(t)$. As $\mathbi{d}(t)$ is a purely real-valued quantity, this implies no loss of information at all since the spectrum of a real-valued function has Hermitian symmetry. Also, with a spectrometer, we anyways can only measure positive frequency components.}.

Equation (\ref{eq:theory:dipolespec}) is an integral over infinitely many quantum paths, i.e. triplets of drift momentum, $\mathbi{p}$, ionization times, $t_\rmi$ and recombination times, $t$, which makes its evaluation in general very costly.

\vspace{6pt}
\paragraph{Saddle-point approximation.\\}\label{sec:saddlepoint}
The quintuple integral in (\ref{eq:theory:dipolespec}) can be drastically simplified, and the analogy to the classical model can at the same time be driven further, by realizing that those contributions for which the phase
\begin{equation}
	\tilde{S}[\omega,(\mathbi{p},t_\rmi,t)]=\omega t + S(\mathbi{p},t_\rmi,t)
\label{eq:theory:saddlepoint-action}
\end{equation}
is stationary with respect to the variables $(\mathbi{p},t_\rmi,t)$ will largely dominate, whereas a rapidly varying phase will make the contributions of most quantum paths interfere destructively. In analogy to the classical principle of stationary action, one can thus find three equations, corresponding to the derivative of $\tilde{S}$ with respect to the variables $\mathbi{p}$,  $t_\rmi$ and $t=t_\mathrm{r}$, at constant $\omega$:
\begin{align}
	\int_{t_\rmi}^{t_\mathrm{r}}[\mathbi{p}+\mathbi{A}(t')]\:\rmd t'& = 0 \label{eq:theory:saddlepointp}\\
	\frac{[\mathbi{p}+\mathbi{A}(t_\rmi)]^2}{2} + \Ip & = 0 \label{eq:theory:saddlepointti} \\	
	\frac{[\mathbi{p}+\mathbi{A}(t_\mathrm{r})]^2}{2} + \Ip & = \omega \label{eq:theory:saddlepointtr}
\end{align}
Solving these three coupled equations yields triplets, 
$(\mathbi{p}^\mathrm{s},t_\rmi^\mathrm{s},t_\mathrm{r}^\mathrm{s})$, defining \emph{saddle point trajectories}. Note that these are complex trajectories due to tunnel ionization being classically forbidden: (\ref{eq:theory:saddlepointti}) can only be fulfilled with purely imaginary initial velocities. The trajectories can be visualized, e.g., by plotting the real parts of ionization and recombination times as a function of the XUV photon energy, shown in figure \ref{fig:sfa+class-ti-tr}---the connection to the classical trajectories is obvious. The simple classical model turns out to be in reasonable agreement but, obviously, the more rigorous quantum-mechanical calculation yields a more precise description, notably in the cut-off region.

\myfigure{1}{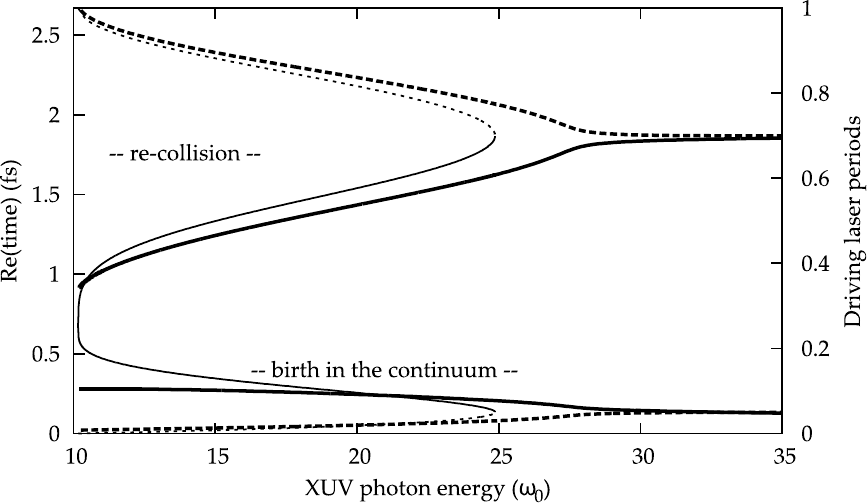}{Real part of ionization and recombination times of saddle-point trajectories as a function of the harmonic order, i.e. the emitted XUV photon energy in units of the driving laser photon energy, obtained by solving the coupled (\ref{eq:theory:saddlepointti}) to (\ref{eq:theory:saddlepointtr}) for HHG in N$_2$, $\Ip=15.6\:$eV, with an 800-nm laser and an intensity of \mbox{$I=1.2\times10^{14}\:$W cm$^{-2}$}. The driving laser field has a cosine time-dependence, i.e. time zero marks the field maximum. Full and dashed lines correspond to the short and long trajectories, respectively. The thin lines shown again the results of the classical calculation from figure \ref{fig:class-ti-tr}.}

The saddle-point trajectories are a finite number of quantum paths contributing to each frequency component of the atomic/molecular dipole and thus of the XUV emission. The different quantum paths are ordered according to the real part of the continuum electron excursion duration, $\tau^\mathrm{s}_n=t^\mathrm{s}_{\mathrm{r},n} - t^\mathrm{s}_{\rmi,n}$, and the first two of these, $\tau_1$ and $\tau_2$, shown in figure \ref{fig:sfa+class-ti-tr}, can be identified as the short and long trajectories found in the classical treatment. With the finite number of saddle-point trajectories, (\ref{eq:theory:dipolespec}) can be re-written as a discrete sum \cite{Lewenstein1995,Sansone2004nonadiabatic}:
\begin{align}
	\epsXUV(\omega) =& \sum_n \epsXUV_n(\omega) \nonumber \\
			\propto&  -\rmi \sum_n  (\rmi \tau_n^\mathrm{s} /2 )^{-3/2}[\det(\tilde{S}'')]^{-1/2}\, \mathbi{d}_\mathrm{rec} [(\mathbi{p}^\mathrm{s},t_\mathrm{r}^\mathrm{s})_n]   \nonumber\\
		&\times  \exp\{\rmi\omega t_{\mathrm{r},n}^\mathrm{s} +\rmi S[(\mathbi{p}^\mathrm{s},t_\rmi^\mathrm{s},t_\mathrm{r}^\mathrm{s})_n]\} \:{d^\mathrm{L}_\mathrm{ion}} [(\mathbi{p}^\mathrm{s},t_\rmi^\mathrm{s})_n]   \,.
\label{eq:theory:saddlepointdipolespec}
\end{align}
The first pre-factor, containing the excursion duration, $\tau^\mathrm{s}_n$ is a result of the integration over $\mathbi{p}$ around the saddle point and expresses EWP spreading, which reduces contributions from trajectories that spend longer time in the continuum. The second pre-factor is the result of the saddle point integration over both $t_{\rmi}$ and $t_{\mathrm{r}}$ and involves the determinant of the $2\times2$ matrix of the second derivatives of $\tilde{S}$ with respect to these two variables evaluated at the saddle point. Note that the imaginary part of the stationary action, $ S[(\mathbi{p}^\mathrm{s},t_\rmi^\mathrm{s},t_\mathrm{r}^\mathrm{s})_n]$, provides the ADK tunnelling rate (\ref{eq:ADKrate}) accounting for the first step of the HHG process \cite{Ivanov1996,Ivanov2005Anatomy}. 

A detailed derivation and discussion of the saddle-point approximation refined for HHG in molecules can be found in \cite{Chirila2006sfa,Elmarthesis,Etches2010inducing}. This includes additional quantum trajectories in which the active electron is ionized at one atomic center within the molecule and recombines at another.

Note that it is the saddle point approximation that ultimately gives a physical meaning to the {\em individual} DMEs rather than to the the mean value of the electron acceleration, and through (\ref{eq:theory:saddlepointtr}) associates a given radiated frequency to an electron scattering wave with a well defined energy, thus recovering the energy conservation relation $\omega=k^2/2+I_p$ of the intuitive three-step model.

\subsubsection{Macroscopic high harmonic emission} \label{sec:macroscopic}
The XUV light that is measured and used in experiments is obviously not radiated by a single molecule but by an  HHG medium consisting of many emitters with a certain density profile $\rho(\bfr)$. This medium interacts with a focused laser beam with a transverse and longitudinal intensity distribution $I(\bfr)$. All emitters radiate according to the local laser intensity and phase, and the laser and XUV fields propagate in a dispersive medium. The \emph{macroscopic} XUV spectrum $\EXUV(\omega)$ is obtained by solving Maxwell's wave equation with a source term $\propto \rho(\bfr)\,\epsXUV(\omega)$. This calculation corresponds essentially to coherently summing up the contributions of all single-emitters in the medium. The macroscopic field can thus be obtained as~\cite{LHuillier1991}:
\begin{multline}
 \EXUV(\omega,\bfr') \propto \sum_n \int \frac{\exp\left[\rmi \omega c(\omega)^{-1} \vert\bfr'-\bfr\vert \right]}{\vert\bfr'-\bfr\vert} \\
	\times  \rho(\bfr)\:\epsXUV_n[\omega,I(\bfr)] \,\rmd^3\bfr \:\,
 \label{eq:theory:macrofield}
\end{multline}
where $ c(\omega)$ is the light phase velocity in the dispersive HHG medium. Interference is constructive mostly in the forward (i.e. driving laser propagation) direction and significant amplitude in the far field is obtained when the wave front mismatch between the newly generated field $\epsXUV$ and the phase front of the propagating field $\EXUV$ is minimized at each point in the medium. Much theoretical and experimental effort has been invested into approaching this condition and studying the effects caused by deviations from it, see e.g. \cite{Constant1999,LHuillier1991,Durfee1999,Balcou1992,Salieres1995,[***][]Gaarde2008Macroscopic,Ruchon2008Macroscopic}.

Since for increasing $n$ in (\ref{eq:theory:saddlepointdipolespec}), i.e. for classes of trajectories with increasing durations, the phase $\tilde{S}[\omega,(\mathbi{p},t_\rmi,t)]$ varies more and more rapidly with the laser intensity, phase matching is increasingly hard to achieve. A number of studies have shown that, consequently, the contribution of only a \emph{single} trajectory class can be retained in the macroscopic emission if phase matching is optimized for this class (see, e.g., \cite{Salieres2001}). Then, one term of the sum (\ref{eq:theory:macrofield}) will completely dominate. This is the most important and most pleasant effect of macroscopic phase matching: it cleans up the mess at the single emitter level created by the several interfering trajectory classes contributing to each spectral component of the molecular dipole. 

In self-probing experiments, one ultimately wants to access information on the single-molecule level. If phase matching were \emph{perfect} for the shortest trajectory class throughout the HHG medium and for the full XUV bandwidth, the macroscopic field, $\EXUV$, could be considered an `amplified true replica' of the single-molecule emission, \emph{restricted to the shortest trajectory}, i.e. $\EXUV\propto\epsXUV_1$. In realistic conditions, it is possible to achieve very good phase matching over a large bandwidth and, as recently shown theoretically in \cite{Jin2009,Jin2011medium}, single-emitter information can indeed be extracted from macroscopic HHG spectra. In particular the XUV spectral phases measured by different groups, including us, are generally in very good agreement with single-emitter theory, restricted to the shortest trajectory and calculated for some effective intensity close to the peak intensity of the driving laser pulse \cite{Mairesse2003Attosecond, Doumy2009,[**][]Varju2005,Goulielmakis2008SingleCycle}---at least this can be said for HHG in atoms, where theory is well-proven.

In section \ref{sec:phasematch}, we will motivate a few experimental strategies to ensure good phase matching in the experiment. A very instructive general review of macroscopic effects in HHG is given by Mette Gaarde in \cite{[***][]Gaarde2008Macroscopic}.

\subsection{Improving the dipole matrix elements} \label{sec:theoimprovements1}

The points to be briefly discussed in this and the following section are no minor details or concerns of purists---rather, the very base of the decoding methods discussed in section \ref{sec:decoding} depends upon them. 

\subsubsection{Improved description of the continuum}\label{sec:continuum}

As shown in paragraph \ref{sec:saddlepoint}, the saddle point resolution of the SFA equations provides, at least in principle, a one-to-one a correspondence between individual DMEs and the spectral components of the radiated dipole, thus conferring to the DME {\em the} leading role in most of the interpretations and exploitations of high harmonic spectra. Unfortunately, the DME expressions directly inherit from the most (in)famous yet necessary approximation of the SFA, namely the plane-wave approximation for the continuum. If the $\ket{\bfk}$ were exact eigenstates of the laser-field-free system, the {\em individual} DMEs in their different forms would verify \footnote{With $\hat{\mathcal{H}}\ket{\bfk}=\varepsilon_k$, $\hat{\mathcal{H}}\ket{\psi_0}=-\Ip$, and the commutator $[\hat{\mathbf{p}},\hat{\mathcal{H}}]=-\rmi \nabla V(\hat{\bfr})$, we can write: $\rmi\braket{\psi_0\vert-\nabla V(\hat{\bfr})\vert\bfk} = \braket{\psi_0\vert-\rmi\nabla V(\hat{\bfr}) + \hat{\mathcal{H}}\hat{\mathbf{p}}\vert\bfk}-\braket{\psi_0\vert\hat{\mathcal{H}}\hat{\mathbf{p}}\vert\bfk} = \braket{\psi_0\vert \hat{\mathbf{p}}\hat{\mathcal{H}}\vert\bfk}-\braket{\psi_0\vert\hat{\mathcal{H}}\hat{\mathbf{p}}\vert\bfk} = \varepsilon_k \braket{\psi_0 \vert \hat{\mathbf{p}} \vert \bfk} +\Ip\braket{\psi_0 \vert \hat{\mathbf{p}} \vert \bfk} = \omega \braket{\psi_0 \vert \hat{\mathbf{p}} \vert \bfk}$. The next step is analogous, using $[\hat{\mathbf{r}},\hat{\mathcal{H}}]=\rmi\hat{\mathbf{p}}$.}:
\begin{equation}
\braket{\psi_0\vert-\nabla V(\hat{\bfr})\vert\bfk}=-\rmi\omega\braket{\psi_0 \vert \hat{\mathbf{p}} \vert \bfk}=\omega^2\braket{\psi_0 \vert \hat{\bfr} \vert \bfk}.
\end{equation}
This however indeed fails with free-electron states, i.e. plane waves, confirming the plane-wave approximation as one of the most obvious weaknesses of the SFA, which is additionally very challenging to improve on.

This issue has been tackled in particular by Chii-Dong Lin, Robert Lucchese and coworkers: Their \emph{quantitative rescattering theory} \cite{Le2009QRS,[**][]Lin2010review} consists in improving the SFA result by replacing the recombination DME $\mathbi{d}_\mathrm{rec}(\bfk)$ by the complex conjugate of accurate photo-ionization DMEs from elaborate stationary scattering calculations. The probability amplitude terms for ionization and continuum propagation are calculated with the SFA (alternatively, the molecular ADK theory \cite{Tong2002moadk,*Zhao2011effect} is used for ionization). The results compare well with TDSE calculations for H$_2^+$ and allow to reproduce many experimental result for CO$_2$ and N$_2$. This approach is based on the detailed-balance principle, stating that photo-recombination is time-reversed photo-ionization and the corresponding matrix elements are complex conjugates of each other.

The applicability of detailed balance is, however, questioned by Smirnova, Ivanov et al.. As explained in \cite{[**][]Sukiasyan2010}, matrix elements from field-free \emph{stationary} scattering theory contain, in the language of electron trajectories, also complex multiple scattering events evolving over longer times. These will be particularly sensitive to the strong laser field and will probably not at all be accurately described by a field-free calculation. Very importantly, as seen in section \ref{sec:saddlepoint}, HHG is very selective to specific quantum trajectory classes and usually favours the \emph{shortest} ones. Therefore, the very beneficial ``filtering'' effect of phase matching in a macroscopic medium will likely also remove sharp continuum resonances and complex scattering trajectories. While this intuition is very much in favour of a simplified structure of the continuum, implementing it in a theoretical model is very difficult and a present field of research---see, e.g. \cite{Sukiasyan2010,Smirnova2007Coulomblaser}.

\myfigure{1}{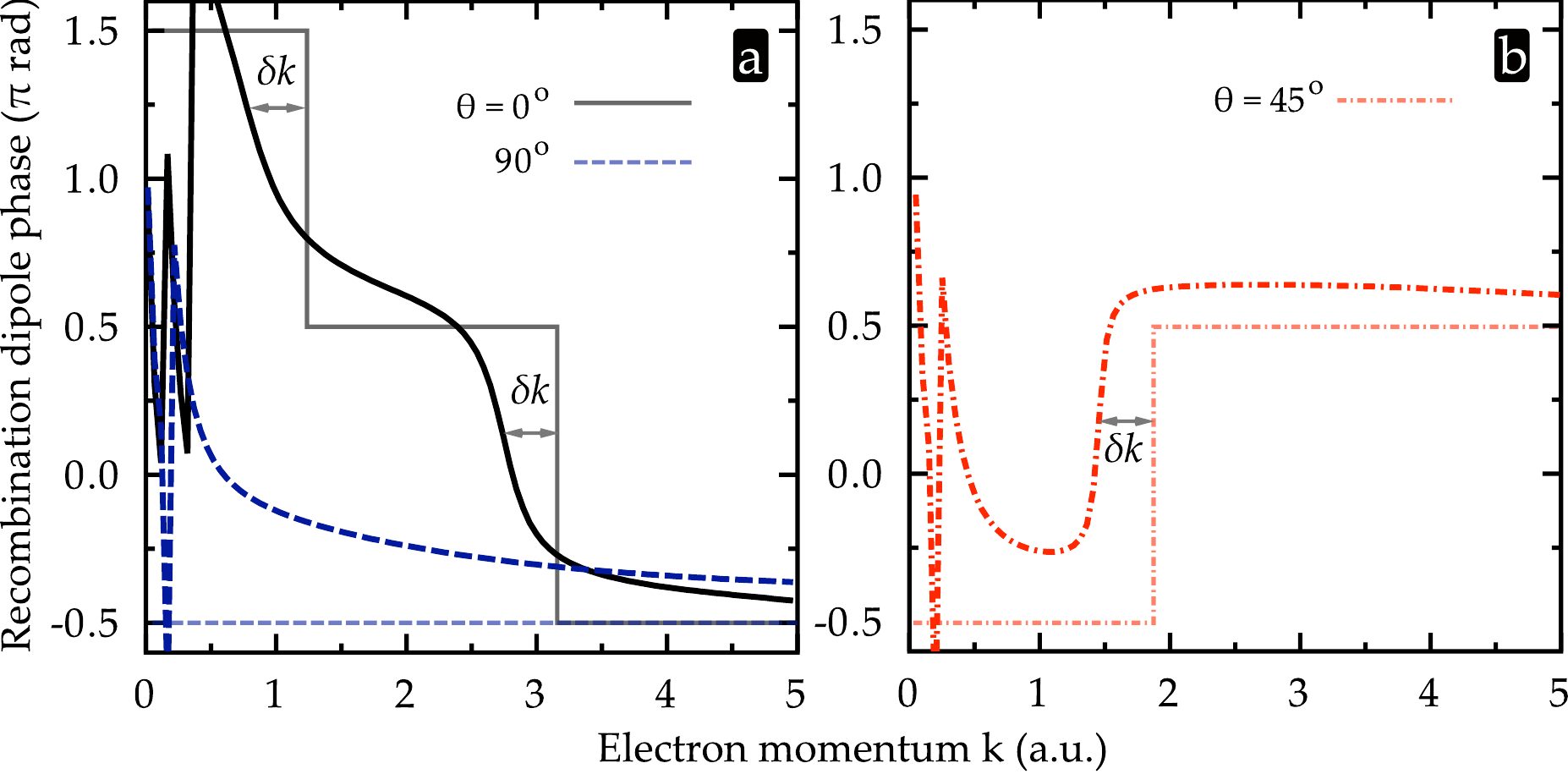}{Phases of the HOMO recombination dipole computed with Coulomb waves, at three different orientations $\theta=0^\circ,90^\circ$ (a) and  $\theta=45^\circ$ (b), versus (asymptotic) electron momentum $k$. Two values of the effective charge are considered for the Coulomb waves: $Z=0$, corresponding to plane waves (thinner, lighter lines), and $Z=1$, corresponding to the asymptotic charge of N$_2^+$ (thicker, darker lines).}
Our contribution to this effort was targeted at the influence of the ionic potential on the phase of the recombination DME \cite{Haessler2010tomo}. We calculated the (length form) recombination DME,  $\mathbi{d}_\mathrm{rec}(\bfk)$, using for $\psi_0$ the HOMO of N$_2$ from a Hartree-Fock calculation and for $\ket{\bfk}$ Coulomb waves, i.e. exact scattering states for the hydrogen atom \footnote{While the bound states of the hydrogen atoms are derived in virtually any quantum physics text book, it is not so common to find the scattering states. One text book that does treat them is \cite{[{*}][]Bransdenbook}}. This can be seen as a first order improvement on the plane-wave description, since asymptotically, the N$_2^+$ ion acts on the recolliding electron just like a proton. Figure \ref{fig:N2-HOMO+coulomb-waves} compares the phase of the recombination dipole resulting from this calculation to the phase of the corresponding plane-wave dipole for three different angles, $\theta$, of $\bfk$ relative to the molecular axis. The plane-wave dipole has a phase that is a multiple of $\pi/2$ for all $k$, i.e. it is purely imaginary valued. Sudden $\pi$ phase jumps indicate sign changes. Switching to Coulomb waves for an effective ion charge of $Z=1$, the phase of the dipole is completely ruined for $k\lesssim0.8\:$a.u. (corresponding to kinetic energies $\lesssim9\:$eV). The very rapid phase variation at these low momenta is a direct imprint left by the Coulomb waves, as can be seen from their partial wave expansion \cite{Bransdenbook}, where each angular momentum $\ell$ contributes with a phase $\arg [\Gamma(\ell+1+\rmi Z/k)]$. At higher momenta, including values typical for HHG, both series of curves show similar patterns, up to a phase shift with slow $k$-dependence and a translation $\delta k\approx0.4\:$a.u..

This is very good news: in particular the phase jumps, which are a direct manifestation of spatial structure of the HOMO (see section \ref{sec:imaging}), are clearly retained in the Coulomb-wave result. The global phase shift corresponds to the scattering phase. The observed momentum shift of the phase features translates to an energy shift of $\approx15\:$eV in the spectral region of ($0.9\:$a.u.$\lesssim k\lesssim2\:$a.u.) relevant to most of the experiments reported so far using 800-nm driving lasers---an energy shift that is very similar to typical bound state energies or ionization potentials of small molecules. This nicely corresponds to the idea that the recolliding electron, as it approaches the core, experiences an additional acceleration by the ionic potential \cite{Lein2007Molecular}. The typical energy gain due to this acceleration seems to be on the order of $\Ip$. This is why in the interpretation of self-probing experiments, as a first correction for the errors introduced by the plane wave approximations, one often modifies the relation between the recolliding electron wave number $k$ \emph{at the recollision instant} and the measured XUV photon energy $\omega$. The SFA-equation (\ref{eq:theory:saddlepointtr}), $\omega=k^2/2+\Ip$, is correct for the \emph{asymptotic} electron wavenumber far away from the core. To take into account the increase of the electron energy as it approaches the core, one just strikes out $\Ip$ to obtain the \emph{heuristic} relation:
\begin{equation}
	\omega=k^2/2.
\label{eq:heuristic}
\end{equation}

A very similar study has been done by Lein \emph{et al.} \cite{Ciappina2007Influence}, who calculated recombination dipoles with the H$_2^+$ bound state and either plane waves or two-center Coulomb waves, i.e. scattering states for the two-center Coulomb potential of H$_2^+$. The same differences as in our calculation were found: the Coulomb-waves lead to smoothed and slightly shifted phase jumps.

So far our discussion of an appropriate description of the continuum was focused on the (un)ability to use {\em exact} continuum eigenstates $\ket{\bfk}$ of the system, which would be selected by the energy conservation equation (\ref{eq:theory:saddlepointtr}). Now, the energy criterium is not sufficient to select properly the (even exact) continuum states, as the latter are infinitely degenerate in 3D. This raises the question of orientation, or angular momentum, distribution of the wave-packet within a given energy level $k$. The SFA with saddle-point approximation tells us, through (\ref{eq:theory:saddlepointp}), that the {\em mean} orientation of $\bfk$ must be parallel to the laser polarization. This is what intuition tells us too, but this information is not enough to fully characterize the wave-packet.

One can now choose to describe the continuum EWP with only a single $\bfk$-orientation, i.e. to include only the scattering states with asymptotic momenta parallel to the laser field. This (plus the plane-wave approximation), are basic assumptions needed to for the direct information retrieval schemes discussed in details in section \ref{sec:directretrieval}. There are, however, pathological cases where this limitation must fail to give a reasonable description of the EWP. When strong-field-ionizing a molecule from an antisymmetric orbital with a nodal plane parallel to the laser field, the continuum EWP will remain antisymmetric all along propagation and keep the nodal plane. Such a wavepacket will \emph{not} contain any scattering states with asymptotic momenta parallel to the field, although the \emph{mean} orientation of the momenta is indeed along this direction (this is also discussed in \cite{Smirnova2009pnas}). 

One possibility to find which $\bfk$ orientations / angular momenta to include in the EWP would be again to rely on the detailed balance principle and assume that that EWP expansion is given by the corresponding single-photon ionization probability amplitudes \cite{Le2009QRS}. Here again, this assumption is questionable owing to the fundamental differences between photoionization and strong field (tunnel) ionization---sudden and selective transition against quasistatic and laser driven propagation, see e.g. the discussion in \cite{Walters2008}. Ref. \cite{[*][]Higuet2011cooper} presents a very interesting and elaborate related study, based on the observation of a systematic shift between the spectral position of the Cooper minimum \cite{Cooper1962} in photoionization and HHG emission from argon atoms, explained by the particular shape of the recolliding EWP, thus formalizing experimentally the limitations of the detailed balance principle.

For the purpose of what follows, we conclude that the models working with a description of the continuum EWP containing only $\bfk$ parallel to the laser will work only if ionization is \emph{not} along a nodal plane in the initial bound state. For the most general case, models would have to include a far more involved description of the EWP.

\subsubsection{Multi-electron effects}\label{sec:multielectron}

The vast majority of models in strong-field physics make use of the single-active-electron approximation, which has given very satisfactory results for atoms. For example, in the SFA, from the very start, i.e. already in the ansatz (\ref{eq:theory:sfaansatz}), only one-electron wavefunctions are considered and the only interaction treated is that between the continuum electron and the laser field, while the remaining electrons are assumed to be completely frozen. For molecules, it is now frequently invoked that ``multi-electron effects'' will play a significant role. This refers to a rich variety of different phenomena, all demanding involved treatment of multi-electron dynamics, but arising essentially from two simple facts: 
\begin{itemize}
\item[\emph{(i)}] The laser field acts on \emph{all} electrons of the molecule, i.e. also on those that remain bound. It can thus polarize both the bound wavefunction in the neutral molecule as well as the molecular ion \cite{Jordan2008Corepolarization}; and it may excite the molecular ion on a sub-cycle time scale, changing the electronic configuration of the ion between ionization and recombination in HHG \cite{[*][]Mairesse2010Multichannel};

\item[\emph{(ii)}] The electrons of the molecule, including the one being ionized, interact among each other through Coulomb repulsion and they are indistinguishable, i.e. they are correlated. As an electron is being tunnel ionized, it interacts with the remaining electrons and may thereby electronically excite the ion \cite{Spanner2009oneelectron,Walters2010correlation}. When the continuum electron returns, it will again interact with the bound electrons in the ion, possibly exciting it \cite{[][]Sukiasyan2009,Sukiasyan2010}.
\end{itemize}
Since the computational cost of directly solving the TDSE increases exponentially with the dimensionality of the problem, such calculations are still limited to a maximum of 2 active electrons \cite{Tchitchekova2011molecular}. Of the methods able to deal with more electrons, multi-configuration time-dependent Hartree-Fock (MCTDHF) may be the most promising one, as it fully includes correlation but also allows to gradually ``switch it on and off'' by including more or less configurations \cite{[{* A comprehensive introduction to MCTDHF and the problem of treating correlated multi-electron systems is given in }][]Caillat2005correlated,[{* Results from a more recent 3D-version of the \mbox{MCTDHF} code are presented in }][]Jordan2008Corepolarization}. To date, strong-field theorists are just starting to gain qualitative and quantitative understanding of multi-electron phenomena. 

Here, we shall only briefly discuss how we should adapt the formulation of the recombination DME for multi-electron systems, and identify the approximation which leads back to the single active electron expression---a fair approximation for atoms and molecules presenting little relaxation upon ionization. A first major step was presented in 2006 simultaneously by Patchkovskii et al. \cite{Patchkovskii2006High,[**][]Patchkovskii2007} as well as Santra and Gordon \cite{[*][]Santra2006} with an insightful generalization of the SFA three-step model for HHG to multi-electron systems. In order to keep the notation from becoming cluttered, we leave away the spin-coordinate everywhere---the remainder of this section should be taken as a sketch of the essentials instead of a rigorous derivation, which can instead be found in different versions in \cite{Patchkovskii2007,Santra2006,Sukiasyan2010}. The general expression for the multi-electron recombination DME reads:
\begin{equation}
	\mathbi{d}_\mathrm{rec} = \braket{\Psi_0\vert { \sum_{m=1}^f }\hat{\mathbf{d}}_m \vert\Psi_\bfk},
\label{eq:multieldrec}
\end{equation}
where $\ket{\Psi_0}$ and $\ket{\Psi_\bfk}$ are the $f$-electron ground and ionized states, respectively---the latter is labelled by the released electron's asymptotic momentum $\bfk$.

We further assume the ionized state to be factorized in the form $\Psi_\bfk(\bfr_1,\ldots,\bfr_f)= \hat{A}\Phi^+(\bfr_1,\ldots,\bfr_{f-1})\xi_\bfk(\bfr_f)$, where $\Phi^+(\bfr_1,\ldots,\bfr_{f-1})$  is the $(f-1)$ electron wavefunction of the ionic core, $\xi_\bfk(\bfr)=\braket{\bfr | \bfk}$ is the associated continuum electron's wavefunction, and the operator $\hat{A}$ adds all antisymmetric permutations of the electron coordinates. 

Note that factorizing  $\ket{\Phi^+}$ and $\ket{\bfk}$ implies that the continuum electron is \emph{de}correlated from the ionic core. This approach is reasonable once the electron has reached the continuum with a relatively high energy, but it cannot account for the interaction of continuum electron and ionic core during tunnel ionization or recombination. The anti-symmetry of the total wavefunction  ensures, however, that exchange correlation is included (i.e. indistinguishability and the Pauli exclusion principle are properly accounted for). Also, at this point, $\ket{\bfk}$ is taken as an exact mean-field scattering state with energy $\varepsilon_k=k^2/2$, i.e. it accounts for interaction with the mean potential due to the ionic electrons.

Using permutation antisymmetry of the wavefunctions and the symmetry of the multi-electron dipole operator, the recombination DME can now be written as:
\begin{equation}
	\mathbi{d}_\mathrm{rec} = \braket{\psi^\mathrm{D} \vert \hat{\mathbf{d}}_f \vert \bfk} + \mathbi{d}^\mathrm{ex}.
\label{eq:tomo:multiDME}
\end{equation}
Here, $\mathbi{d}^\mathrm{ex}$ is an (at this point somewhat opaque) ``exchange correction'' term and
\begin{multline}
\psi^\mathrm{D}(\bfr_f)=\sqrt{f}\int d\bfr_1\ldots\int d\bfr_{f-1} \\
\left[\Psi_0(\bfr_1,\ldots,\bfr_f)\right]^\star \,\Phi^+(\bfr_1,\ldots,\bfr_{f-1}) \label{eq:dysondef}
\end{multline}
is the \emph{Dyson-orbital}---the scalar product of the multi-electron wavefunctions of the neutral and the ionic core, where integration runs over the first $(f-1)$ electron coordinates. Calculating the Dyson orbital means projecting out the difference between the neutral and the ionic core, which can be seen as a \emph{hole in the ion} \cite{Smirnova2009pnas,Smirnova2009co2,Haessler2010tomo}. For an alternative, more general, definition of the hole (density) see \cite{Luennemann2010ultrafast,Breidbach2005}. 

The exchange correction term becomes more transparent in the Hartree-Fock framework \cite{[*][]Bransdenbook,[*][]Piela2007book}, where a multi-electron wavefunction is expressed as a single antisymmetrized product of orthonormal single-electron wavefunctions, i.e. orbitals \footnote{A more compact way of writing this antisymmetrized product is a Slater determinant.}:
\begin{align}
	\Psi_0 (\bfr_1,\ldots,\bfr_f) &= \hat{A}\psi_1(\bfr_1) \ldots \psi_f(\bfr_f) \nonumber\\
	\Phi^+ (\bfr_1,\ldots,\bfr_{f-1}) &= \hat{A}\phi_1(\bfr_1) \ldots \phi_{f-1}(\bfr_{f-1}).  \nonumber
\end{align}
Note that this implies that the scattering state $\ket{\bfk}$ is orthogonal to the orbitals $\ket{\phi_m}$ forming the ionic core state. The exchange correction term can now be written as
\begin{equation}
	\mathbi{d}^\mathrm{ex}=\sum_{m=1}^f  (-1)^{f+m} \braket{\Phi^+ \vert \hat{\mathrm{d}_1}\vert \Psi^{(m)}_0} \braket{\psi_m\vert\bfk}, \label{eq:hfexchangeterm}
\end{equation}
where
\begin{multline}
\Psi^{(m)}_0 (\bfr_1,\ldots,\bfr_{f-1}) =  \\
\hat{A}\psi_1(\bfr_1) \ldots \psi_{m-1}(\bfr_{m-1})\psi_{m+1}(\bfr_{m}) \ldots \psi_{f}(\bfr_{f-1}) \nonumber 
\end{multline}
is the $(f-1)$-electron Hartree-Fock wavefunction of an unrelaxed ion, obtained by simply removing an electron from the $m^\mathrm{th}$ orbital of the \emph{neutral} molecule.

What we learn from (\ref{eq:hfexchangeterm}) is that the exchange correction depends on the dipole matrix elements between the actual ion and the unrelaxed ion, as well as on the overlap of the continuum electron state, $\ket{\bfk}$,  with all bound orbitals, $\ket{\psi_m}$, of the \emph{neutral}. Knowing that $\ket{\bfk}$ is orthogonal to the $\ket{\phi_m}$, which build the \emph{ion}, we can conclude that if the $\ket{\phi_m}$ are very similar to the $\ket{\psi_m}$, the overlap $\braket{\psi_m\vert\bfk}$ will be very small and the exchange terms will be negligible. In other words, the degree of orbital relaxation upon ionization decides about the relevance of exchange terms. See also \cite{Sukiasyan2010} for a discussion of the physical interpretation of exchange terms.

In Koopmans' approximation \cite{Koopmans1934,Bransdenbook}, i.e. neglecting orbital relaxation upon ionization such that $\ket{\Phi^+}$ is built with the same orbitals as the neutral but $\ket{\psi_f}$ from which the continuum electron has been ionized, the exchange terms rigorously vanish, since $\ket{\bfk}$ is orthogonal to all $\ket{\psi_m}$. Also, the Dyson orbital becomes simply $\psi_f(\bfr)$, and the multi-electron recombination dipole (\ref{eq:tomo:multiDME}) becomes the same as the single-active-electron dipole (\ref{eq:drec}).

%
%
%
\subsection{Aspects relevant to HHG in molecules} \label{sec:theoimprovements2}

While the issues treated in the last section already exist for HHG in atoms, the following discusses some aspects of HHG in molecules, where there are nuclear degrees of freedom and the larger spatial extent of the binding potential leads to energetically closer lying bound states than in atoms.

\subsubsection{Nuclear dynamics}\label{sec:nuclear}
Vibrational and rotational dynamics in molecules typically take place on the $\sim$100-fs or ps time scale, respectively, which means that they are so slow compared to the electron dynamics of the HHG process, that nuclei are normally safely approximated as fixed. The lightest nuclei, such as protons or deuterons, can, however, be expected to move significantly \emph{during} the continuum electron excursion duration, $\tau$, in HHG, with typical values of half a driving laser period, e.g. $\tau\approx1.3\:$fs for an 800-nm driving laser.

It turns out to be fairly straightforward to include nuclear dynamics into the SFA model when the Born-Oppenheimer (BO) approximation is adopted. Essentially, the molecular dipole (\ref{eq:theory:sfadipole}) is modulated by the nuclear overlap integral \cite{Lein2005}:
\begin{equation}
	C(\tau) = \int \chi_0(\mathbf{R}) \chi(\mathbf{R},\tau)\:\rmd\mathbf{R},
\label{eq:nucloverlap}
\end{equation}
where $\chi_0(\mathbf{R})$ is the nuclear part of the ground state molecular wavefunction which is ionized in HHG, and $\chi(\mathbf{R},\tau)$ is the nuclear part of the wavefunction of the molecular ion, which has evolved during the excursion duration $\tau$. The coordinate $\mathbf{R}$ represents the nuclear configuration---in the H$_2$ model system studied in \cite{Lein2005}, this is simply the internuclear distance $R$. 

\myfigure{.55}{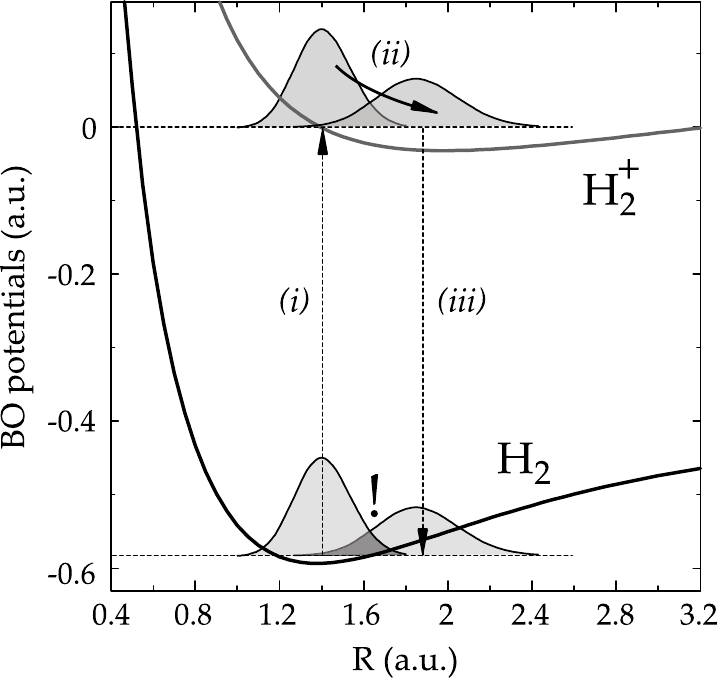}{Nuclear dynamics taking place while the electron undergoes the three steps of HHG in H$_2$, according to the model described in \cite{Lein2005}. The overlap of the initial and evolved nuclear wavefunctions at the recombination instant modulates the HHG amplitude.}
The appearance of the overlap integral (\ref{eq:nucloverlap}) can be understood in the following way, illustrated in figure \ref{fig:nucl-dyn}: 

\noindent \emph{(i)}~As the laser ionizes the molecule, a nuclear wavepacket is launched on the electronic (ground state) potential surface of the molecular ion, simultaneously with the continuum EWP. \emph{(ii)}~During the continuum excursion of the electron, the nuclear wavepacket evolves as well. \emph{(iii)}~At recollision, the system recombines with a certain probability into the ground state and emits an attosecond burst of XUV light. For \emph{coherent} emission, recombination has to lead back to the initial state~\footnote{The coherence of the whole HHG process is crucial so that many molecules in a macroscopic medium emit high harmonic radiation coherently, their contributions adding up to a macroscopic signal. Of course, the HHG process could, for instance, start with the molecule in the ground state and end with a vibrationally excited molecule. This excited state would, however, have an arbitrary phase relative to the continuum electron, which is `phase-locked' to the ground state and the light emission would consequently be incoherent and not participate to the HHG spectrum detected in experiments.}, 
the nuclear part of which is the vibrational ground state of the neutral molecule. The probability amplitude of this transition depends upon the overlap of the initial and evolved nuclear wavefunctions, i.e. recombination will be all the less likely the more both nuclear wavefunctions have become different.

This model has since been extended to more complex molecules, such as CH$_4$, and it has been shown that even for some molecules with heavier nuclei, significant modulation of HHG may occur \cite{[]][]Patchkovskii2009,[]][]Madsen2010}. Also, the longer the driving laser wavelength, the longer the excursion durations, the more time there is for the nuclei to move, and thus the stronger the effect to be expected. With mid-IR drivers, the nuclear motion during $\tau$ will thus in general have to be taken into account.

\subsubsection{Multi-orbital/channel contributions.}\label{sec:multiorbital}
Here, we remain in the multi-electron formulation of section \ref{sec:multielectron}. So far, ionization from a single orbital, $\psi_f$, has been considered---and if we say that the orbitals $\psi_1$ to $\psi_f$ are ordered by their energy, then we have actually just considered ionization from the HOMO. Or, in proper multi-electron terms, we have considered the ionic wavefunction, $\Phi^+$, to be the electronic ground state of the ion. Due to the exponential dependence of the tunnelling rate on $\Ip$, this had initially been thought to give the dominating contribution to HHG in any case. However, it turns out that valence orbitals are energetically much closer in molecules than in atoms, and therefore \emph{(i)}~the relative contribution of energetically lower lying orbitals, referred to as \mbox{HOMO-1,} \mbox{HOMO-2, \ldots~} can become significant; \emph{(ii)}~ionization from the HOMO can be strongly suppressed when the laser is polarized parallel to an orbital nodal plane, further reducing its dominance \cite{Smirnova2009co2,Smirnova2009pnas}; \emph{(ii)}~electron correlation can lead to valence excitation of the ion during tunnelling \cite{Spanner2009oneelectron,Walters2010correlation}. Excited states of the ion correspond, in the Hartree-Fock framework, to ions with an electron ``missing'' in energetically lower lying orbitals. Such \emph{channels}, usually named after the corresponding state of the ion (A, \mbox{B, \ldots}~ for the first, \mbox{second, \ldots}~ excited state) may thus indeed give contributions that are significant compared to the X-channel involving the ionic ground state. The different channels take the system from the same initial to the same final state---the neutral ground state---via different states of the ion. In the total amplitude of the HHG process, the individual channel amplitudes interfere.

In the multi-orbital/channel case, the $f$-electron state of the ionized system, $\ket{\Psi_\mathrm{ion}}$ is a superposition of ionic states \emph{each with their associated continuum electron}, e.g.:
\begin{equation}
	\ket{\Psi_\mathrm{ion}(t)} = \sum_c b_c \ket{\Phi_{c}^+}\otimes\ket{\bfk_c},
\label{eq:fullionwp}
\end{equation}
where the sum runs over the participating channels $c=\mathrm{X, A, B,}\ldots$. The amplitudes $b_{(c)}$ only contain the relative weight of the channels, initially set by the tunnelling process and possibly time-dependent if the channels are coupled, e.g. by the laser. Time-dependence arises from the various channel-dependent continuum states $\ket{\bfk_c}$ as well as from the ionic stationary states $\ket{\Phi_{c}}$ (with energies $\varepsilon_c$)---each of them oscillating with a phase $\exp[-\rmi\varepsilon_{c} t]$.
The recombination DME consequently writes as a sum over channels:
\begin{equation}
	\mathbi{d}_\mathrm{rec}(t) =  \sum_c b_{c} \left[\braket{\psi^\mathrm{D}_{c} \vert \hat{\mathbf{d}}_f \vert \bfk_{c}} + \mathbi{d}_{c}^\mathrm{ex}\right].
\label{eq:tomo:multichannelDME}
\end{equation}
What does the superposition in (\ref{eq:fullionwp}) imply for our idea of the hole in the ion? A rigorous answer requires analysis of the entanglement between between the ionic core and the continuum electron together with the measurement process, which is relevant to the degree of coherence in the ion. For a discussion of these aspects , we refer the reader to refs. \cite{Schoffler2008,Smirnova2009pnas,Pabst2011decoherence}. 

An intuitive picture can be found when ignoring entanglement and assuming the different continuum states, $\ket{\bfk_c}$, to have a large overlap, so that they can be approximated by the some average state $\ket{\bfk}$ for all channels. This does not seem unreasonable for the energetically very broad EWPs in HHG, but becomes questionable, e.g., when ionization channels involve orbitals with opposite symmetries with respect to the laser polarization direction \cite{Smirnova2009pnas}. Neglecting exchange terms in the DME (\ref{eq:tomo:multichannelDME}), we then find
\begin{equation}
	\mathbi{d}_\mathrm{rec}(t) = \Big \langle   \sum_c  b_c\psi^\mathrm{D}_{c} \Big\vert \hat{\mathbf{d}}_f \Big\vert \bfk \Big\rangle,
\label{eq:theo:holeDME}
\end{equation}
i.e. a recombination DME between a continuum electron, in state $\ket{\bfk}$, and a bound ionic wavepacket formed by the channel-specific Dyson-orbitals, $\ket{\psi_\mathrm{hole}}=\ket{\sum_c b_c\psi^\mathrm{D}_c}$---a \emph{time dependent hole in the ion}.  This picture provides a means to graphically represent the wavepacket (\ref{eq:fullionwp}) and to imagine it as a rapidly evolving hole in the ion, as done in refs. \cite{Smirnova2009co2,Haessler2010tomo,Mairesse2010Multichannel}.  Let us remind the reader, though, that this picture does not follow easily from the rigorous DME  (\ref{eq:tomo:multichannelDME}) and its validity can still be debated.

When laser-induced coupling of the channels can be neglected, the $b_c$ are constant and the hole-wavepacket dynamics are simply given by the free evolution of the coherent superposition of ionic eigenstates. In the simplest multichannel case where the superposition is restricted to two states separated by an energy difference $\Delta\varepsilon_c$, this results in a beating with a half period $\tau=\pi/\Delta\varepsilon_c$; i.e., for $\Delta\varepsilon_c>2\:$eV, this becomes shorter than a femtosecond and the hole density in the ion thus moves on an attosecond timescale!

The next level of sophistication is then to include coupling between the channels participating in HHG \cite{[*][]Mairesse2010Multichannel}, so that the hole in the ion is no longer evolving freely but is influenced by the laser field or possibly by the continuum electron. 

\section{Decoding the HHG signal} \label{sec:decoding}
Based on the theoretical understanding gained in section \ref{sec:theory}, we will now proceed to discuss how information on the molecule can be extracted from measurements of the properties of its high harmonic emission.

\subsection{Obtaining temporal resolution} \label{sec:decode:temporal}
%

\subsubsection{Pump--probe with laser pulses}
The most obvious way of making a time-resolved measurement is a classical pump--probe experiment with a first laser pulse (femtosecond IR or even shorter XUV) initiating some dynamics in the molecule and a second, delayed laser pulse driving HHG. The achievable time-resolution is obviously given by the duration of both laser pulses---in any case it will remain on the scale of a few cycles of the IR driving laser. This clearly allows to follow few-femtosecond dynamics by probing transient configurations with the HHG driving pulse; as reported, e.g., for nuclear dynamics in \cite{[**][]Wagner2006Inaugural,Li2008,Mairesse2008Polarizationresolved,[**][]Woerner2010}. Such experiments often pose the challenge that the pump laser pulse in general launches the dynamics only in a more or less important fraction of all molecules contributing to HHG. It is, however, possible to arrange for a preferential detection of the harmonic emission from the excited species. Ways to achieve this, will be mentioned in section \ref{sec:exp:highcontrast}. 

\subsubsection{Sub-laser-cycle pump--probe} \label{sec:decode:chirpencoded}
%
\paragraph{Ionization-induced dynamics.\\} 
Another regime of temporal resolution is entered if one considers the three steps of HHG as \emph{(i)} pump,\emph{(ii)} delay line, and \emph{(iii)} probe. This means that the dynamics are launched in the ion by the ionization process, which releases the continuum EWP for HHG. In this case, all molecules contributing to the HHG signal undergo the dynamics and the experimenter has no contrast problem at all. The pump-probe delay is set by the mean duration of the continuum electron trajectories associated with a certain electron energy span, and the probe pulse is the recolliding EWP. This concept was pioneered by Niikura at al. \cite{Niikura2003Probing} who traced the expansion of the D$_2^+$ molecular ion in its electronic ground state immediately after ionization with $\approx1\:$fs resolution. In this experiment, the probe process was not recombination and XUV emission---which is what this tutorial shall focus on---but excitation and subsequent dissociation of D$_2^+$ by electron impact.

If the driving laser pulse has a cosine-carrier wave, as considered in section \ref{sec:class}, there are two control knobs for the pump-probe delay here: the driving laser intensity and its carrier wavelength $\lambda_0$. Tuning the intensity only, the delay is finely adjustable, but in very limited range only: between $\approx0.4$ and 0.65 laser periods (see figures \ref{fig:class-ti-tr} and \ref{fig:sfa+class-ti-tr}). Niikura et al. thus used 4 different driver wavelengths between 800~nm and 1850~nm, which is possible with optical parametric amplifiers (OPA) \cite{Cerullo2003ultrafast}.

As for the pump-step, another ionization mechanism than tunnelling can be imagined. It obviously has to happen ``suddenly'' on a sub-laser-cycle timescale, so that attosecond XUV pulses come to mind---attosecond pulse \emph{trains} will work, if the train-periodicity is a multiple of that of the laser which drives the continuum electron dynamics. Theoretical studies have shown that, in principle, attosecond pulse trains could be used for controlling the first step of the HHG process \cite{Schafer2004Strong}, but only a few experimental attemps have been done so far \cite{Biegert2006control,Gademann2011attosecond}.

\vspace{6pt}
\paragraph{Chirp-encoded recollision.\\} 
A special sub-category of the above concept for sub-cycle pump-probe measurements is chirp-encoded recollision \cite{[**][]Baker2006Probing,*[***][]Bucksbaum2006pacerNV,[**][]Lein2005,[**][]Smirnova2009co2,*[***][]Vrakking2009newsviews,Niikura2005Mapping,[**][]Mairesse2003Attosecond}. One simply shrinks the ``certain electron energy span'' mentioned above to a sharp energy value and considers the excursion duration associated with a sharp spectral component of the EWP corresponding to the emission of a given harmonic order. When the experimental conditions are such that the HHG emission is strongly dominated by a single trajectory class, the inherent chirp of the re-colliding EWP (see section \ref{sec:class}) implies that there is a \emph{unique mapping} of the XUV frequency on the excursion duration, i.e. on the pump-probe delay. For example, for the short electron trajectories, higher harmonic orders are associated with longer electron excursion durations.

Each harmonic order thus provides a ``frame for the attosecond movie'' \cite{[**][]Smirnova2009co2,*[***][]Vrakking2009newsviews}: in a single EWP recollision, a whole number of frames is shot for a movie lasting about half a laser cycle. To increase the length of the movie or to move around the instants when the frames are shot, one has to modify the continuum electron trajectories by, e.g., changing the driving laser wavelength or intensity \cite{[**][]Torres2010revealing}.

This way of exerting control over the continuum electron trajectories and thus, e.g., the excursion duration associated with a certain return energy, can be generalized to \emph{laser waveform shaping} \cite{[*][]Goulielmakis2007,[*][]Kitzler2005, Mauritsson2009subcycle,Chan2011synthesis,[**][]Chipperfield2009}, i.e. the sculpting of the electric-field cycles below the envelope of ultrashort laser pulses. Changing the carrier wavelength is one fundamental way to do so, which is generalized to Fourier synthesis from several colour components. To date, this technology is still in its infancy, but it can be expected to become crucial for future progress in self-probing. This will become even clearer when we re-mention the potential of waveform shaping for several other issues in the remainder of this tutorial.

Note that when including the long trajectories, e.g. by separately analyzing the XUV beam on axis (vastly dominated by the short trajectory class) and off axis (dominated by the long trajectory class if phase matched), the ``length of the attosecond movie'' can almost be doubled.

\subsection{Indirect retrieval---making experiments and simulations converge}
When models are made more and more sophisticated, they quickly become (at least mathematically) so complex that one no longer finds human-friendly analytical expressions that connect an experimentally measurable quantity (or a set thereof) with the sought-for information on the molecule. In order to nonetheless retrieve the information, one will then try to reproduce measured data---at least qualitatively---with model calculations that result from either hypotheses on the model parameters or some sort of fitting procedure. The central term for this way of analyzing experiments is \emph{consistency} between experimental data and the model predictions. This becomes all the more convincing, if there is redundancy in the experimental data, i.e. if the same model parameter set allows to reproduce the dependencies of an experimentally measured quantity on different experimental parameters.

One has to realize, though, that this way of retrieving information either demands quite elaborate and accurate \emph{ab-initio} models, or sufficient \emph{a-priori} knowledge about the system under study, in order to keep the number of free parameters at a minimum and thus avoid the ``fitting an elephant''-effect \cite{Mayer2010elephant}. A very interesting general discussion of the task of ``extraction of physically relevant model parameters from the measured raw data'' is part of the essay \cite{[***][]Schwarz2006}.

\subsubsection{Retrieving dynamics of systems with known (static) spatial structure}  \label{sec:decode:olga}
Recent work of Smirnova, Mairesse et al. provides a first good example that will breathe some life into the last---admittedly very abstract---paragraph. The authors study the interference of multiple HHG channels as introduced in section \ref{sec:multiorbital}. Qualitative features in their experimental data, such as the spectral position of HHG intensity minima as a function of molecule alignment angle or driving laser intensity (for CO$_2$ \cite{[**][]Smirnova2009co2,*[***][]Vrakking2009newsviews}), or the harmonic orders and molecule alignment angles where the strongest ellipticity of the harmonics is observed (for N$_2$ \cite{Mairesse2010Multichannel}), are shown to be reproduced by their fairly elaborate model. While the spatial structure of the involved bound and continuum states are taken as known from theory, this agreement suggests that the model manages to describe with good accuracy the involved multi-channel dynamics, i.e. it gives realistic relative amplitudes and phases of the involved channels.

There is only one parameter that is considered as more or less free, or unknown, in the calculations: the relative phase, $\Delta\varphi$ of the channels acquired during the tunnelling process. It turns out that the measurements are only well reproduced when this ionization phase difference is $\Delta\varphi=0$ for the X- and B-channel in the case of CO$_2$, whereas $\Delta\varphi=\pi$  for the X- and A-channel in N$_2$. This result is much more interesting than it might seem at first sight, because $\Delta\varphi$ decides on the shape of the multi-channel Dyson orbital, i.e., since the channel-specific Dyson orbitals are known from theory, one now knows the shape of the hole in the ion created by tunnel ionization! 

\subsubsection{Extracting sub-laser-cycle time-dependent nuclear wavepackets} \label{sec:decode:nucl}
Another example for iterative retrieval based on a mathematically complex modelling of the measured quantity can be found in the work of Lein, Baker et al. \cite{Lein2005,Baker2006Probing,*[***][]Bucksbaum2006pacerNV}.  They reconstructed the expansion of the H$_2^+$ molecular ion in its electronic ground state immediately after ionization with $\approx75\:$as resolution over a time window of  $\approx600\:$as.

The analysis of the experiment is based on the extended SFA model mentioned in section \ref{sec:nuclear} and the nuclear overlap integral, $C(\tau)$, which modulates the molecular dipole and thus the single-molecule complex XUV spectrum. The \emph{intensity} of the macroscopic harmonic emission that is dominated by a single trajectory class is thus proportional to $\vert C(\tau)\vert^2$, which monotonically decreases with $\tau$ when the molecular ion expands. Making use of chirp-encoded recollision, one can now map the harmonic frequency to a pump-probe delay $\tau$. Since for the short trajectory, higher harmonic frequencies are mapped onto longer $\tau$, the harmonic spectral intensity measured for H$_2$ molecules thus  decreases faster with frequency than if there were no expansion of the ion. 

The conceptual difficulty of how to isolate the effect of $\vert C(\tau)\vert^2$ on the spectral intensity from all other factors shaping the measured macroscopic harmonic spectrum is solved by normalizing with a heavier isotope. In the BO approximation, the electronic parts of the wavefunctions of H$_2$ and D$_2$ molecules are identical. Taking the ratio of the harmonic spectral intensities measured first with D$_2$ and then with H$_2$ under equal conditions thus removes all the influence of this electronic part as well as the instrument spectral response. The measured ratio should thus be equal to $\vert C_\mathrm{D_2}(\tau) / C_\mathrm{H_2}(\tau)\vert^2$. With D$_2$'s nuclei being twice as heavy, the nuclear dynamics will be slower, and $\vert C_\mathrm{D_2}(\tau) / C_\mathrm{H_2}(\tau)\vert^2$ will be a monotonically increasing function of $\tau$, as long as the H$_2^+$ ion expands (the H$_2^+$ ground state is bound, i.e. the molecular ion will not dissociate but vibrate with a half-period of $\approx8\:$fs, and the dynamics observed here is merely the very first initial expansion).  In the experiments, this was indeed found to be the case.

In order to retrieve the nuclear dynamics from this measured ratio, a genetic algorithm was applied.  The algorithm iteratively finds a BO potential in which the evolution of the nuclear wavepacket is such that $\vert C_\mathrm{D_2}(\tau) / C_\mathrm{H_2}(\tau)\vert^2$, calculated by solving the TDSE with this BO potential, matches the measured ratio of harmonic spectral intensities. Once the algorithm has converged, one thus has reconstructed the time-dependent nuclear wavepacket in H$_2$ (and D$_2$). 

The same experiment was done for CH$_4$ and CD$_4$ molecules, with consistent results. The genetic algorithm was, however, not yet applied because the treatment of nuclear dynamics in such more complex molecules is more involved and computationally costly \cite{Patchkovskii2009,Madsen2010}.

The precision of the frequency--time mapping is of course of crucial importance for extracting information from the measured spectra. In all the above-mentioned studies, this mapping was calculated using the classical model of section \ref{sec:class}, assuming that the continuum electron dynamics remains unaffected by the nuclear dynamics in the ion. We have later confirmed this, at least partly, by measuring the recombination times in H$_2$ and D$_2$ that were found very similar \cite{Haessler2009H2D2}. \stef{In \cite{Haessler2009H2D2}, we also point out that $ C(\tau)$ is a complex valued quantity and thus also contributes a phase to the HHG amplitude \cite{[][]Diveki20xx}.}

\subsection{Direct retrieval} \label{sec:directretrieval}
In some (rare) cases, the underlying model allows to directly calculate the information on the molecule from measured data. This was already the case in the last paragraph if one was satisfied with knowing the squared ratio of nuclear overlap integrals---it is only to get the more tangible BO potentials and nuclear wavepackets that one finally has to resort to the iterative retrieval. In this section, we will show how the SFA or Lewenstein model discussed in section \ref{sec:sfa} leads to direct retrieval schemes for electronic structure and dynamics.

\subsubsection{Factorizing the HHG amplitude} \label{sec:factorize}
A first instructive insight can be gained from a calculation for laser field-free conditions \cite{mythesis,[*** ][]vanderZwan2008Molecular,Elmarthesis}, i.e. we ignore where the EWP comes from and how it got accelerated, but only consider a dipole formed by the interference of a bound part with energy $-\Ip$:
\begin{equation}
 \psi_0(\bfr,t) = \psi_0(\bfr) \rme^{\rmi \Ip t}\,,
\end{equation}
and a continuum plane-wave packet moving along the x-direction only:
\begin{equation}
 \psi_\mathrm{c}(\bfr,t)=\int_{-\infty}^{+\infty} \rmd k\: a(k) \rme^{\rmi[kx - (k^2/2)t]}\,,
\end{equation}
with complex-valued amplitudes $a(k)$, including, e.g., a chirp. This is analogous to the SFA-ansatz (\ref{eq:theory:sfaansatz}). Neglecting, as discussed in connection with (\ref{eq:theory:sfadipole}), bound-bound and continuum-continuum matrix elements, one finds an XUV spectrum
\begin{align}
	\tilde{\epsXUV}(\omega) \propto&  \:\mathcal{F}_{t\rightarrow\omega}\left[ \braket{\psi_0(\bfr,t)\vert \hat{\mathbf{d}} \vert \psi_\mathrm{c}(\bfr,t)} + \mathrm{c.c.}\right] \nonumber \\
	 =& \int\hspace{-2pt} \rmd t\: \rme ^{\rmi \omega t} \hspace{-2pt}\int\hspace{-2pt} \rmd k\: a(k)\, \rme^{-\rmi(k^2/2 + \Ip)t}  \braket{\psi_0(\bfr)\vert \hat{\mathbf{d}} \vert \rme^{\rmi kx} } \nonumber \\ 
	&+  \int\hspace{-2pt} \rmd t\: \rme ^{\rmi \omega t} \hspace{-2pt}\int\hspace{-2pt} \rmd k\: a^*(k)\, \rme^{\rmi(k^2/2 + \Ip)t}  \braket{\rme^{\rmi kx} \vert \hat{\mathbf{d}} \vert \psi_0(\bfr) }  \nonumber \\ 
	=&\: 2\pi \hspace{-2pt}\int\hspace{-2pt} \rmd k\: a(k)\, \braket{\psi_0(\bfr)\vert \hat{\mathbf{d}} \vert \rme^{\rmi kx}} \,\delta_\mathrm{D}(\omega - k^2/2 - \Ip) \nonumber\\
	&+  2\pi \hspace{-2pt}\int\hspace{-2pt} \rmd k\: a^*(k)\, \braket{\rme^{\rmi kx} \vert \hat{\mathbf{d}} \vert \psi_0(\bfr) }  \,\delta_\mathrm{D}(\omega+k^2/2 + \Ip) 
\end{align}
Obviously, the second term, which results from the complex conjugate in the first line, stands for the negative frequency components and we can omit it as we did in (\ref{eq:theory:dipolespec}). For the first term, the Dirac-$\delta$ function picks out two $k$-values from the integration interval $[-\infty,+\infty]$, namely $k=\pm\sqrt{2(\omega-\Ip)}$ and we obtain:
\begin{multline}
	{\epsXUV}(\omega) \propto  a(k)\braket{\psi_0(\bfr)\vert \hat{\mathbf{d}} \vert \rme^{\rmi kx}} + a(-k)\braket{\psi_0(\bfr)\vert \hat{\mathbf{d}} \vert \rme^{-\rmi kx}}, \\
\text{with } \omega=k^2/2 + \Ip\,.
\label{eq:simplefactorize}
\end{multline}
The complex XUV amplitude for frequency $\omega$ is thus a sum of two terms, describing EWP recollisions with $k$ and $-k$, i.e. from opposite sides of the molecule, each of the terms factorized into a complex EWP spectral amplitude and the recombination DME.

Obtaining a corresponding expression from a rigorous SFA-treatment, which includes the laser field, is possible via the saddle point approximation. In section \ref{sec:saddlepoint}, we have shown how the integrals in the full SFA expression for the single-molecule complex XUV spectrum can be made to collapse, to obtain a sum over (saddle-point) trajectories. We have then argued in section \ref{sec:macroscopic} that phase matching in a macroscopic medium will allow to select only one trajectory class to dominate the macroscopic emission; i.e. the XUV field reaching our detector will be proportional to only one term out of the sum in (\ref{eq:theory:saddlepointdipolespec}), usually the one for the short trajectories, $n=1$. We thus conclude that with the mentioned approximations, the expression for the measurable XUV field factorizes into two terms: the recombination dipole matrix element $\mathbi{d}_\mathrm{rec}$, and a complex continuum EWP amplitude:
\begin{equation}
\alpha(\bfk) =-\rmi (\rmi \tau_n^\mathrm{s} /2 )^{-3/2}[\det(\tilde{S}'')]^{-1/2} \exp[\rmi\tilde{S} ]\,  {d^\mathrm{L}_\mathrm{ion}}. 
\label{eq:factorizealpha}
\end{equation}
The Lewenstein model thus lets us derive an expression analogous to (\ref{eq:simplefactorize}). Note that in (\ref{eq:factorizealpha}), the argument $\bfk$ designates the electron wavevector at the recollision instant, $\bfk=\mathbi{p}^\mathrm{s} + \mathbi{A}(t^\mathrm{s}_\mathrm{r})$, whereas in the ionization DME, the electron wavevector at the ionization instant has to be used: $\bfk'=\bfk + \mathbi{A}(t^\mathrm{s}_\rmi) - \mathbi{A}(t^\mathrm{s}_\mathrm{r})$. Note also that in order to factor out the recombination dipole matrix element, it is actually sufficient to make the saddle-point approximation for the momentum, $\mathbi{p}$ and the  recombination time, $t_\mathrm{r}$, only, as shown in \cite{[*** ][]vanderZwan2008Molecular,Elmarthesis}. 

In order not to miss the second term for the recollision from the opposite side of the molecule, one has to be careful of a little stumbling block. When thinking of the three steps of the self-probing paradigm, one often thinks about a single recollision only. This is also what we did when discussing the solutions of the saddle-point equations (\ref{eq:theory:saddlepointti}) to (\ref{eq:theory:saddlepointtr}), visualized in figure \ref{fig:sfa+class-ti-tr}: we have only considered trajectories starting in one laser \emph{half}-cycle, i.e. $t_\rmi\in[0,T_0/2]$. Though, to grasp a valid picture, one obviously has to consider ionization over at least a full laser period, $T_0$. For the most common situation of a linearly polarized laser pulse with cosine-shaped carrier wave, a search for saddle points will find \emph{two} short trajectories: $(\mathbi{p}^\mathrm{s},t_\rmi^\mathrm{s},t_\mathrm{r}^\mathrm{s})$ and $(-\mathbi{p}^\mathrm{s},t_\rmi^\mathrm{s}+T_0/2,t_\mathrm{r}^\mathrm{s}+T_0/2)$. These correspond to the EWP recollisions with $\mathbi{p}^\mathrm{s}+\mathbi{A}(t_\mathrm{r}^\mathrm{s})=\bfk$ and $-\mathbi{p}^\mathrm{s}+\mathbi{A}(t_\mathrm{r}^\mathrm{s}+T_0/2)=-\mathbi{p}^\mathrm{s}-\mathbi{A}(t_\mathrm{r}^\mathrm{s})=-\bfk$, i.e. from opposite sides of the molecule.

Combining the factorization possible via the saddle point approximation with our conclusions from sections \ref{sec:continuum} and \ref{sec:multielectron}, we can write for the measurable complex XUV spectrum:
\begin{multline}
	\epsXUV(\omega) \propto \alpha(\bfk) \braket{\psi^\mathrm{D}\vert \hat{\mathbf{d}} \vert \bfk } \,+\alpha(-\bfk) \braket{\psi^\mathrm{D}\vert \hat{\mathbf{d}} \vert -\bfk },\\
\text{with } \omega=k^2/2\,.
\label{eq:factorizedsignal}
\end{multline}
For the recombination DME, we have used the single-channel multi-electron formulation (\ref{eq:tomo:multiDME}) and neglected exchange terms, which is justified whenever Koopmans' approximation is valid.
We approximate the continuum wavefunction by a packet of plane waves with spectral amplitude $\alpha(\bfk)$, i.e.  $\ket{\bfk}=\exp[\rmi\bfk\cdot\bfr]$ in (\ref{eq:factorizedsignal}). The error introduced by the plane wave approximation is reduced by using the heuristic relation (\ref{eq:heuristic}) of XUV frequency $\omega$ and electron wavenumber $k$.

The elephant in the room, pointed out in 2008  by Van der Zwan et al. \cite{[*** ][]vanderZwan2008Molecular,Elmarthesis}, is that this expression is not yet fully factorized but a sum of two factorized terms. There are two cases where full factorization can be obtained: \emph{(i)}~Either, the Dyson orbital is (un)gerade. As orbitals can always be chosen to be real valued, $\braket{\psi^\mathrm{D}\vert\hat{\mathbf{d}}\vert\bfk}= [\braket{\psi^\mathrm{D}\vert\hat{\mathbf{d}}\vert-\bfk}]^*$ in length form. For (un-)gerade symmetry, the length-form-DME is purely (real) imaginary valued due to the symmetry properties of the Fourier transform, resulting in $\epsXUV(\omega) \propto [\alpha(\bfk) - \alpha(-\bfk)] \braket{\psi^\mathrm{D}\vert \hat{\mathbf{d}} \vert \bfk }$ for gerade and $\epsXUV(\omega) \propto [\alpha(\bfk) + \alpha(-\bfk)] \braket{\psi^\mathrm{D}\vert \hat{\mathbf{d}} \vert \bfk }$  for ungerade $\psi^\mathrm{D}$. For a multi-cycle driving laser with symmetric carrier wave, one can readily show with (\ref{eq:factorizealpha}) \footnote{Use the Fourier-Transform symmetry properties for the ionization DME, replace $t_\mathrm{r}^\mathrm{s}\rightarrow t_\mathrm{r}^\mathrm{s}+T_0/2$ and note that $S$ depends on $k^2$ and is thus the same for both recollision directions.}, that the two EWPs re-colliding from either side and \emph{time-delayed} by half a laser period, are related by $\alpha(-\bfk)=\pm\alpha(\bfk)$ for \mbox{even/odd} harmonic orders and gerade $\psi_\mathrm{D}$, which finally leads to \mbox{$\epsXUV\propto-2\alpha(\bfk)\braket{\psi^\mathrm{D}\vert\hat{\mathbf{d}}\vert\bfk}$} for odd and $\epsXUV=0$ for even harmonic orders. The reader may verify by him/herself that for ungerade orbitals, the signs of both terms are swapped, giving the same final expression for $\epsXUV$. Using velocity form only affects the recombination dipole, but for the same final result.  \emph{(ii)}~Or, recollision can be restricted to one side of the molecule only, such that either $a(\bfk)$ or $a(-\bfk)$ vanishes. This is possible with few-cycle driving laser pulses, where the highest electron energies are only obtained during a single half-cycle, or with an asymmetric laser waveform, which can, e.g., be obtained by combining the driving laser with its second harmonic \cite{Mauritsson2009subcycle}. Note that when considering the emission of a macroscopic ensemble, the molecules will have to be \emph{oriented} in the laboratory frame in order to probe them all from the same side (see section \ref{sec:alignment}).


The factorization of the complex XUV spectrum in the recombination DME and a complex continuum EWP amplitude, the latter containing both the result of tunnel-ionization and the continuum EWP acceleration, has first been used---in a somewhat ad-hoc way---by Itatani et al. \cite{[*** ] [] Itatani2004Tomographic} in 2004. It has later been more firmly established by Le, Lin et al. \cite{Le2008Theory,Le2008Extraction} for rare gas atoms and the simplest of all molecules, H$_2^+$, and provides the basis for their ``quantitative rescattering theory'' \cite{Le2009QRS,[**][]Lin2010review}, already mentioned in section \ref{sec:continuum}. Another derivation is demonstrated in \cite{Frolov2011analytic}.

\subsubsection{Measuring the recombination dipole} \label{sec:measdip}
%
\vspace{6pt}
\paragraph{Calibrating for the EWP.\\}
Suppose now, that the conditions are such that the measurable complex XUV spectrum (\ref{eq:factorizedsignal}) can indeed be simplified to a single factorized term. If one knew the EWP amplitude $\alpha(\bfk)$, measuring amplitude, phase, and polarization state of the HHG spectrum, $\epsXUV(\omega)$, would constitute a measurement of the complex vector quantity of the recombination DME for $\bfk$.

\begin{figure}
	\includegraphics[width=.5\columnwidth]{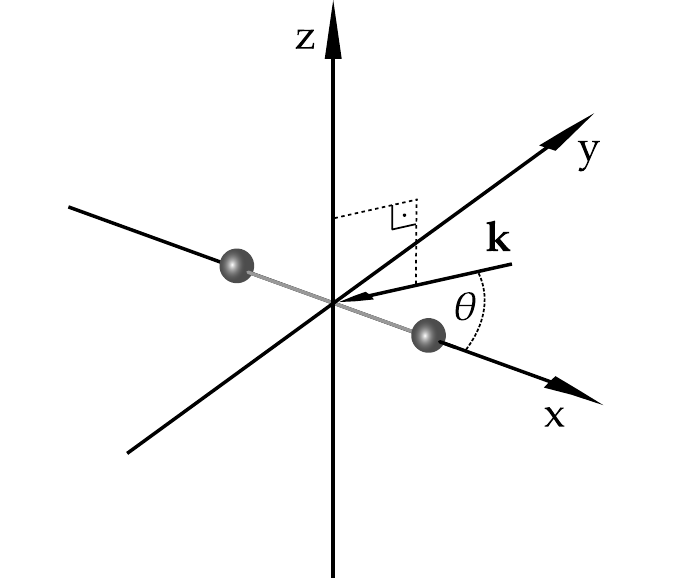}
	\caption{ \label{fig:tomo-scheme1} Coordinates in the description of the scheme for measuring the recombination DME. The two spheres mark the nuclei of a simple linear molecule---the internuclear axis is thus along the x-axis. This axis could as well be some other distinct axis of a more complex molecule. The recolliding electron wave vector $\bfk$, confined to the $(x,y)$-plane, makes an angle $\theta$ with the internuclear axis. The angle $\theta$ is known and can be varied since we can align the molecule in the lab frame.}
\end{figure}

This directly leads to a scheme, proposed by Itatani et al. \cite{Itatani2004Tomographic}, where $\alpha(\bfk)$ is extracted from measurements in a suitable known reference system. `Known' means that the corresponding DME can be calculated accurately, as is the case for rare gas atoms. Dividing experimental HHG spectra generated in different rare gases by calculated recombination DME, Levesque et al. \cite{Levesque2007High} could show that the EWP amplitude, $\alpha(\bfk)$, essentially depends on the driving laser field and the medium ground state energy, but \emph{not} on the precise structure of the ground state, except for a $k$-independent scaling factor. The same conclusion was reached from theoretical studies in atoms and molecules by Le et al. \cite{Le2008Theory,Le2008Extraction}, also including macroscopic phase matching \cite{Jin2009}. In molecules, the scaling factor depends on the ionization angle, $\theta_\rmi$. The latter is defined analogously to the recollision angle (see figure \ref{fig:tomo-scheme1}, i.e. as the angle between the $x$-axis and the electron wavevector at the \emph{ionization instant}, which is anti-parallel to the driving laser field at that instant). The recollision and ionization angle are linked via the continuum trajectories, and in a linearly polarized driving laser field, $\theta_\rmi=\theta$.

Thus, using a reference with the same ionization potential as the studied molecule (so that for both media, $\omega$ corresponds to the same continuum electron momentum $k$) in the same experimental conditions (so that driving laser field and phase matching are the same), the EWP amplitudes, $\alpha(\bfk)$ can indeed be calibrated---up to a scaling factor $\eta(\theta_\rmi)$. The reason for this is easy to understand: in the ionization step, the main factor is the ADK tunnelling rate (\ref{eq:ADKrate}) depending only on $\Ip$ and the laser intensity. The amplitude and phase factors introduced by the excursion in the continuum are only determined by the laser field. This implies that the $k$-dependence of $\alpha(\bfk)$ is approximately the same for the reference atom and the molecule at all angles, i.e. the tunnel ionization step acts as a strong spatial filter (this filtering effect can be clearly shown in models, see, e.g., \cite{Ivanov2005Anatomy,[**]Mairesse2008Electron,Murray2011tunnel}). The difference in $\alpha(\bfk)$ for the reference and the molecule at the varying angles due to the dependence of the tunnelling amplitude on the orbital geometry is then reduced to the $k$-independent scaling factor $\eta(\theta_\rmi)$. This suggest the interpretation of $\eta(\theta_\rmi)$ as a ``normalized tunnelling amplitude''. Note that $\eta(\theta_\rmi)$ can change its sign, which describes the ``EWP's memory of the phase of the orbital lobe it was tunnelling from''---in other words: the fact of \emph{self}-probing rather than probing with an external probe (see section \ref{sec:signchanges}). Note that, of course, the ``absolute'' sign (or, more generally, ``absolute'' phase) of $\eta(\theta_\rmi)$ has no physical sense since orbitals have an arbitrary absolute phase. Only the variation with ionization angle is physical and relevant to experiments. 

Consequently, taking the ratio of the measured complex XUV spectra of the studied molecule and the reference, the electron wavepacket amplitudes $\alpha(\bfk)$ will almost cancel out---we only keep the scaling factor $\eta(\theta_\rmi)$:
\begin{equation}
	\frac{\EXUV^\mathrm{mol}(\omega)}{E_{\mathrm{\tiny{XUV}}}^\mathrm{ref}(\omega)} = \eta(\theta_\rmi) \frac{\braket{\psi_\mathrm{mol}^\mathrm{D} \vert \hat{\mathbf{d}} \vert \bfk}}{\braket{\psi_\mathrm{ref}^\mathrm{D}\vert \hat{\rmd}_{\parallel \bfk} \vert k}}\,.
\end{equation}
Note that we divide here by a scalar: the recombination dipole for an atom and hence the emitted XUV field is always polarized parallel to the electron recollision direction, and we normalize by this component $\parallel\bfk$.

Let $(x,y,z)$ be the coordinates of the molecular reference frame, with the internuclear axis (or any other distinct axis on a more complicated molecule) along $x$. The most-polarizable axis of molecules can be oriented in the laboratory frame (see section \ref{sec:alignment}), i.e. $(x,y,z)$ can also be used for the laboratory frame. We can arrange that the driving laser and XUV propagation direction is along $z$---the EWP movement and hence $\bfk$, controlled by the laser field, are thus confined to the $x$-$y$-plane. As long as we are using a linearly polarized driving laser, $\bfk$ is simply parallel to the laser polarization direction; otherwise the recollision direction can be calculated or calibrated in some way (see e.g. \cite{Shafir2009} for a calibration method). Together with the ``energy conservation'' relation $\omega=k^2/2$, the $(x,y,z)$-coordinates of the continuum electron wavevector $\bfk$ associated with $\omega$ are thus determined. This situation is depicted in figure \ref{fig:tomo-scheme1}, where the recollision angle, $\theta$ is defined as the angle between $\bfk$ and the $x$-direction.

Representing the two measurable polarization components, $x$ and $y$, of the XUV emission from the molecule by
\begin{equation}
	{\EXUV^\mathrm{mol}(\omega,\theta)}_{x/y}=A^\mathrm{mol}_\mathrm{x/y}(\omega,\theta) \exp[\rmi\varphi^\mathrm{mol}_\mathrm{x/y}(\omega,\theta)]\,,
\end{equation}
and the XUV emission from the reference atom, polarized parallel to $\bfk$, by
\begin{equation}
	E_{\mathrm{\tiny{XUV}}}^\mathrm{ref}(\omega)=A^\mathrm{ref}(\omega) \exp[\rmi\varphi^\mathrm{ref}(\omega)]\,,
\end{equation}
where all amplitudes, $A$, are positive and real valued, we can calculate the $x$ and $y$-component of the recombination DME, $\mathbi{d}$, for the molecule as:
\begin{multline}
	d_\mathrm{x/y}=\braket{\psi_\mathrm{mol}^\mathrm{D}\vert \hat{\mathbf{d}} \vert \bfk}_\mathrm{x/y} =  \frac{1}{\eta(\theta_\rmi)}  \frac{A^\mathrm{mol}_\mathrm{x/y}(\omega,\theta)}{A^\mathrm{ref}(\omega)} \\
		\times \exp\left[\rmi\varphi^\mathrm{mol}_\mathrm{x/y}(\omega,\theta) - \rmi\varphi^\mathrm{ref}(\omega)\right]  \braket{\psi_\mathrm{ref}^\mathrm{D}(\bfr)\vert \hat{\rmd}_{\parallel k} \vert k}\,.
\label{eq:exp:moldip}
\end{multline}

\noindent What has to be measured/known is thus:
\begin{itemize}
	\item the spectral intensity, $(A^\mathrm{ref}(\omega))^2$, and phase, $\varphi^\mathrm{ref}(\omega)$, of the XUV emission from a suitable known reference atom (no alignment is necessary),
	\item the spectral intensity $(A_\mathrm{mol}(\omega,\theta))^2$, and phase, $\varphi^\mathrm{mol}(\omega,\theta)$, of the XUV emission from the molecule, separately for the two polarization components, and with known recollision angle $\theta$,
	\item for the scaling factor, $\eta(\theta_\rmi)$, we actually do not need to know its exact values but only its $\theta_\rmi$-dependence. A separate dedicated experiment could measure the  $\theta_\rmi$-dependence of the tunnelling \emph{probability}, which gives information on $\eta(\theta_\rmi)^2$. However, detecting possible sing changes will be a major problem---see the following section \emph{b} for possible solutions.
\end{itemize}

Looking at (\ref{eq:exp:moldip}) and assuming that the reference DME is a slowly varying function of $k$ and does not contain any rapid phase changes, we can say that the modulus of the molecular DME is esentially given by the square root of the ratio of harmonic intensities measured for the molecule and the reference atom as well as by the angular dependence of the tunnelling rate for the molecule. The phase of the molecular DME is esentially given by the spectral phase difference of the harmonic emission from the molecule and the reference atom, as well as possible sign changes of $\eta(\theta_\rmi)$.

\vspace{6pt}
\paragraph{Mind the --self-- in self-probing!\\} \label{sec:signchanges}

The self-probing paradigm may easily be misunderstood suggesting the image of an external EWP probing the molecule. However, the fact that the EWP recolliding with the parent ion is created by ``tearing'' the outermost part of the bound state wavefunction through a narrowed potential barrier, has important implications. The EWP thus keeps a sort of ``phase-memory'' of the orbital lobe it originated from.

This already appears in the SFA-framework, although it does not include an accurate description of the tunnelling process. Take the gerade and ungerade version of the same Dyson orbital, i.e. $\psi^\mathrm{D}_\mathrm{g}$ and $\psi^\mathrm{D}_\mathrm{u}$, with $(\psi^\mathrm{D}_\mathrm{g})^2=(\psi^\mathrm{D}_\mathrm{u})^2$, and assume that our linearly polarized driving laser field leads to a single recollision from one side of the molecule only. We will compare the emitted complex XUV spectra, $\epsilon^+(\omega)$ and $\epsilon^-(\omega)$, resulting from recollision with $\bfk$ and $-\bfk$, respectively, which we can achieve by simply flipping the sign of our driving laser field. As mentioned earlier, for the gerade orbital, the length form DME $\braket{\psi^\mathrm{D}_\mathrm{g}\vert\hat{\bfr}\vert\bfk}=-\braket{\psi^\mathrm{D}_\mathrm{g}\vert\hat{\bfr}\vert-\bfk}$, while for the ungerade orbital $\braket{\psi^\mathrm{D}_\mathrm{u}\vert\hat{\bfr}\vert\bfk}=\braket{\psi^\mathrm{D}_\mathrm{u}\vert\hat{\bfr}\vert-\bfk}$. For \emph{both} cases, we thus find
\begin{align*}
 \epsilon^-(\omega)=&-\rmi (\rmi \tau_n^\mathrm{s} /2 )^{-3/2}[\det(\tilde{S}'')]^{-1/2} \exp[\rmi\omega t^\mathrm{s}_\mathrm{r}+\rmi S] \\
	&\times  (-\mathbi{E}(t_\rmi)\cdot \braket{-\bfk\vert\hat{\bfr}\vert\psi^\mathrm{D}_\mathrm{g}}) \,\braket{\psi^\mathrm{D}_\mathrm{g}\vert\hat{\bfr}\vert-\bfk}\\
	=&-\epsilon^+(\omega).
\end{align*}
What we can read from this example is, that for a gerade orbital when switching to the opposite recollision direction, we observe a $\pi$ phase flip of the emitted XUV spectrum---which is as expected: the EWP moves over the gerade orbital in the opposite direction so the resulting dipole moment has the opposite phase. However, for the ungerade orbital, the same $\pi$ phase flip occurs, which is simply because the created EWP has opposite phases for both recollision directions. With a linearly polarized laser, we can thus not directly see the difference between a gerade and an ungerade version of an orbital in the symmetry of the measured $\epsilon(\omega)$. Similar examples can be constructed with the separate XUV field components, to show that any orbital nodal planes that cross the origin of the coordinate system defined in figure \ref{fig:tomo-scheme1} will not show up in the measurable $\epsilon(\omega)$ as long as $\theta_\rmi=\theta$, as illustrated in figure \ref{fig:co2homo+ewp}. In other words: the measured $\epsilon(\omega)$ will always look as if it was coming from a gerade orbital with $\sigma$-symmetry.
\myfigure{.6}{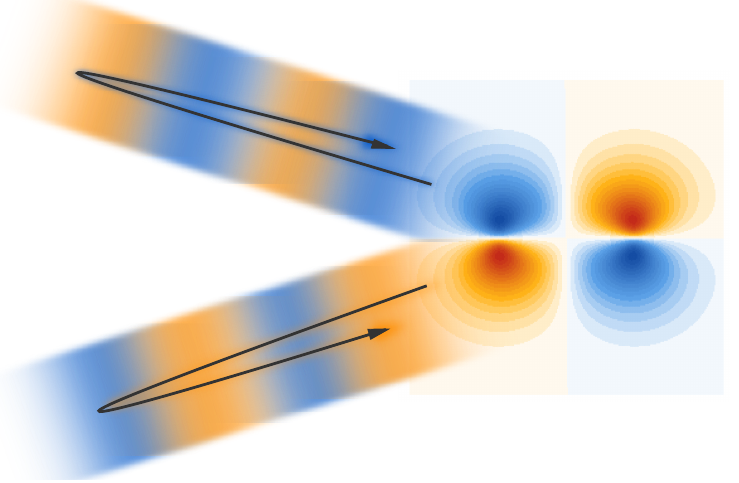}{Illustration of the ``phase memory'' of the continuum EWP. Shown is the HOMO of CO$_2$, which has $\pi_\mathrm{g}$-symmetry; blue and orange colours indicate opposite phases. With a neutral probe, the horizontal component of the complex XUV spectrum would have opposite phases for the two indicated recollision directions---in \emph{self}-probing driven with a linearly polarized laser, though, no phase change occurs between the two. This is because the EWP has opposite phase at the two angles.}

This ``phase memory'' of $\alpha(\bfk)$ is taken into account by defining the ionization amplitude, $\eta^\mathrm{mol}(\theta_\rmi)$, as a real valued quantity that can change signs (which is sufficient as long as orbitals are real valued). This sign is very difficult to measure. If the orbital symmetry is known in advance, the same symmetry can be imposed on the DME via this sign, as we did in \cite{Haessler2010tomo}, also discussed in section \ref{sec:tomoexp}. In the general case, a reliable tunnelling theory could provide us $\eta^\mathrm{mol}(\theta_\rmi)$. MO-ADK theory \cite{Tong2002moadk,Zhao2011effect}, many-body S-matrix theory \cite{Becker2005,MuthBoehm2000suppressed,JaroniBecker2003Dependence}, or the approach of refs. \cite{Spanner2009oneelectron,Murray2010partial,Murray2011tunnel} could be used for this purpose.

A purely experimental way of dealing with the ``phase memory'' is of course desirable. One possibility are measurements using polarization-shaped driving laser \emph{fields} that allow to have different recollision and ionization angles \cite{[**][]Kitzler2007,Kitzler2005,[**][]Shafir2009,[*]Niikura2010symmetry}. Illustratively speaking, one can then take the continuum electron from one orbital lobe and steer it to recollide at another. Starting from this idea, generalized schemes could be devised that probe the molecular ion from different directions while keeping the ionization angle approximately constant, thus approaching the situation of the continuum EWP as an ``external probe''.  In this case, we might as well omit the $\eta$-factors in (\ref{eq:exp:moldip}). \stef{Another recently proposed way of measuring the orbital symmetry is based on characteristic structures in ``high harmonic polarization maps'' \cite{[][]Hijano2010Orbital}.}

\subsubsection{Interference in the recombination dipole} \label{sec:imaging}
Which obvious features can be expected to appear in those molecular DMEs and how can they be easily related to the molecular structure and/or dynamics?
\vspace{6pt}
\paragraph{Structural interference.\\}
Consider a simple, diatomic, mono-nuclear molecule, for which we can make Koopmans' approximation and the X-channel dominates HHG. Then, the recombination dipole relevant to HHG is simply that between the continuum electron and the HOMO. Suppose further that the HOMO can be written as an antisymmetric combination of two atomic orbitals: $\psi_0 = \phi_0(\bfr - \mathbi{R}/2) - \phi_0(\bfr+\mathbi{R}/2)$, where $\mathbi{R}$ is the internuclear distance vector, making an angle $\theta$ with the driving laser polarization direction and thus with the recolliding electron wave vector $\bfk$. The recombination DME in velocity form then reads:
\begin{equation}
	\braket{\psi_0\vert -\rmi\nabla_\bfr \vert \rme^{\rmi \bfk\cdot \bfr}}  = 2\rmi\bfk \sin\left(\frac{kR}{2} \cos\theta \right) \braket{ \phi_0(\mathbf{r}) \vert \rme^{\rmi \bfk\cdot \bfr}}\,.
	\label{eq:exp:twocenter1}
\end{equation}
This result is simply a consequence of the Fourier shift theorem. A sign change of the sine corresponds to destructive quantum interference and happens for
\begin{equation}
	R \cos\theta = n \lambda_e,
	\label{eq:twocenter-}
\end{equation}
where n is an integer and $\lambda_e=2\pi/k$ is the electron de Broglie wavelength. Destructive interference thus occurs if the recollinding electron wavelength is equal to the internuclear distance projected on the recollision direction. The molecule thus behaves like a two-point emitter whose emissions are dephased due to \emph{(i)} the path difference between the centers, and \emph{(ii)} the symmetry of the orbital. 

If, instead, one considers a symmetric combination of atomic orbitals, $\psi_0 = \phi_0(\bfr - \mathbi{R}/2) + \phi_0(\bfr+\mathbi{R}/2)$, one finds along the same lines destructive interference for:
\begin{equation}
	R \cos\theta = \left(n- \frac{1}{2} \right) \lambda_e\,,
	\label{eq:twocenter+}
\end{equation}
i.e.  if \textit{half} the recolliding electron wavelength is equal to the internuclear distance projected on the recollision direction. The latter relation together with the heuristic dispersion relation $\omega=k^2/2$ predicts an interference position in the harmonic spectrum that agrees well with that obtained from TDSE simulations~\cite{Kamta2005Threedimensional,Ciappina2007Influence} for H$_2^+$.

\myfigure{.75}{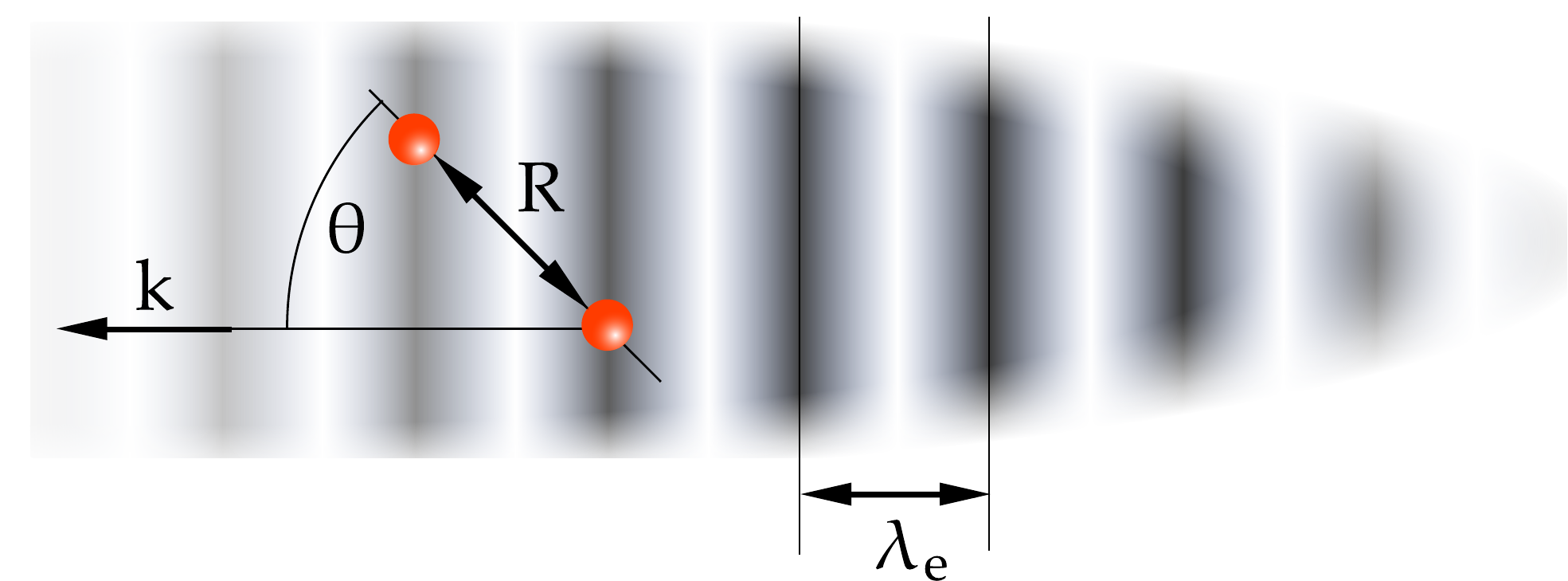}{Two-center-interference. Depending on the relative sign of the molecular orbital at its two centers (orange dots), the interference is either constructive or destructive when the recolliding electron de Broglie wavelength equals the distance of the two centers projected on the recollision direction.}
Such destructive interference, i.e. a sign change in the recombination DME (or, in complex representation, a DME going through zero and changing phase by $\pi$), should leave a clear trace in the high harmonic spectrum of aligned molecules. . This was first observed in numerical experiments \cite{Lein2002Role,Lein2002Interference}, where the solution of the TDSE for H$_2^+$ revealed minima in the HHG spectral intensity and phase jumps of $\approx\pi$ value at the positions predicted by equation (\ref{eq:twocenter+}). 

Obviously, constructive interference occurs as well, but does not leave such clear signatures in the observable HHG spectrum.

Note that the recombination DME in length form can be expressed in a similar, yet more complicated form as (\ref{eq:exp:twocenter1}). It also presents a sign change, but not necessarily at the same position as in velocity form, which is due to the error introduced by using plane waves in the model (cp. the discussion of different forms of the DME in section \ref{sec:sfa}).

This effect is commonly referred to as ``two-center interference'' and can be generalized to linear orbitals combinations that are not purely (anti-)symmetric \cite{Odzak2009}. It is not only relevant to the simplest, homonuclear diatomic molecules, but is is a prototype for ``structural interference'' in general, i.e. interference structures due to the multi-center nature of molecular orbitals. The most general formulation of structural interference is found when recognizing the recombination DME with the plane wave approximation as the \emph{spatial Fourier transform}, $\mathcal{F}_{\bfr\rightarrow\bfk}$, of $\bra{\psi^D \hat{\mathbf{d}}}$.  Interference thus occurs at \emph{characteristic spatial frequencies} of a channel-specific (i.e. static) Dyson orbital.

In this generalized picture, it becomes clear that the nuclear configuration of the molecule does \emph{not} directly play a role. In (\ref{eq:exp:twocenter1}) to (\ref{eq:twocenter+}), the internuclear distance, $R$, only appeared because the positions of the nuclei are natural sites to place the basis functions when constructing a molecular orbital. Obviously, we are free to change the way we represent our orbitals and where to center our basis functions. In general, the quantity $R$ thus stands for the distance of these \emph{centers} of the molecular orbital---i.e. a purely electronic property.

\vspace{6pt}
\paragraph{Dynamic interference.\\} As already mentioned in section \ref{sec:multiorbital}, if several channels contribute significantly to HHG, they will interfere in the total recombination dipole. For two different channels, the ions with energy difference $\Delta\varepsilon$ accumulate a phase difference $\Delta\phi=\Delta\varepsilon\tau$ during the electron excursion duration, $\tau$. Strictly speaking, the continuum electron trajectory associated with one XUV frequency, $\omega$, is also not exactly the same for two different channels---so, do we need to take another relative phase into account?

Let us use a variational approach to find the answer to this question. We consider a single channel and calibrate our energy axis such that the ion has zero energy---the neutral bound state energy, $-\Ip$, appearing in the SFA model thus becomes the ionization potential associated with this channel. Then, all the phase accumulated by the ionized system during the continuum excursion is contained in the quasi-classical action $\tilde{S}$ of the continuum electron, given by (\ref{eq:theory:saddlepoint-action}) and (\ref{eq:theory:sfaaction}). Within the stationary phase approximation, and for constant $\omega$, we find the action to change with varying ground state energy by
\begin{align}
	\left. \frac{\rmd \tilde{S}}{\rmd \Ip} \right\vert_{\omega}  \Delta\Ip =&  \left[(t_\mathrm{r}-t_\rmi) \frac{}{}\right.\nonumber\\
		&\:+ \underbrace{\frac{\partial \tilde{S}}{\partial t_\mathrm{r}}}_{=0} \frac{\partial t_\mathrm{r}}{\partial \Ip}   + \underbrace{\frac{\partial \tilde{S}}{\partial t_\rmi}}_{=0} \frac{\partial t_\rmi}{\partial\Ip}  + \underbrace{\frac{\partial \tilde{S}}{\partial p}}_{=0} \left.\frac{\partial p}{\partial\Ip}\right]\Delta\Ip \nonumber\\
		=&\Delta\Ip\tau \,,
\label{eq:dSdIp}
\end{align}
which is valid for $\Delta\Ip\ll p^2/2$. Since varying the \emph{ground state} state energy is equivalent to varying the \emph{ion} energy (the difference is a mere recalibration of the energy axis which does of course not change \emph{relative} phases), this result also readily tells us the phase variation for varying \emph{ion} energy, i.e. the relative phase of two channels in HHG, and thus confirms the above expression for $\Delta\phi=\Delta\varepsilon\tau$. The value of $\tau$ is that corresponding to the saddle point found for the value of $\Ip$ around which we have linearized $\tilde{S}$---a good value would thus simply be the channel averaged-excursion duration.

Due to the mapping between XUV frequency, $\omega$, and electron excursion duration, $\tau$, this time-dependent interference shows up as frequency dependent interference in the  measurable XUV spectra---just as structural interference. The fact that by changing the driving laser field, one can change the $\omega$--$\tau$-mapping and thus the spectral position of this \emph{dynamic} interference allows to separate it from the \emph{structural} interference discussed above, which is time-independent.

This makes it possible to tease apart both types of interference \cite{Torres2010revealing}. They are, however, simply two aspects of one and the same physics: the recombination  DME of HHG results from interference of a continuum EWP and the Dyson orbital. If multiple channels contribute, the Dyson orbital is a time-dependent wavepacket and the spatial structure of the wavepacket components leads to structural interference while their relative phases lead to dynamic interference. Both combined contain the spatial information on the time-dependent Dyson orbital at the recollision instant. Of course, this ``instant'' can only be an average over the spectral components of the continuum EWP, and the temporal resolution will be given by the considered spectral range of the recolliding EWP.

\subsection{Molecular orbital tomography} \label{sec:tomo}
In the last section, it has become clear that the recombination DME encodes the spatial structure of the Dyson orbital. The question that immediately arises is: how to retrieve the Dyson orbital? The key to the answer is the plane-wave approximation for the recolliding EWP, which makes the recombination DME, $\braket{\psi^\mathrm{D}(\bfr)\vert \hat{\mathbf{d}} \vert \rme^{\rmi\bfk\cdot\bfr}}$, a Fourier transform, $\mathcal{F}_{\bfr\rightarrow\bfk}$. Itatani et al. \cite{Itatani2004Tomographic} first made use of this fact when they proposed a scheme for tomographic imaging of electrons in molecules.

	\subsubsection{Concept}
	\label{sec:tomoconcept}
The recombination DME, $\mathbi{d}$, is measurable as described in section \ref{sec:measdip}. With the same coordinates and definitions, the $q$-component ($q=x,y$) of the matrix element in \emph{length form} then writes 
\begin{align}
	d^\mathrm{L}_q(\bfk)&=\braket{\psi^\mathrm{D}(\bfr)\vert q \vert \rme^{\rmi\bfk\cdot\bfr}} \nonumber \\
		&= \iint \left[q\int [\psi^\mathrm{D}(x,y,z)]^* \rmd z \right] \rme^{\rmi (k_x x + k_y y)}\:\rmd x \rmd y.
	\label{eq:tomo:dqlength}
\end{align}
Each component of $\mathbi{d}$ thus contains the Fourier transform of 
\begin{equation}
q\tilde{\psi}^\mathrm{D}(x,y)=q \int [\psi^\mathrm{D}(x,y,z)]^* \rmd z\,,
\end{equation}
i.e. $q$ times the Dyson orbital projected onto the plane perpendicular to the laser propagation direction. This implies that an orbital odd in $z$ will not contribute. For orbitals even in $z$, this projection contains the complete information.

Measuring $d_q$ for one recollision angle $\theta$ thus yields data points in Fourier space of the object $q\tilde{\psi}^\mathrm{D}(x,y)$---points at the coordinates $(k_x,k_y)$ that all lie on a line, given by the recollision angle, $\theta$, and the length of the electron wave vectors, $k$, of the recolliding EWP components. These are associated with the harmonic photon energy, $\omega$, via energy conservation: $k^2/2=\omega$. A whole spectrum of XUV frequencies, mapped onto points $(k_x,k_y)$ consequently yields a slice through one quadrant of Fourier space, as illustrated in figure \ref{fig:tomo-scheme2}. Repeating the measurement for more $\theta$-values, slice per slice of Fourier space is collected until it is sufficiently well sampled. What ``sufficiently'' means has yet to be figured out in simulations, discussed in section \ref{sec:sampling}.
\myfigure{.6}{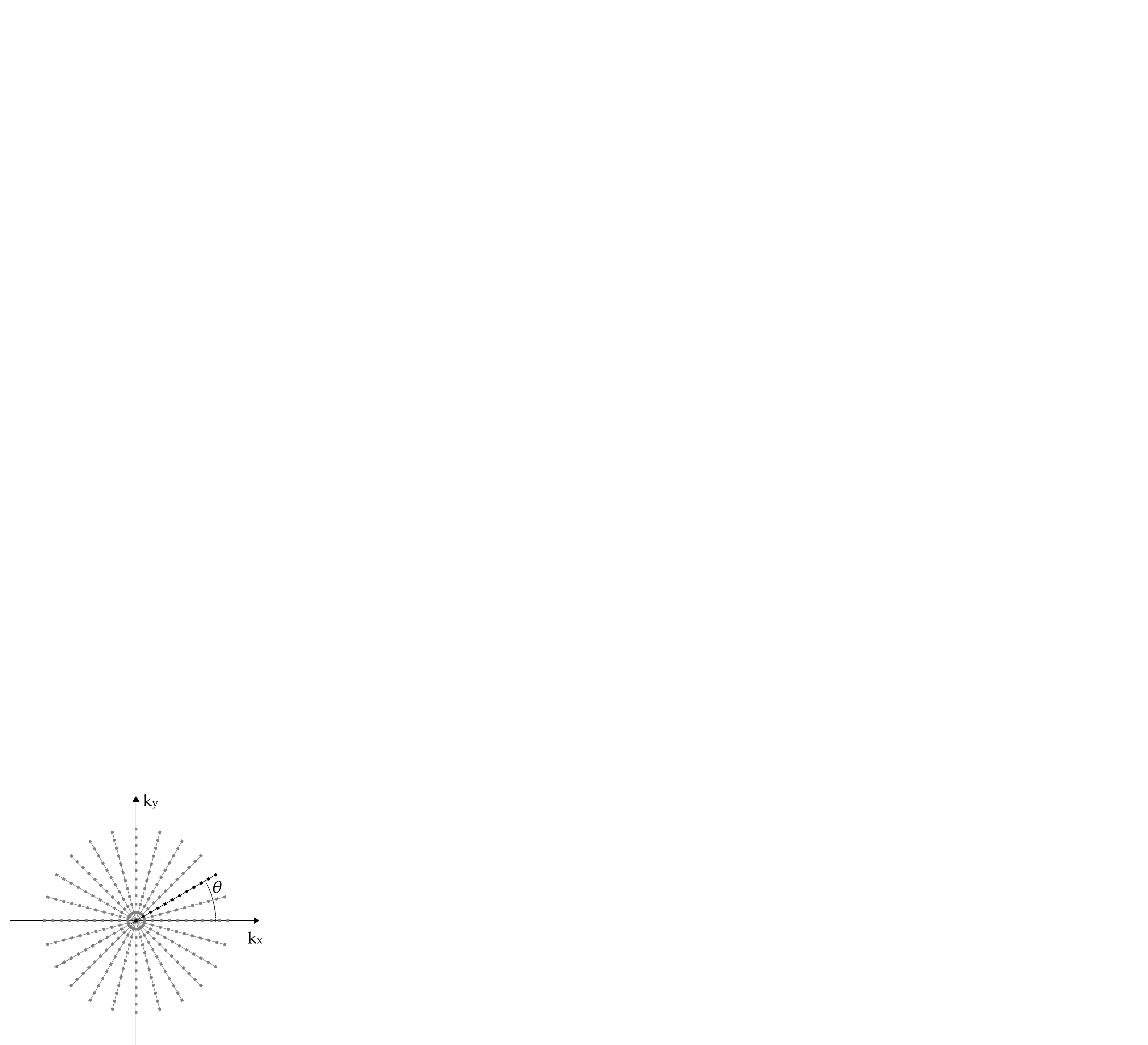}{Fourier space of the object $q\tilde{\psi}^\mathrm{D}(x,y)$. A measurement for one recollision angle $\theta$ yields a line of data points (shown in black). For a linearly polarized driving laser, where $\theta$ is independent of $k$, this line is straight, while the line may be bent if some sort of polarization shaping is employed. Repeating the measurement for more $\theta$-values, slice per slice of Fourier space is collected. If $q\tilde{\psi}^\mathrm{D}(x,y)$ has a known symmetry, one can limit the $\theta$-range to one quadrant and complete the Fourier space data according to this symmetry. Note that, in principle, the points along each line will not be equidistant since they are associated to equidistant harmonic energies and thus $k^2/2$ values.}

The inverse 2D Fourier transform, $\mathcal{F}_{\bfk\rightarrow\bfr}$, applied to the so-obtained data, thus yields $q\tilde{\psi}_\mathrm{mol}(x,y)$ in real space, and the sought-for molecular orbital (projection) can be reconstructed as
\begin{equation}
	\tilde{\psi}^\mathrm{D}_q(x,y) = \frac{\mathcal{F}_{\bfk\rightarrow\bfr}[d^\mathrm{L}_q(k_x,k_y)]}{q}.
\label{eq:tomo:recon-int}
\end{equation}
An explicit, discretized version of this equation is given as equation 2 in ref. \cite{Haessler2010tomo,*[***][]Smirnova2010NV} (using a different notation, though).

From both DME components, $x$ and $y$, i.e. parallel and perpendicular to the molecular axis, the \textit{same} orbital can in principle be reconstructed. Due to the limited discrete sampling in Fourier space, they will, however, most likely not give the same result and one can define
\begin{equation}
\tilde{\psi}^\mathrm{D}(x,y) = \frac{1}{2} \left(\tilde{\psi}^\mathrm{D}_x(x,y)+ \tilde{\psi}^\mathrm{D}_y(x,y)\right)
\label{eq:tomo-halfsum}
\end{equation}
for the \textit{reconstructed molecular orbital} (projection) so that distortions will hopefully average out. 

%
The same scheme can be written based on the \emph{velocity form} of the recombination DME, i.e. with $\hat{\mathbf{d}}=-\rmi\nabla_\bfr$. Equation \ref{eq:tomo:dqlength} is then replaced by 
\begin{align}
	d^\mathrm{V}_q(\bfk)=&-\rmi \braket{\psi^\mathrm{D}(\bfr)\vert \frac{\partial}{\partial q} \vert \rme^{\rmi\bfk\cdot\bfr}} \nonumber\\
		 &= k_q\int\int \left[\int [\psi^\mathrm{D}(x,y,z)]^* \rmd z \right] \rme^{\rmi (k_x x + k_y y)}\:\rmd x \rmd y .
	\label{eq:tomo:dqvelocity}
\end{align}
It follows that the orbital can be obtained via (\ref{eq:tomo-halfsum}) using
\begin{equation}
	\tilde{\psi}^\mathrm{D}_q(x,y) = \mathcal{F}_{\bfk\rightarrow\bfr}\left[\frac{d^\mathrm{V}_q(k_x,k_y)}{k_q} \right].
\label{eq:tomo:recon-int-velocity}
\end{equation}
Again, neither this nor (\ref{eq:tomo:recon-int}) is \textit{a priori} superior to the other and they may give different results. There may, however, be technical reasons to prefer one or the other form: if $\tilde{\psi}^\mathrm{D}$ has a nodal plane containing the $x$- or $y$-axis, one will run into numerical problems when dividing by $x$ or $y$ in (\ref{eq:tomo:recon-int}).

\subsubsection{Criticism}
		\label{sec:tomoissues}
Someone who claims to be able to measure an orbital risks to be attacked with rotten tomatoes by many physicists and chemists who come to the defense of the facts that  \emph{quantum mechanics says that wavefunctions are not observable},  and \emph{electron orbitals are nothing but an artifact of the independent particle approximation and do not exist in nature.}

The second objection is easy to dispel: the orbital accessible by self-probing with HHG is the Dyson orbital, which is  uniquely (up to a global phase) defined with \emph{exact} multi-particle wavefunctions. Of course, the Dyson orbital is also no more than a theoretical concept associated with a certain physical process---the reader may make up his/her own mind about the extent to which a Dyson orbital ``exists'' and perhaps seek advice in E. Schwarz's instructive essay \cite{[***][]Schwarz2006}.

As for the first objection: Other than the impossibility of a $\psi$-meter, i.e. an apparatus that would register a \emph{definite value} of the wavefunction of a system \stef{\emph{in a single measurement}}, there is no law ruling out the possibility of inferring a wave function from a set of measured data. In fact, this is done by many physicists with great success---see e.g. \cite{Raymer1997,[][]Lundeen2011Direct,*Hosten2011nv}. The fundamental question of full characterization of waves is not restricted to quantum mechanics. For example when characterizing laser pulses, we have no detector that would be fast enough to directly track the electric field oscillations. This restriction, however, does not prevent us from using interference with a fully characterized reference wave to recover the phase of the light field---even the carrier-envelope phase (CEP) \cite{Holzwarth2000optical,Jones2000carrier}. Here, we are doing the same---our reference is the continuum EWP. Even if we did not make the plane-wave approximation but knew the exact scattering wave packet and could retrieve the Dyson orbital in some iterative procedure, the continuum EWP is, as any quantum wavefunction, defined up to a global phase only. The same is then always true for our retrieved Dyson orbital. This is the only difference to retrieval procedures for `physical fields', such as the electric field of a light pulse, and this is why this tomography scheme does not violate any quantum mechanics principle. Again, we recommend the essay \cite{[***][]Schwarz2006} for further discussion of such ``so called measurements'' .

What is provocative about this scheme is rather the claim to possess a fully (up to a global phase) characterized reference wave---which we do not actually have. In fact, we \emph{approximate} the continuum EWP as a packet of plain waves. 
While this approximation is certainly crude, it is accepted practice to base the analysis of experiments on idealized models in order to obtain a comprehensible understanding of the physics---or, as in this case, to obtain a simple direct retrieval of information from a measurement. The plane wave approximation is responsible for the surprisingly simple direct reconstruction by means of an inverse Fourier transform. If one measures a DME between two states---bound and continuum---one cannot avoid making an educated guess about one of the two if one wants to \textit{directly} retrieve the other. In an indirect retrieval procedure, this guess could be improved iteratively, as sketched by Patchkovskii et al. \cite{Patchkovskii2007}.

Numerical experiments that use exact scattering states and then apply the tomographic reconstruction procedure based on plane waves give mixed results. Van der Zwan et al. \cite{vanderZwan2008Molecular}, who have calculated harmonic spectra by solving the single-active-electron TDSE, tend to be optimistic and obtain very good orbital reconstructions. Walters et al. \cite{Walters2008}, on the other hand, who use complex conjugated accurate photoionization DMEs from a specialized quantum chemistry code, report significant distortions of the orbitals reconstructed using electron recollision energies typical for HHG experiments (i.e. below, say, 100 eV). Note that this approach has the same vulnerability as the quantitative rescattering theory, discussed ins section \ref{sec:continuum}.

\subsubsection{Sampling Fourier space}
	\label{sec:sampling}
\myfigure{.9}{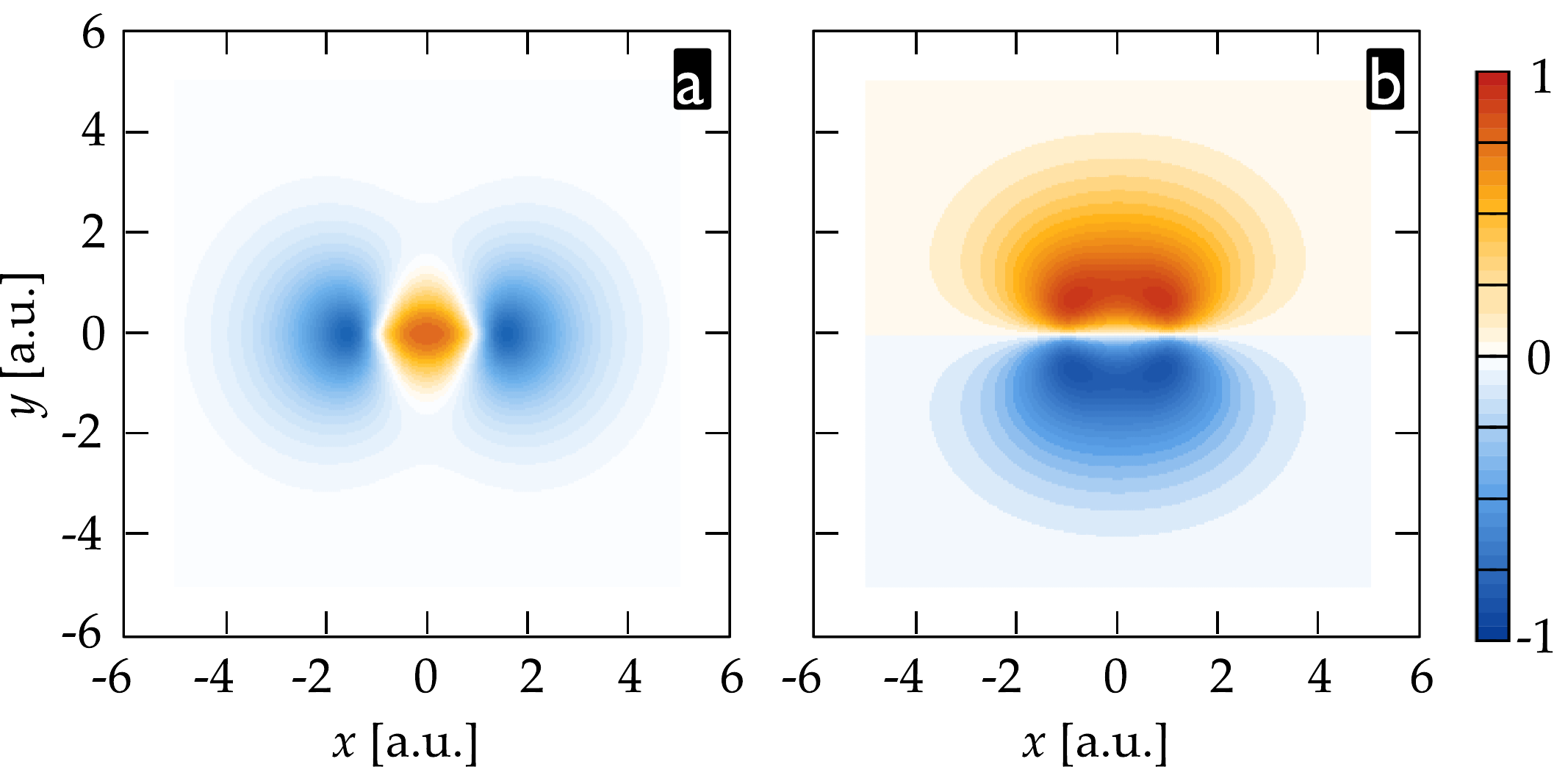}{(a) HOMO and (b) HOMO-1 of N$_2$ from a Hartree-Fock calculation, implemented in the GAMESS package \cite{gamess}.}
 The simulations to be discussed in the following do not aim at testing the approximations at the basis of the orbital tomography scheme itself but rather suppose that the above model is accurate and address the question of how precise the discrete sampling has to be and how well we need to separate the polarization components. This issue will be studied in an model case: we consider the N$_2$ molecule in the Hartree-Fock and Koopmans' approximations, and disregard all but the X-channel. This means that the Dyson orbital is equal to the HOMO. The N$_2$ HOMO, shown in figure \ref{fig:full_n2orbs}a, is a good candidate for such tests as it has very distinct features besides its $\sigma_g$ symmetry that can serve as a reference, such as the nodes at the nuclei positions (at $x=\pm1\:$a.u.) and the diamond-shaped central lobe. We thus calculate the DME vector, $\mathbi{d}$, with the Hartree-Fock HOMO, computed with the GAMESS code package \cite{gamess}, and  plane-waves. We do this at $\bfk$-points corresponding to the odd harmonic orders of an 800 nm laser ($\omega_0=0.057\:$a.u.) with $\omega=k^2/2$, and an angular step $\Delta\theta$. All simulations will consider the length form only, i.e. the DME will be calculated in length form and the reconstructions will be based on (\ref{eq:tomo:recon-int}) and (\ref{eq:tomo-halfsum}).

\myfigure{.9}{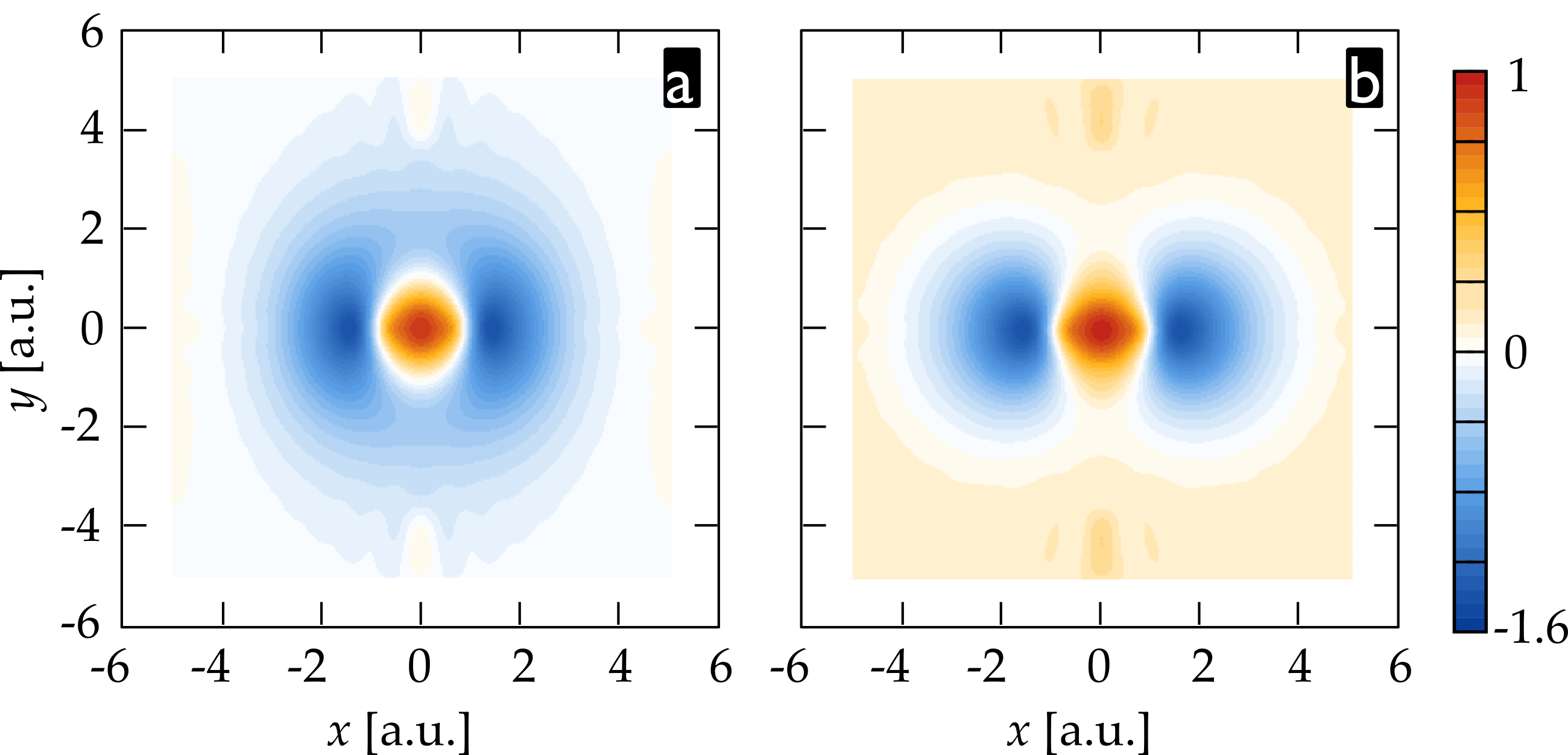}{Simulation of a tomographic reconstruction, sampling $k$-points corresponding to harmonics 1 to 991 and an angular step of $\Delta\theta=10^\circ$, (a) with and (b) without restricting the DME to its parallel component, $d_\parallel$, only.}
	The first question to be addressed is whether it is necessary to measure the DME \textit{vector}---i.e. to separately measure its $x$ and $y$-components. This question arises because the major component of the DME will usually be parallel to $\bfk$. With a linearly polarized driving laser, the easiest experiment will then be to only measure the XUV polarization component parallel to the driving laser polarization, to which $\bfk$ is parallel, and then neglect the DME component perpendicular to $\bfk$ in the analysis. We simulate this by projecting the computed DME onto $\bfk$, obtaining $d_\parallel= (\bfk/k)\cdot\mathbi{d}$, and then using $d_x=d_\parallel\cos\theta$ and $d_y=d_\parallel\sin\theta$. Only for very large spectral ranges like that considered in figure \ref{fig:orb-h1-991}, this approximation induces clear distortions. With the gigantic spectral width of harmonic 1 to 991 (i.e. $k=0.33\:$a.u. to $k=10.6\:$a.u.), the reconstruction is close to perfect if the full vector DME is considered, whereas the approximation of using only $d_\parallel$ causes the outer part of the orbital in figure \ref{fig:orb-h1-991}a to be more spherical than the exact N$_2$ HOMO. For strongly restricted spectral widths, this distortion appears as well but the one caused by limited sampling is largely dominant. 

\myfigure{.66}{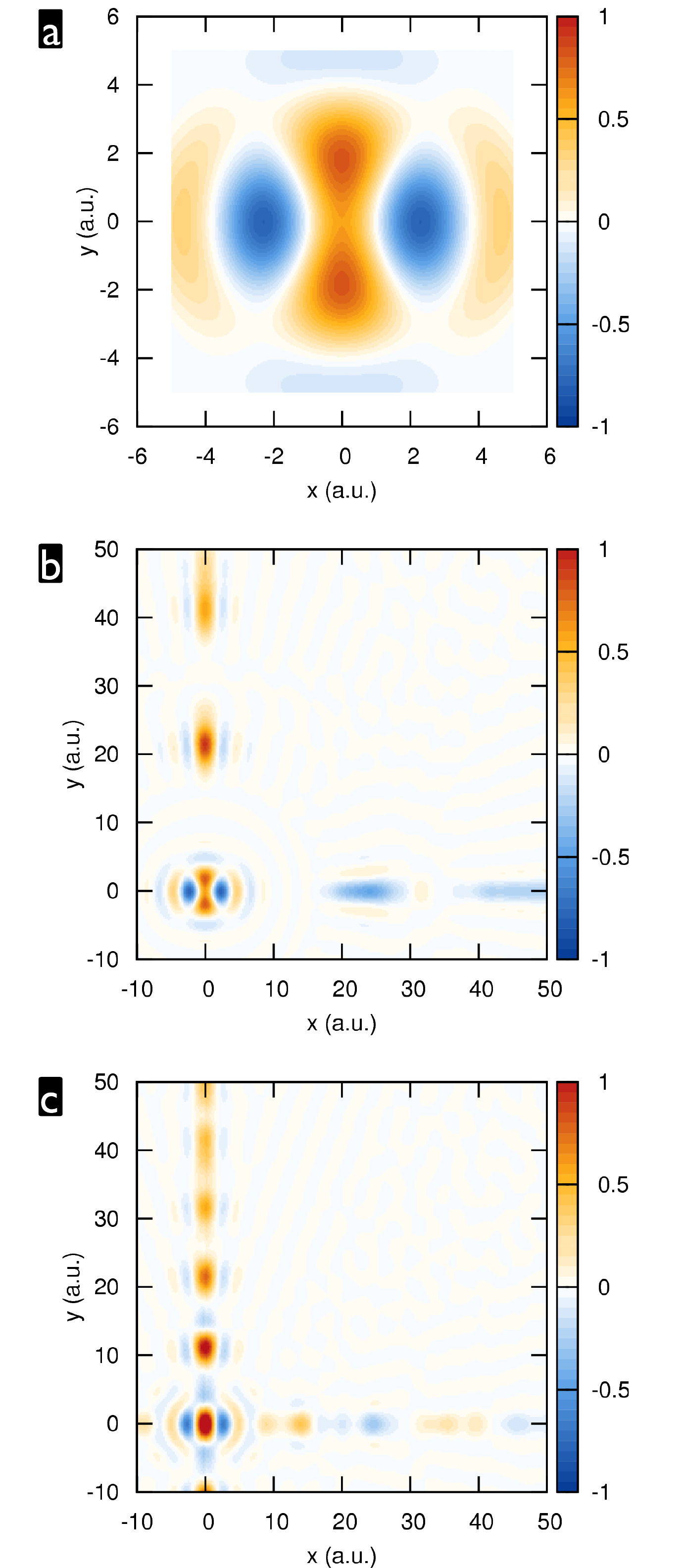}{Simulation of a tomographic reconstruction, based on the parallel component, $d_\parallel$, of the DME only, sampling $k$-points corresponding to harmonics 17 to 31 and an angular step of (a) $\Delta\theta=5^\circ$, (b) $\Delta\theta=10^\circ$ and (c) $\Delta\theta=20^\circ$.}
\begin{figure*}
	\includegraphics[width=\textwidth]{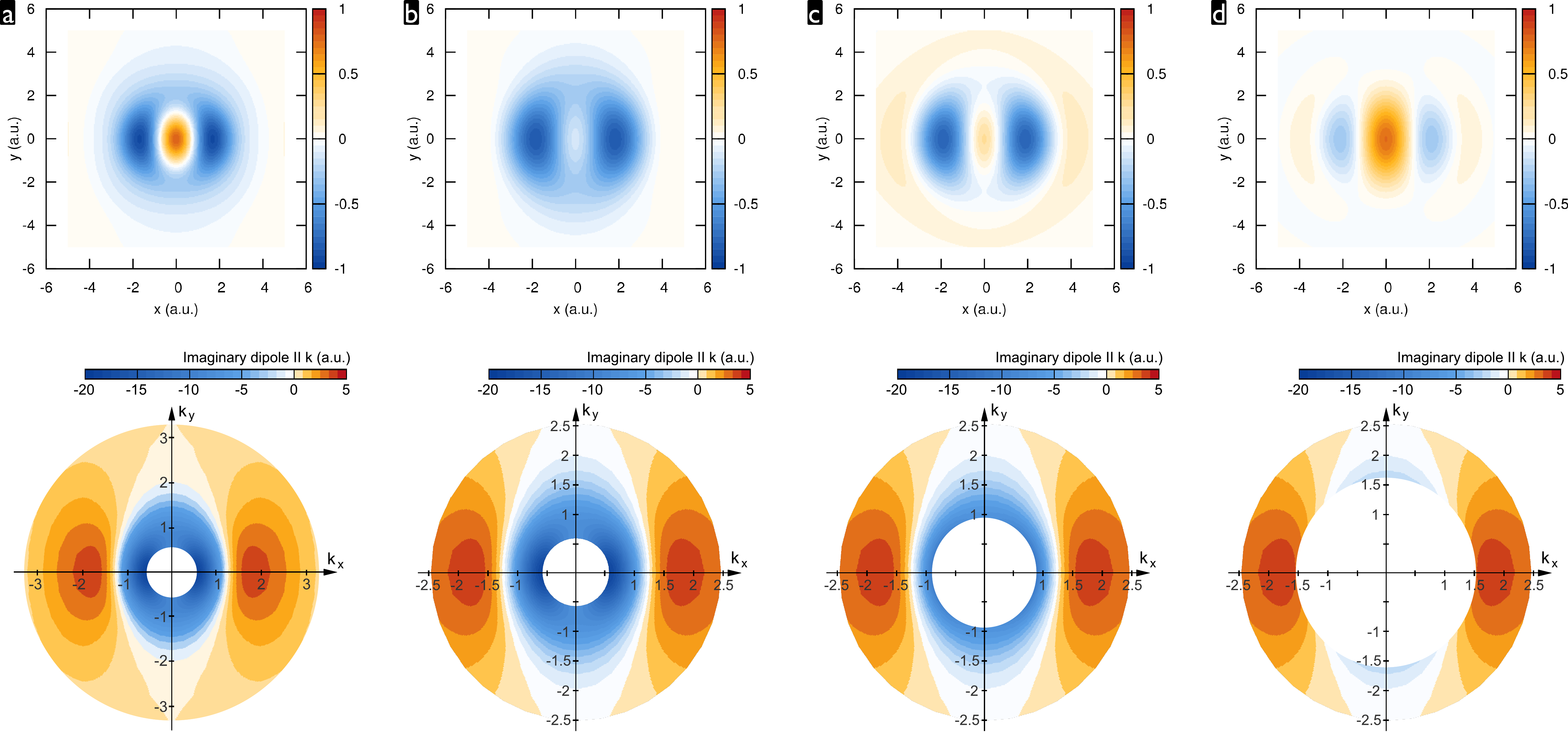}
	\caption{ \label{fig:krange1} Tomographic reconstructions and the slices of the DME, $d_\parallel$, that have been used. The sampling corresponds to the odd harmonic orders of an 800-nm laser and an angular step of $\Delta\theta=10^\circ$. The considered harmonic range is (a) H3--H99, (b) H3--H53, (c) H7--H53  and (d) H21--H53. Note that due the symmetry of the N$_2$ HOMO, the corresponding length-form DME is purely imaginary valued.}
\vspace{12pt}
\end{figure*}
Concerning the sampling, there are essentially two questions to be answered: Which $k$-range in the recombination DME has to be taken into account for a reasonably good reproduction of the orbital and with which density does this range have to be sampled?

\begin{figure*}
	\includegraphics[width=\textwidth]{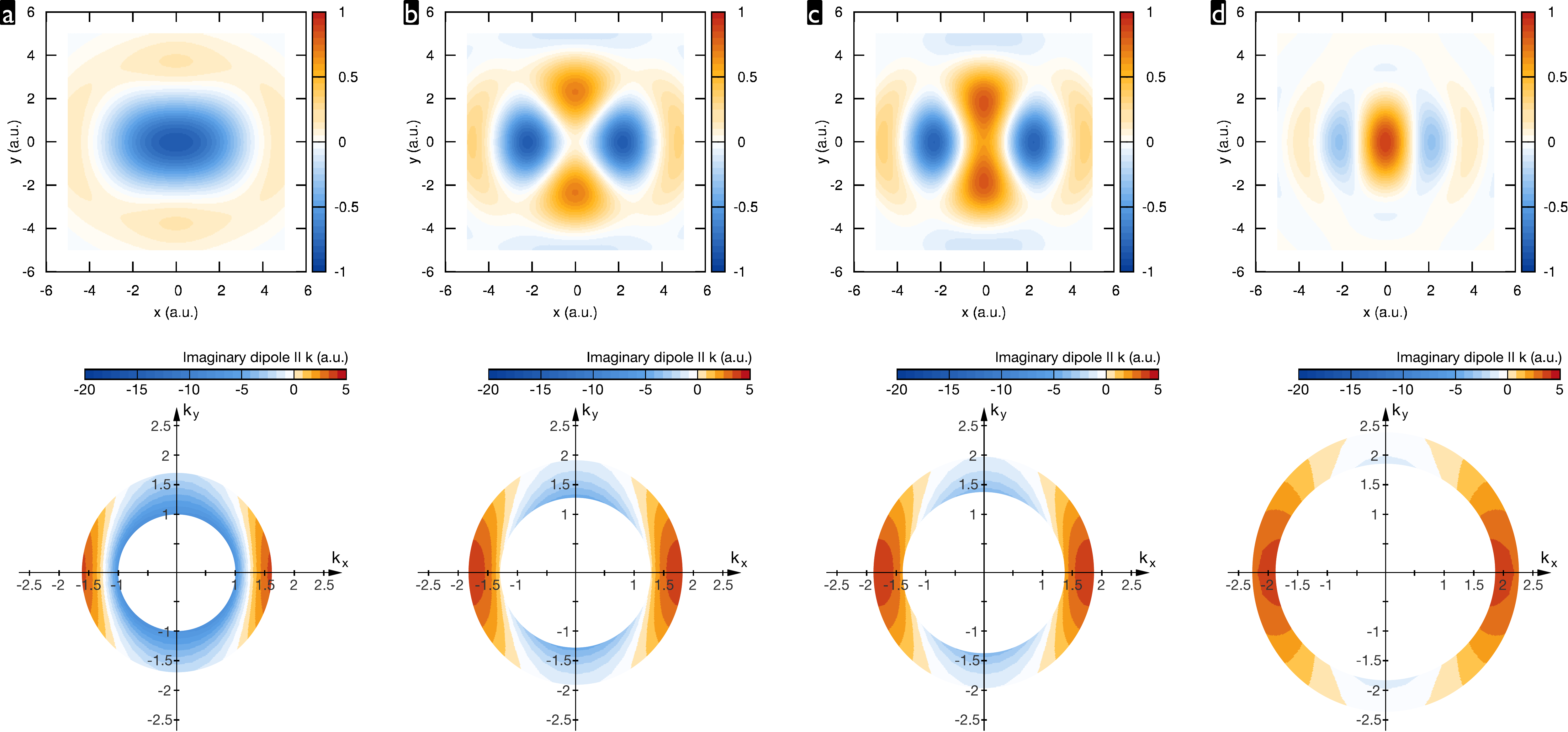}
	\caption{ \label{fig:krange2} Tomographic reconstructions and the slices of the DME, $d_\parallel$, that have been used. The sampling corresponds to the odd harmonic orders of an 800-nm laser and an angular step of $\Delta\theta=10^\circ$. The considered harmonic range is (a) H9--H23, (b) H15--H29, (c) H17--H31  and (d) H31--H45.}
\end{figure*}

In figure \ref{fig:sampling}, only $d_\parallel$ is used for the reconstructions and $\Delta\theta$ is varied, considering only a very restricted spectral range (harmonics 17 to 31), motivated by the experiment presented in \cite{Haessler2010tomo}. Between $\Delta\theta=5^\circ$ (figure \ref{fig:sampling}a) and $\Delta\theta=10^\circ$ (figure \ref{fig:sampling}b and, larger, in figure \ref{fig:krange2}c), no clear difference is visible and the principle features of the HOMO are quite well reproduced. The distortions with respect to figure \ref{fig:orb-h1-991} are due to the limited spectral range. For  $\Delta\theta=20^\circ$, additional distortions appear. Zooming out from the orbitals, another effect of the discrete sampling becomes apparent. If the sampling of the DME were done with an equidistant grid in $\bfk$-space, the result would be a periodic repetition of the reconstructed orbital in real space, thus imposing a certain minimal sampling density. In the experiment, we sample points equidistant on a \mbox{$k^2$-scaling,} along lines with an angular step $\Delta\theta$. This leads to the effect shown in figure \ref{fig:sampling}b,c, were the reconstructed orbital is repeated on the $x$ and $y$ axes, with a period inversely proportional to the sampling steps in $\bfk$-space. Due to the non-equidistant sampling, the repetitions become smeared out more and more as the distance from the origin increases. With $\Delta\theta=20^\circ$, the first repetition gets dangerously close to the actual reconstructed orbital, whereas $\Delta\theta=10^\circ$ turns out to be sufficiently small.

We have seen above that the narrow experimental spectral range allows to recover the principle structure of the N$_2$ HOMO. How sensitive is this to the exact position of the narrow spectral window and how fast does the reconstruction improve if the spectrum is enlarged? Looking at the N$_2$ HOMO, one can already guess that there is some characteristic spatial frequency that should be included in the $k$-range if the essential shape of the orbital should be reproduced. This frequency is $k=2\pi/L=1.75\:$ a.u., and corresponds to the distance $L\approx3.6\:$ a.u. of the two negative lobes (see figure \ref{fig:full_n2orbs}a). What other frequencies are important?

Figure \ref{fig:krange1} shows reconstructions using $\Delta\theta=10^\circ$ and different spectral ranges. The reconstruction is still fairly close to the exact HOMO using harmonics 3 to 99, which, in comparison with figure \ref{fig:orb-h1-991}a, shows that the improvement by including almost 900 more harmonic orders is rather marginal---the spectral amplitudes are simply very low at those high $k$-values. Note that, as expected, the DME exhibits sign changes close to the characteristic spatial frequency mentioned above.
Cutting the highest orders further and including only harmonics 3 to 53 leads to the disappearance of the positive central lobe and ruins the reconstruction. It is due to the very dominant low-frequency negative-amplitude components that the characteristic shape of the HOMO is lost. Cutting some of these, as done in figures \ref{fig:krange1}c,d, quickly allows to recover the characteristic shape of the N$_2$ HOMO. If one is constrained to limit the used spectral range, it should thus be cut on the low frequency side as well as on the high frequency side around the characteristic spatial frequency. 

	Can spectra be even narrower? In our experiments, we have so far been limited to 8 harmonic orders of an 800-nm driving laser, from H17 to H31, in the difficult phase measurements---and we have already seen in figure \ref{fig:sampling}, that these may contain sufficient information. Figure \ref{fig:krange2} shows reconstructions using only 8 odd harmonic orders. Clearly, when the considered spatial frequencies are too low so that the above mentioned characteristic spatial frequency is not contained in the range and negative-amplitude components are strongly dominating, the orbital structure is not reproduced. Around the characteristic frequency, the exact position of the narrow spectral window is crucial---one really has to hit the `sweet spot' of the DME. Using harmonics 17 to 31 indeed seems to be very close to this optimal situation. With only too high frequencies which almost only contribute with positive amplitudes, the reconstruction turns out less satisfactory again, similar to the example in figure \ref{fig:krange1}d.  It should again be noted that these simulations are based on length-form DME and $\omega=k^2/2$ is used. The conclusions drawn here can thus only be qualitative and there is significant uncertainty in the link of the harmonic orders considered here to those observed in the experiment. 

These simulations show that it is realistic to acquire experimental data that contain sufficient information to recover the shape of the active orbital beyond its essential symmetry. When using very narrow spectra, though, tomographic reconstruction becomes a game of chance about just hitting the essential part of the DME. Roughly, this essential part should contain a characteristic spatial frequency and a `well balanced' amount of the positive and negative spectral amplitudes. If the DME has the same sign all over the filtered slice, chances are large that one simply reconstructs an object without any particular structure other than the imposed symmetry. This sign is of course directly related to the \emph{phase} of the DME---our observations thus underline the importance of phase measurements.

For a reliable extraction of an a priori unknown orbital, the used experimental spectra clearly have to be rather large. Then, one also has to include both XUV polarization components, $\parallel\bfk$ and $\perp\bfk$---or measure an XUV polarization component \emph{in the molecular frame}, i.e. one that has always the same orientation with respect to the molecule, not with respect to the recollision direction. One such component in principle already contains all information on the orbital (see (\ref{eq:tomo:recon-int}) and (\ref{eq:tomo:recon-int-velocity})).

However, even with spectral widths that are quite challenging to achieve experimentally (although mid-IR lasers will help in extending the harmonic spectrum considerably), the reconstructions of \emph{static} orbitals will probably not be precise enough to be considered a benchmark for calculations. In a system where a single channel dominates HHG, we would probably always have a hard time to decide whether a reconstruction more resembles simply a Fourier-filtered Hartree-Fock orbital or a much more accurate simulation based on a properly calculated Dyson orbital. On the other hand, for the observation of dynamics, the attainable spatial resolution should be sufficient for most cases in order to provide useful comparison to theory---it will thus be the \emph{temporal resolution} that makes molecular orbital tomography particularly relevant to scientific applications. An example will be described in section \ref{sec:tomoexp}.

\subsection{Conclusion: What has to be measured?}\label{sec:todolist}
We take tomography as the basis of our argumentation because it demands full characterization of our observable ``electric field of the XUV emission'' and is a very illustrative and instructive example for discussing which information is needed and which role it plays for the extraction of information. For any other specific model and derived method for extracting information, the demands might be relaxed, but of course, a certain redundancy in the information can always help to make the results more convincing.

For tomography, we have to measure the recombination DME for the molecule via the observables XUV spectral intensity and spectral phase of (at least) one polarization component \emph{in the molecular frame}, as described at the end of section \ref{sec:measdip}. A first particular difficulty comes from the fact that we sample Fourier space ``line by line'' and in order to assemble the Fourier space of our Dyson orbital (see figure \ref{fig:tomo-scheme2}), we need to relate those measured lines or ``slices''.

For the spectral intensities, this is no problem at all since we can measure \emph{absolute} values here. We do not even need to elaborately calibrate our spectrometer, since any instrument response function will be divided out when normalizing by the spectral intensity measured for the reference atom.

Phase measurements are different in the sense that there are no ``absolute'' phases, but only phases relative to some reference. For the assembled Fourier space, we would like to have only an uncertainty about a \emph{global} phase, i.e. constant for all $k$ and $\theta$, i.e. all measured phases should be relative to the same reference. This could be achieved by measuring the complete phase of the XUV light, including the CEP, for every slice of Fourier space \footnote{Note that even then, we would keep an arbitrary global phase from the theoretical Dyson orbital in our calculated reference DME.}; but to date simply no methods exist to do so. However, all the information we really need is the phase-\emph{change} from one slice to the next and from one point in the slice to the next---i.e. two phase derivatives. These will be sufficient to obtain the phase $[\varphi^\mathrm{mol}_\mathrm{x/y}(\omega,\theta) - \rmi\varphi^\mathrm{ref}(\omega)]$ as a function of the two parameters $k$ and $\theta$, or $k_\mathrm{x}$ and $k_\mathrm{y}$, up to a global constant. In section \ref{sec:phasemeas}, we will discuss different phase measurement techniques for derivatives of the phase with respect to $\omega$ and $\theta$, i.e. the phase change in radial and angular direction in the Fourier space shown in figure \ref{fig:tomo-scheme2}.

What is left to be measured is the normalized ionization amplitude, $\eta^\mathrm{mol}(\theta_\rmi)/\eta^\mathrm{ref}$, which could be paraphrased as the $\theta$-dependence of the ``intensity'' and the ``sign'' of the recolliding EWP. As mentioned in the end of section \ref{sec:measdip}, these issues are removed when one manages to keep the ionization angle approximately constant (at least limit it to a relatively small range). This will also be beneficial for experiments where one aims at observing dynamics launched by tunnel ionization. This dynamics would certainly not be exactly the same for different ionization angles. If one is dealing with symmetric orbitals, the prior knowledge of their symmetry \cite{Haessler2010tomo} or a measurement of it \cite{Shafir2009,Niikura2010symmetry} will allow to impose the corresponding symmetry on the DME.

Finally, let us remind the readers of the other experimental conditions to be provided for an \emph{ab-initio} tomographic reconstruction of a Dyson orbital: The molecular frame has to be held fixed in the laboratory frame, i.e. molecules have to be oriented (for symmetric molecules, it is sufficient to align, see section \ref{sec:alignment}). Recollision has to be limited to one side only (except if known symmetry can be exploited). Phase-matching should be excellent throughout the measured XUV spectral range.

\section{Experimental techniques} \label{sec:exp}

Apart, of course, from a state-of-the-art driving laser, HHG is experimentally fairly simple: It comes down to focusing a laser pulse, sufficiently energetic and short to reach the required intensity, into a gas cloud at  $\sim100\:$mbar provided either by a cell with static pressure or by a (pulsed) gas jet. The latter has the advantage of creating a rotationally cold gas sample as is required for molecular alignment (see section \ref{sec:alignment}). The generated laser-like coherent XUV beam is then sent onto a suitable detector downstream.


\subsection{Holding the molecules in the laboratory frame}\label{sec:alignment}
The first experimental prerequisite is to hold the molecules in a certain orientation in the laboratory frame, where the EWP movement directly follows the driving laser field. Aligning and orienting molecules is a formidable experimental challenge and an active area of research with a long history \cite{Friedrich1991review,Stapelfeldt2003,Seidemann2005,Kumarappan2007}. As illustrated by figure \ref{fig:mol-align-orient}, \emph{alignment} and anti-alignment conventionally refer to head-on  ($\updownarrow$) versus broadside  ($\leftrightarrow$) localization of some particular axis of a molecule, whereas \emph{orientation} refers to control of the up ($\uparrow$) and down ($\downarrow$) directions of an aligned molecule.
\myfigure{1}{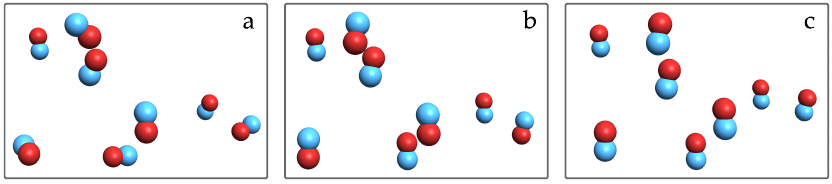}{Illustration of a randomly aligned (a), aligned (b), and oriented (c) ensemble of molecules.}

If one wants to control the orientation of a molecule, one wants to control the angular part, $\chi(\theta,\phi)$, of its nuclear wavefunction, which is conveniently expressed on the basis of the spherical harmonics $\ket{J,M}=Y_{JM}(\theta,\phi)$. In order to make the probability $\abs{\chi(\theta,\phi)}$ actually peak into a single direction, one has to \emph{coherently} populate a wavepacket of $\ket{J,M}$, which is typically done via laser-induced stimulated $J\rightarrow J\pm2$ Raman transitions. The rotational temperature comes into play because we are dealing with a macroscopic number of molecules in \stef{an} ensemble with a thermal (Boltzmann-)distribution of the $\ket{J,M}$-states. In each molecule, starting from its definite $\ket{J,M}$-state, a \emph{coherent} wavepacket, $\chi(\theta,\phi,t)$, is created. The total angular probability function, $P(\theta)$, for the macroscopic ensemble is then an incoherent average of coherent wavepackets over the initial thermal distribution. The more ``incoherence'' there is in the total wavepacket, the less sharp it can peak into a certain direction and hence the less good the achievable alignment/orientation can be.

On the one hand, very high degrees of orientation generally demand very low rotational temperature of the molecules. On the other hand, HHG requires relatively high ($\sim10^{17}\:$cm$^{-3}$) medium densities. Since very cold super-sonic gas jets \cite{Even2000Cooling,Hillenkamp2003Condensation} employing (several) skimmers and possibly some means of state-selection \cite{Holmegaard2009} provide only orders of magnitude lower densities, a trade-off has to be found and the lowest temperatures are sacrificed for a sufficient molecule density. While today, in very low-density gas samples, nearly perfect alignment and excellent degrees of orientation can be achieved \cite{Ghafur2009Impulsive,Guerin2008Ultimate,Holmegaard2009,Holmegaard2010}, the aforementioned trade-off for HHG media has so far only been found for the one-dimensional \emph{alignment} of linear, symmetric top and asymmetric top molecules \cite{Torres2007Probing,Kajumba2008Measurement}.

The non-adiabatic method used in these experiments as well as for the demonstrations of orbital tomography reported so far \cite{Itatani2004Tomographic,Haessler2010tomo} was pioneered by Seidemann \cite{Seideman1999} and Rosca-Pruna and Vrakking \cite{Rosca-Pruna2001}. A relatively strong femtosecond laser pulse `kicks' the molecules and creates the rotational wave-packet, which evolves freely after the pulse has passed. After a prompt alignment immediately after the laser pulse, and given a certain commensurability of the rotational eigenfrequencies in the wavepacket \footnote{This is assured for linear and symmetric top (2 of the 3 moments of inertia are the same) molecules, were all $J$-level-energies are multiples of the $J=0$-energy.}, it will regularly re-phase and lead to an angular distribution effectively aligned along the laser pulse polarization direction in field-free conditions. Details on this rotational quantum dynamics can be found in \cite{[*][]spannerthesis,mythesis,Torres2005Dynamics,[*]Kajumba2008Measurement,Seidemann2005}. With this technique, significant degrees of alignment in field-free conditions are obtained already with rotational temperatures $\lesssim100\:$K . Alignment quality is commonly described by the ensemble-averaged expectation value $\braket{\cos^2\theta}=\langle\braket{\chi\vert\cos^2\theta\vert\chi}\rangle_\mathrm{therm}$. This measure approaches unity for an angular distribution perfectly peaked along $\theta=0$ and $\pi$, $\braket{\cos^2\theta}=0$ for a disk-shaped distribution peaked along $\theta=\pi/2$, and $\braket{\cos^2\theta}=1/3$ for an isotropic distribution. Figure \ref{fig:align-N2-example} shows results of an example calculation for the alignment of N$_2$ molecule in the conditions of the experiments in \cite{Haessler2010tomo}. The degree of alignment can be improved by using a longer aligning pulse---the optimal duration for N$_2$\stef{, e.g.,} is $120\:$fs---or a higher intensity of the aligning pulse.
\myfigure{.55}{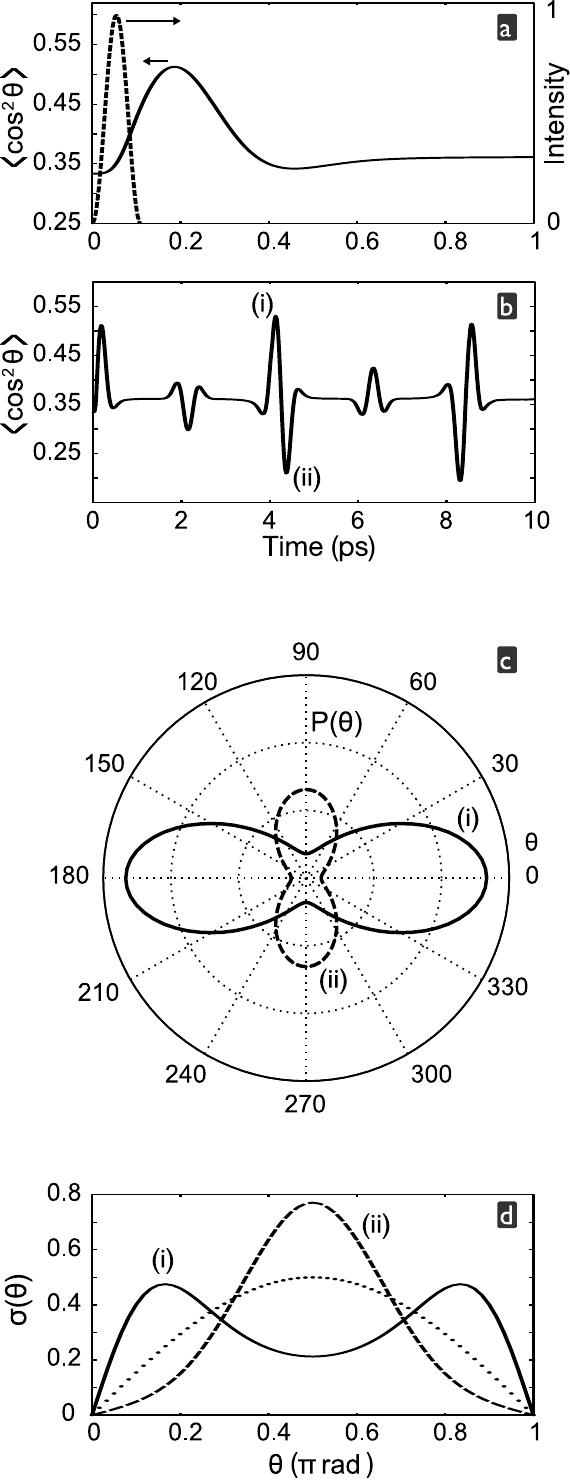}{Results of a calculation for N$_2$ molecules with rotational temperature $T_\mathrm{rot}=90\:$K, interacting with a laser pulse of duration $\tau=55\:$fs, peak-intensity of $5\times10^{13}\:$W/cm$^2$, and linear polarization along z. In panel \textbf{(a)}, the full line shows $\braket{\cos^2\theta}$ during the first picosecond as well as the laser pulse intensity envelope. In panel \textbf{(b)}, the evolution of $\braket{\cos^2\theta}$ is traced over more than one rotational period, $T=8.38\:$ps, of N$_2$. The first recurrence of alignment at the so-called half-revival at $t=4.135\:$ps as well as the immediately following anti-alignment at $t=4.38\:$ps are marked as (i) and (ii), respectively. Panel \textbf{(c)} then shows the angular distributions $P(\theta)$ of the molecules at these times (i) and (ii). Panel \textbf{(d)} contains the same information, but integrated over the azimuthal angle $\phi$: $\sigma(\theta)=2\pi P(\theta) \sin\theta$, which is proportional to the probability of finding a molecule with an angle between $\theta$ and $\theta+\rmd\theta$. The dotted line shows an isotropic distribution for comparison.}

Three-dimensional alignment of more complex molecules can be achieved, e.g., with elliptically polarized laser pulses \cite{Larsen2000} or with series of orthogonally polarized pulses \cite{Lee2006FieldFree,Viftrup2007Holding}.

The temperature--density trade-off is drastically aggravated for the \emph{orientation} of molecules for HHG. Most orientation-schemes break down for initial rotational temperatures $T\gtrsim10\:$K and technological progress towards very cold molecular samples with high densities will thus play a pivotal role for advances in self-probing of molecules. Given sufficiently cold gas jets, high degrees of orientation can be obtained non-adiabatically \cite{Ghafur2009Impulsive} as well as adiabatically \cite{Tanji2005three,Holmegaard2009,Holmegaard2010}, when the up-down-symmetry of the aligning laser field is broken by a relatively weak dc electric field. 
Completely field-free orientation can be obtained, e.g., by means of impulsive excitation by an asymmetric two-colour laser field \cite{Kanai2001,De2009,Oda2010,Tehini2008}, by terahertz half-cycle pulses \cite{Machholm2001} or phase-locked three-colour pulses \cite{Zhdanov2008}.

The optimal technique for an envisaged experiment has to be chosen according to the precise requirements on orientation (how many axes have to be fixed and is it necessary to orient or only align them) and the particular molecular species under study (does it have a permanent dipole moment, what are the orbital symmetries). 

In any practical application, neither alignment nor orientation will be perfect and thus will always contribute a certain error to self-probing measurements. Deconvolving a known angular distribution of the molecular sample from measured data is more complicated than one might think, but it is in principle possible iteratively \cite{Wagner2007Extracting,Yoshii2011Retrieving}.

\subsection{Controlling the EWP trajectories}
Once the molecules are held at in a controlled orientation in the laboratory frame (at least transiently, e.g. for about $100\:$fs around the delay (i) in figure \ref{fig:align-N2-example}), a second, more intense laser pulse can drive HHG in the so-prepared molecular sample. The driving \emph{field}-shape will then directly control the trajectories of the EWP.

The laser field ponderomotive potential scales as \mbox{$U_\mathrm{p}\propto I\lambda_0^2$}, \stef{where $I$ is the intensity}. Increasing the driver wavelength, $\lambda_0$, from the 800 nm of Ti:Sa-based lasers towards few microns available from OPA thus leads to recolliding electrons of higher energy, $\propto\lambda_0^2$ at constant intensity. This will greatly increase the sampled part of Fourier space in orbital tomography (see section \ref{sec:sampling}), but will also be important for the study of molecules with fairly low $\Ip$, implying rather low saturation intensities. At the same time, the electron trajectories become longer, the excursion duration increases $\propto\lambda_0$ and the SFA predicts the emitted XUV spectral intensity to drop $\propto\lambda_0^{-3}$ (cp. (\ref{eq:theory:saddlepointdipolespec})) due to EWP spreading. Solutions of the TDSE \cite{Tate2007} as well as experiments \cite{Shiner2009} brought the unpleasant surprise that this scaling is even worse: $\propto\lambda_0^{-5}-\lambda_0^{-6}$. Indeed, the increase of the cutoff results \emph{for a fixed energy interval} in an additional factor $\propto\lambda_0^{-2}$ \cite{Schiessl2007qpi}. However, as mentioned earlier, macroscopic effects may help to compensate for this drop of the single-molecule dipole \cite{Yakovlev2007,Colosimo2008Scaling, Popmintchev2009}. In general though, experiments using mid-IR drivers are much more challenging than those with 800~nm.

In the simplest case of a linearly polarized driving laser pulse, ionization and recollision will take place in the same direction: that of the  driving laser polarization. Rotating the latter, or alternatively the molecules, then allows to probe the molecule from a range of directions. A few-cycle pulse duration or an assymetric carrier wave generated by the combination of several colour components can limit the electron trajectories to one side of the molecule. The issue already mentioned is that this procedure also varies the ionization direction. 

Multi-colour waveforms can ease some of the issues just mentioned. For example, the ``perfect wave'' recently proposed by Chipperfield et al. \cite{Chipperfield2009}, is asymmetric and also effectively limits recollision to one side of the molecule. Most importantly, it is designed in order to maximise the recollision energy whilst keeping the trajectories as short as possible and thus avoids the dramatic drop in efficiency suffered when increasing the driver wavelength. In calculations, this waveform allowed to increase the maximum recollision energy by a factor 2.5 without losing any efficiency as compared to a monochromatic driver. 

As mentioned earlier, when polarization becomes part of the control parameters of multi-colour waveform shaping, it will be possible to vary the recollision direction whilst keeping the ionization direction approximately constant \cite{Kitzler2007,Kitzler2005,Shafir2009}, and thus, e.g., create the same initial ``hole in the ion'' for each direction from which it is probed after the EWP excursion. The range of recollision angles over which this will be possible with elaborate multi-colour waveforms has yet to be determined. Kitzler et al. showed numerically that with a combination of a fundamental with its equally strong orthogonally polarized second harmonic, the difference between ionization and recollision angle can be as large as $75^\circ$.

\subsection{Achieving phase matched HHG}\label{sec:phasematch}
As mentioned in section \ref{sec:macroscopic}, it is necessary to ensure good phase matching in the experiment in order to obtain a strong macroscopic signal and to be able to infer single-molecule information from it. This means that one has to arrange conditions that minimize the phase difference between the propagating XUV field and the driving polarization over the medium length. 

Essentially three contributions cause a phase mismatch along the medium length: (i) The \emph{Gouy phase shift} of $\pi$ of the driving laser as is goes through its focus \cite{Svelto}. (ii) The intensity dependence of the molecular \emph{dipole phase}, essentially contained in the quasi-classical action of the continuum electron, $S(\mathbi{p},t_\rmi,t)$, given by  (\ref{eq:theory:sfaaction}): this phase varies approximately linearly with intensity~\cite{Varju2005,Salieres2001}. (iii) The \emph{dispersion}, i.e. the difference in phase velocity between the driving laser and the XUV radiation, dominated by the free electrons created by ionization\footnote{Note that these are not the quasi-bound continuum electrons of the three step model, but those permanently detached from the core forming a free electron gas in the HHG medium.}, the density of which obviously strongly depends on the local intensity. Dispersion and absorption from neutral molecules and ions can play a role in long media \cite{Constant1999,Hergott2002Extreme,Rundquist1998phasematched} but can usually be neglected in short gas jets.

Loose focusing can slow down the first two of these variations. In addition, one can arrange that the phase mismatch due to the intensity dependent dipole phase of the short trajectory contribution cancels out that caused by the Gouy phase shift at a short distance after the laser focus~\cite{Salieres1995}. Placing a rather short generation medium at this distance thus allows to approach perfect \emph{on-axis} phase matching.  Significant ionization does, however, cause strong dispersion that rapidly ruins any phase matching. Moreover, it depletes the ground state of the emitters. Phase matching in an ionizing medium becomes a highly dynamical process in both the temporal and
spatial domain. See \cite{Gaarde2008Macroscopic} for a detailed review of macroscopic effects in HHG.

The cold gas jets used in all self-probing experiments provide a medium with an effective length of $<1\:$mm. Ionization is usually kept at very low level. Good phase matching is usually  confirmed by checking for a quadratic dependence of the XUV intensity on the medium pressure. Moreover, the recombination times (also often called emission times) measured for atomic gases \cite{Mairesse2003Attosecond} are usually found to be in very good agreement with the single-atom SFA theory described in section \ref{sec:sfa}.

The phase, 
$S[(\mathbi{p}^\mathrm{s},t_\rmi^\mathrm{s},t_\mathrm{r}^\mathrm{s})_n]$, acquired by the continuum EWP during its excursion (cp. (\ref{eq:theory:saddlepointdipolespec})) varies several times faster with intensity for the long trajectories ($n=2$) than for the short ones ($n=1$). This implies that the radial intensity profile of the driving laser beam translates into an XUV phase front curvature that is larger for the long trajectory contribution. The different divergences thus cause the contributions of short and long trajectories to spatially separate in the far field. In many experiments, one selects the on-axis emission and thus the short-trajectory contribution, which is stronger anyway if one has optimized phase matching for it.

\subsection{Polarization resolution}
\label{sec:exp:polariz}
\myfigure{.9}{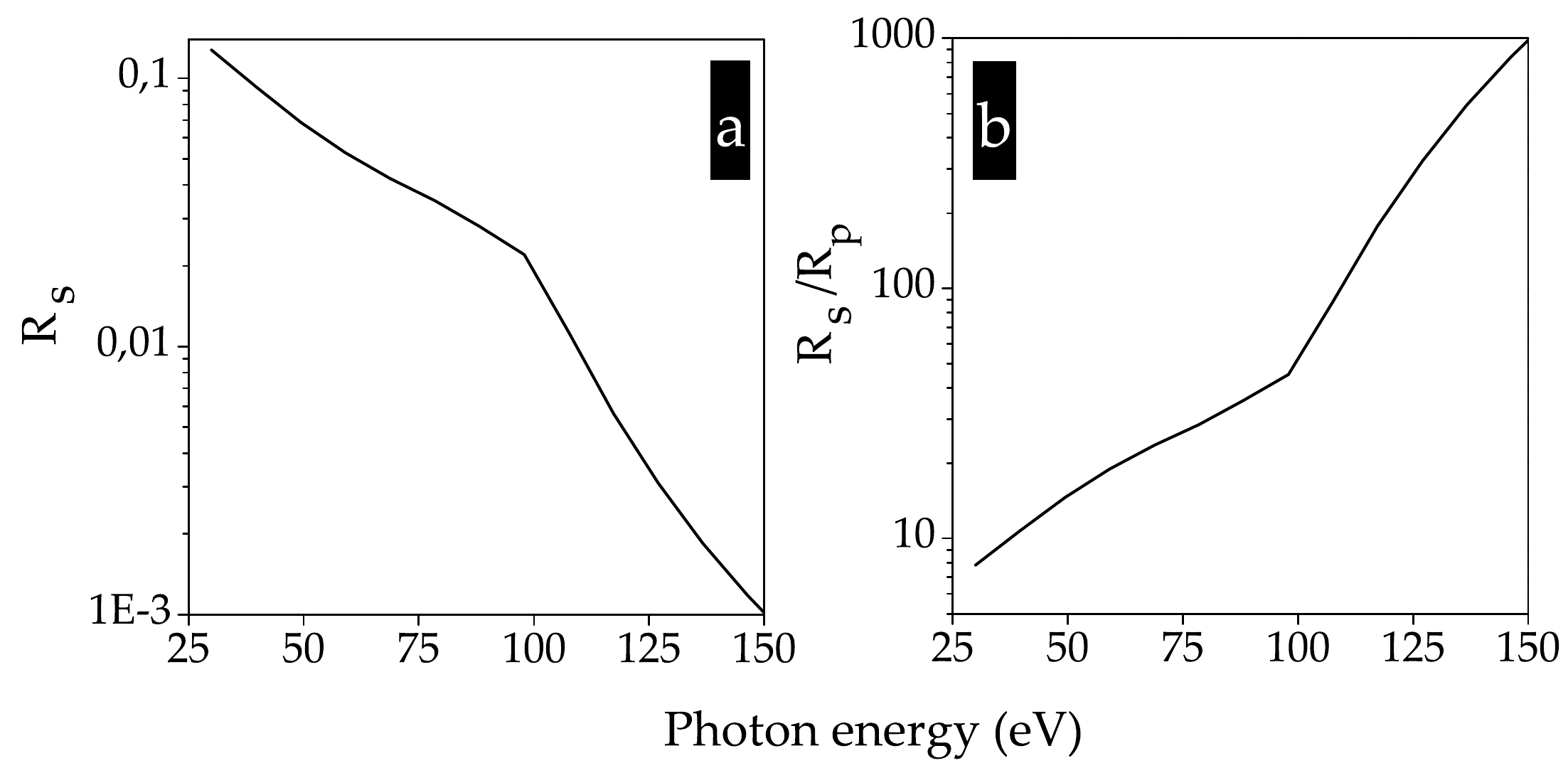}{A silver mirror as an XUV polarizer. (a) Reflectivity under $45^\circ$ incidence for s-polarized light. (b) Ratio of reflectivities for s- and p-polarized light, i.e. the extinction ratio of the polarizer. These data are based on \cite{CXRO}.}
Experiments aiming either at characterizing amplitude and phase of a selected polarization component of the XUV light only \cite{Haessler2010tomo}, or at characterizing the full polarization state of the XUV light \cite{Antoine1997polarization,Levesque2007Polarization,Mairesse2008Polarizationresolved,Zhou2009,Mairesse2010Multichannel} will obviously need to contain a polarization discriminating element---a polarizer. The polarizing element used in all the works just cited was simply a reflection off a bare metal mirror. As figure \ref{fig:Ag-mirror-R} shows, this indeed makes a good polarizer, reflecting s-polarized light about ten times better than p-polarized light. This polarizer is, however,  not convenient for high photon energies above $\approx110\:$eV since the reflectivity drops below 1\%. This behaviour is the same for all metal surfaces. Moving towards grazing incidence obviously allows to reflect with good efficiency much higher photon energies, but at the expense of a decreasing extinction ratio. 

For higher photon energies, either one deals with extremely low XUV flux on the detector, or one switches to multi-layer structures for a polarizer in reflection or transmission \cite{Schaefers1999soft,MacDonald2009polarizer,Imazono2009development}.

Polarization resolution is interesting for several reasons. We have seen in section \ref{sec:sampling} that if we use a wide spectral range for orbital tomography, we will also need polarization resolution. In \cite{Mairesse2008Polarizationresolved}, we have shown how detection of a selected polarization component allows to greatly increase the contrast of the detection of dynamics (see also section \ref{sec:exp:highcontrast}). In \cite{Mairesse2010Multichannel}, the XUV ellipticity served as the observable that is analyzed in order to evidence multi-channel dynamics in the N$_2^+$ ion.

\subsection{Spectral intensity measurement}\label{sec:intensyitymeas}
Measuring the spectral intensity, i.e. doing spectrometry with the XUV light, is the most common measurement and part of virtually any experiment on HHG. For self-probing experiments, spectrometry only needs to be qualitatively accurate, i.e. we only need to know the shape of the XUV spectrum in arbitrary units. The only serious experimental difficulty here is the effective suppression of background signal due to the scattered driving IR laser light which has many orders of magnitude higher flux than the XUV light to be characterized. Metal foils of $\sim100\:$nm thickness are commonly used \cite{Rodrigo2005Alu,Gustafsson2007} as high-pass filters that block the IR light. Other options are grazing incidence reflections off substrates with anti-reflection coating for the IR \cite{Wabnitz2006Generation,Ravasio2009}, or reflections off Si or SiC plates at Brewster's angle for the IR \cite{Takahashi2004Si}.

As opposed to the characterization of the XUV light \emph{on target}, i.e. on the detector, for self-probing, we want to characterize the XUV \emph{at the source}. We thus in principle need to calibrate for the complete optical path from the source to the detector. In data analysis schemes using normalization by some reference, like those introduced in sections \ref{sec:decode:nucl} and \ref{sec:measdip}, the instrument response is, however, divided out anyway.

XUV intensity spectra can either be measured by photoionizing a target gas with known cross-section in an electron spectrometer---which is thus automatically integrated into the phase measurement methods described in section \ref{sec:phasemeas:atto}---or with a grating-based XUV photon spectrometer. Very often, these employ grazing incidence concave flat-field gratings \cite{Harada1999hitachi}. Due to the grazing incidence, the focusing by these gratings is very astigmatic and the focus in the spectral dimension (the tangential focus) lies far before the sagittal focus. The line spacing of the grating is varied over its surface in a way that leads to flat-field conditions, i.e. the spectral (tangential) foci fall on a straight line for the spectral bandwidth the grating is designed for. These two properties lead to an image on a \emph{plane} XUV detector at the distance of the flat-field spectral focus, which gives good spectral resolution in one dimension and at the same time the far-field spatial profile in the perpendicular dimension. This can become very useful, e.g. in the spatial interferometry schemes described in section \ref{sec:phasemeas:spatial}.

While being certainly the easiest observable to measure, the spectral intensity is unfortunately not so easy to interpret. The spectral intensity is also the XUV property which is most prone to being influenced by macroscopic effects \cite{Ruchon2008Macroscopic}. Also, structures such as intensity minima can have a several possible origins (see, e.g., section \ref{sec:imaging} or \cite{Han2010Minimum}). It is thus clear that more observables need to be measured in order to get an as-complete-as-possible set of observations which can then be combined to obtain a consistent physical interpretation.

\subsection{Spectral phase measurement}\label{sec:phasemeas}
\myfigure{.66}{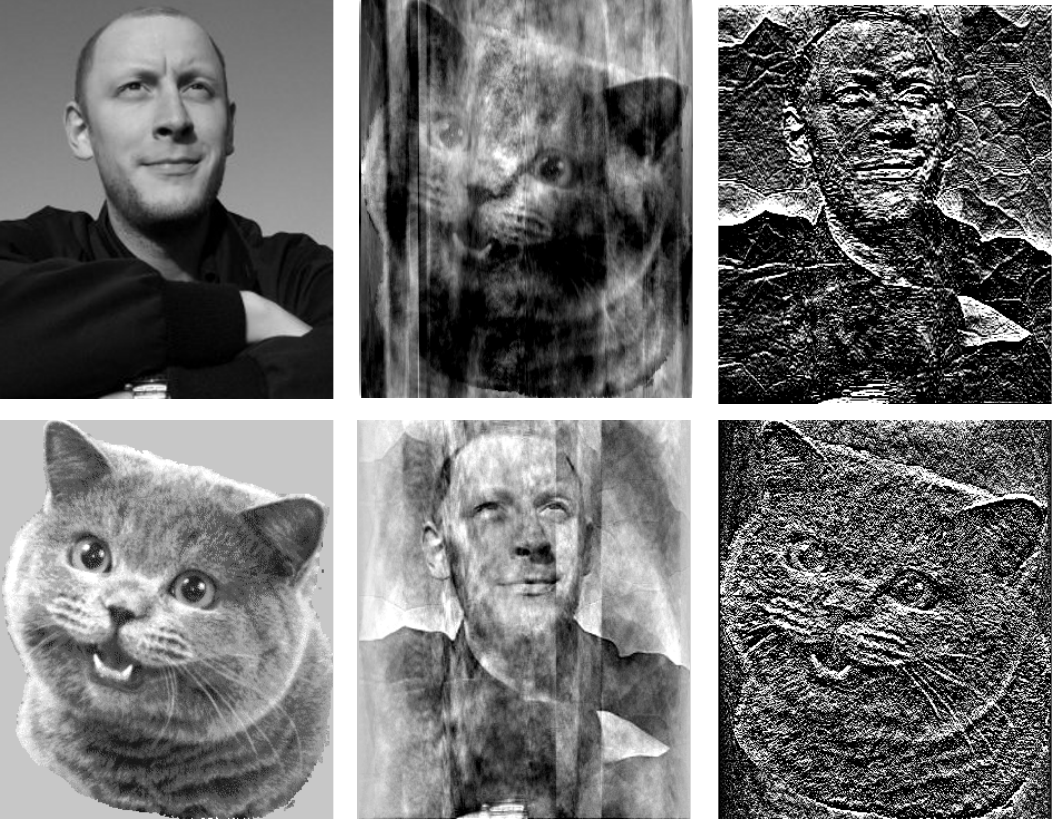}{The importance of the phase for image reconstruction: (Left column) Photographs of one of the authors (S.H.) and a cat. (Middle column) Image reconstructed from the Fourier modulus of the author-image and the phase of the cat-image (upper), and vice-versa (lower). (Right column) Image reconstructed with unity Fourier modulus and the phase of the author/cat-image (upper/lower).}
Figure \ref{fig:fftd-images} shows a little trick inspired by Rick Trebino (see figure 13 of chapter 1 in \cite{TrebinoUFObook}): when an image is created by Fourier transforming a spectrum, it is the spectral \emph{phase} that imposes its ``information content'' on the result. Due to this importance, but also due to the particular difficulty of measuring phases, we will in the following devote special attention to several phase measurement methods for XUV light from HHG.

\subsubsection{From attosecond pulse measurement---spectral interferometry}\label{sec:phasemeas:atto}
Several methods have been developed for the temporal characterization of attosecond light pulses. These have in common to be based on spectral interferometry and to measure the spectral phase, $\varphi(\omega)$, up to a constant, i.e. up the CEP of the attosecond light pulses. Equivalently, one can say that they measure the group delay, $\partial\varphi/\partial\omega$, of the light pulses. 

A method to measure $\partial\varphi/\partial\omega$ has to include some means to make different spectral components interfere with each other, i.e. some means of creating spectrally shifted replicas. Since it is much easier for us to spectrally shift electrons than XUV light pulses, we start by making an ``electron-replica'' of the XUV pulse via photoionization \footnote{This is an idealized picture, neglecting amplitude and phase of the transitions dipole matrix element, which induces of course a deviation from the perfect ``replica'' and has to be compensated for. See section 1.3.4 of \cite{mythesis} for a more detailed description.}. An IR laser field can now be used to shift the spectral components of this electron replica in order to implement spectral interferometry. 

One such method, ``reconstruction of attosecond beating by interference of two-photon transitions'' (RABBIT), is particularly easy to understand in the photon picture. It can be applied for XUV radiation that consists of discrete odd harmonics of the driving IR laser, i.e. in the most common situation of a multi-cycle driver pulse with symmetric carrier wave. When ionizing atoms with this XUV spectrum and a simultaneously present, time-delayed weak ($\sim10^{11}\:$W cm$^{-2}$) IR field of the same frequency, $\omega_0$, as the driving laser, two-color two-photon ionization pathways lead to the appearance of spectral sidebands in the photoelectron spectrum, shifted by one IR photon energy from the odd harmonics.  As illustrated in figure \ref{fig:rabitt-scheme}, the sidebands of two adjacent odd harmonic thus overlap, i.e. the two corresponding ionization paths, $[(q+1)\omega_0 - \omega_0]$ and $[(q-1)\omega_0 + \omega_0]$, interfere in the sideband according to their relative phases. It is very important that the IR probe intensity be kept very low in order to be sure that higher order sidebands involving absorption or emission of $>1$ photon are negligible.
\myfigure{.5}{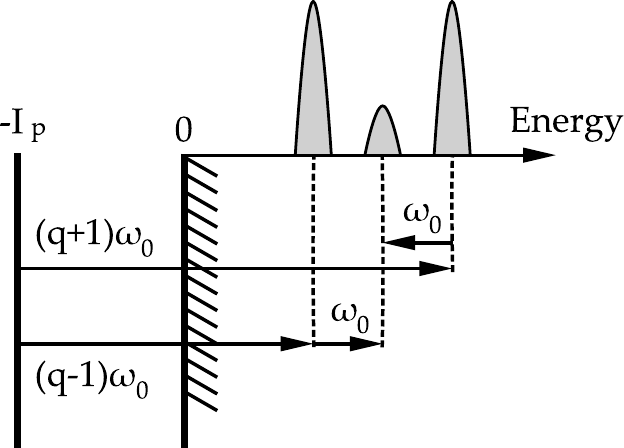}{The two ionization paths leading to the same sideband in the RABBIT scheme. An atom with ionization potential $\Ip$ is ionized by either absorbing an XUV photon, $(q+1)\omega_0$, and emitting an  $\omega_0$-photon, $\omega_0$, or by absorbing a $(q-1)\omega_0$-photon and a $\omega_0$-photon.}

It can be shown that the measurable intensity of sideband $q$ is modulated by an interference term \cite{Paul2001Observation,Muller2002,Veniard1996,mythesis}:
\begin{equation}
 S_q(\tau) = \cos[2\omega_0\tau + \varphi_{q+1}-\varphi_{q-1} - \Delta\phi_\mathrm{at}], 
\label{eq:rabittterm}
\end{equation}
where $\tau$ is the delay of the XUV and the weak IR pulse, $\varphi_{q\pm1}$ is the XUV spectral phase at harmonic $(q\pm1)$, and  $\Delta\phi_\mathrm{at}$ is a small correction term characteristic of the ionized atoms, which can be accurately calculated \cite{Toma2002}. From the phase of the $2\omega_0$-oscillation of the sidebands in a spectrogram, i.e. a collection of spectra for a scanned range of XUV-IR delays, one can thus extract the phase difference for pairs of neighbouring harmonics, which corresponds to a measurement of the group delay, $\partial\varphi/\partial\omega\approx \Delta\varphi/\Delta\omega = (\varphi_{q+1}-\varphi_{q-1} )/ 2\omega_0$.

Now, if we determine the sideband oscillation phase by fitting a cosine with an argument as in (\ref{eq:rabittterm}), the group delay we find depends on how we have fixed the zero on our delay axis and we thus determine the group delay only up to some constant. The delay-axis-zero can be defined ``absolutely'' with an extension of RABBIT, where the weak IR probe beam overlaps not only with the XUV beam in the photoelectron spectrometer, but already with the IR driving beam in the HHG medium \cite{Dinu2003Measurement,Mairesse2003Attosecond}. This will lead to a very small modulation of the HHG driving intensity, oscillating with frequency $\omega_0$ when the probe beam delay is scanned, and the extreme non-linearity of the HHG process will lead to a measurable modulation of the HHG intensity. If we now pick a delay where this modulation is maximum and define it as  $\tau=0$, the measured XUV group delay is determined on a time axis where the zero corresponds to a maximum of the driving laser field---and we have determined an ``absolute'' group delay. Integrating the group delay over $\omega$ thus leads to the XUV spectral phase, $\varphi(\omega)$, up to an integration constant, which is nothing else but the CEPof the XUV pulses.

For a general XUV field without the nice beneficial spectral shape with only odd harmonics, this method can be generalized. The photon picture, illustrated in figure \ref{fig:rabitt-scheme}, then gets quite confusing. However, formulating the two-photon ionization in the strong-field approximation instead, i.e. as a single-XUV photon ionization followed by interaction of the free electron with the IR laser field only, does lead to ways to retrieve the XUV spectral phase \cite{Itatani2002StreakCamera,Mairesse2005,[**][]Quere2005,mythesis}. Then, one can also increase the IR probe intensity if necessary as we do no longer rely on an XUV+\emph{single}-IR-photon model. The most general method recognizes the analogy of the spectrogram, $\omega$ vs. $\tau$, where $\tau$ is scanned over the full overlap of the XUV emission with the IR probe pulse, with the spectrograms recorded in one of the most popular femtosecond pulse characterization methods, ``frequency resolved optical gating'' (FROG) (see chapter 6 of \cite{TrebinoUFObook}). An iterative phase retrieval algorithm can thus be used to extract the spectral phase of an arbitrarily complex XUV pulse---again, up to the CEP. This method has been named ``frequency-resolved optical gating for complete reconstruction of attosecond bursts'' (FROG-CRAB) \cite{Mairesse2005}. The experimental implementation is somewhat more difficult than RABBIT since for FROG-CRAB, the $\tau$-range to be scanned is much longer, the analyzed photo-electrons can only be collected from a small cone around the probe laser polarization direction \cite{Quere2005} and high temporal resolution requires high IR probe intensities which quickly create background-signal problems due to above-threshold ionization. On the bright side, the additional effort for FROG-CRAB is rewarded with its generality.  

Attosecond XUV pulse measurement techniques can thus be applied to measure $\partial\varphi/\partial\omega$ for each recollision angle $\theta$. We performed such measurements in Saclay \cite{Boutu2008Coherent,Haessler2010tomo}.

\subsubsection{Spatial interferometry}\label{sec:phasemeas:spatial}
Two methods to measure the harmonic phase variation with the recollision angle at constant frequency, $\partial\varphi/\partial\theta$, are HHG 2-source interferometry \cite{Smirnova2009co2,Zhou2008Molecular}  and transient grating spectroscopy \cite{Mairesse2008Transient,[**]Mairesse2010Phase,Woerner2010}, both first used in the context of self-probing by Mairesse et al.. The basic concept is the analysis of spatial interference of the XUV emission originating from molecules with different angular distributions.

\myfigure{1}{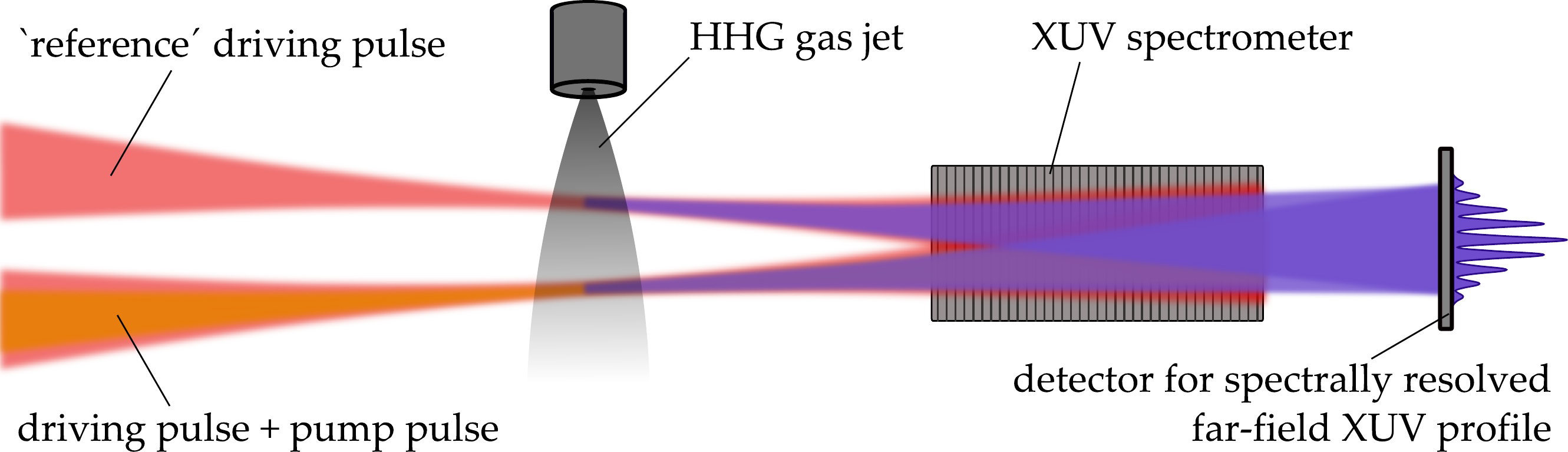}{Scheme for HHG interferometry \cite{Smirnova2009co2} based on two separate HHG sources, one of which serves as phase reference. The XUV spectrometer disperses the spectrum in the direction perpendicular to the page and lets the beam diverge in the other dimension so that the spatial diffraction pattern can be observed.}
HHG 2-source interferometry, schematically shown in figure \ref{fig:hhg-interferometry}, is based on two spatially separated HHG sources, which are phase-locked through the use of the same driving laser. The mutual coherence of the two sources, proven in 1997 at Lund University \cite{Zerne1997phaselocked}, leads to a far-field spatial interference pattern similar to Young's double slit where the fringe position depends on the relative phase of both sources \cite{Zerne1997phaselocked,Descamps2000xuvinterferometry,Corsi2006direct,Merdji2000}. One contribution to the relative phase is of course the delay between the two sources: for a stable interference pattern, the delay fluctuations between the two driving IR beams thus have to be smaller than one period \emph{of the XUV}, i.e. $\sim10\:$as. This is one of the major challenges in the experimental implementation and can be met, e.g., in collinear schemes as those described in the supplementary information to ref. \cite{Smirnova2009co2} or in \cite{Zhou2008Molecular}.

The spatial interference pattern can be observed with simultaneous spectral resolution by using an astigmatic flat-field XUV spectrometer (see section \ref{sec:intensyitymeas}). One of the two sources will provide the phase reference and in the scheme shown in figure \ref{fig:hhg-interferometry}, this is simply obtained by HHG in unaligned molecules. In the second source, some parameter is varied, with respect to which the phase derivative is then measured; this parameter can be the recollision direction, $\theta$, e.g. controlled by a preceding aligning pulse. The spatial fringe shift then directly gives $\partial\varphi/\partial\theta$.

Transient grating spectroscopy, schematically shown in figure \ref{fig:transient-grating}, is based on a grating formed by ``excited'' molecules in an elsewhere ``unexcited'' HHG medium, which can be created by crossing two pump pulses at a small angle in the gas jet. The resulting intensity interference pattern consists of planes separated by a grating period of $\lambda_\mathrm{g}=\lambda_\mathrm{L}/[2\sin(\beta/2)]$, where $\lambda_\mathrm{L}$ is the pump laser wavelength and $\beta$ is the angle between the two pump beams, i.e. typically, $\lambda_\mathrm{g}\sim10\:$\textmu m. This pump-intensity grating will prepare a corresponding excitation-grating in the medium.
\myfigure{1}{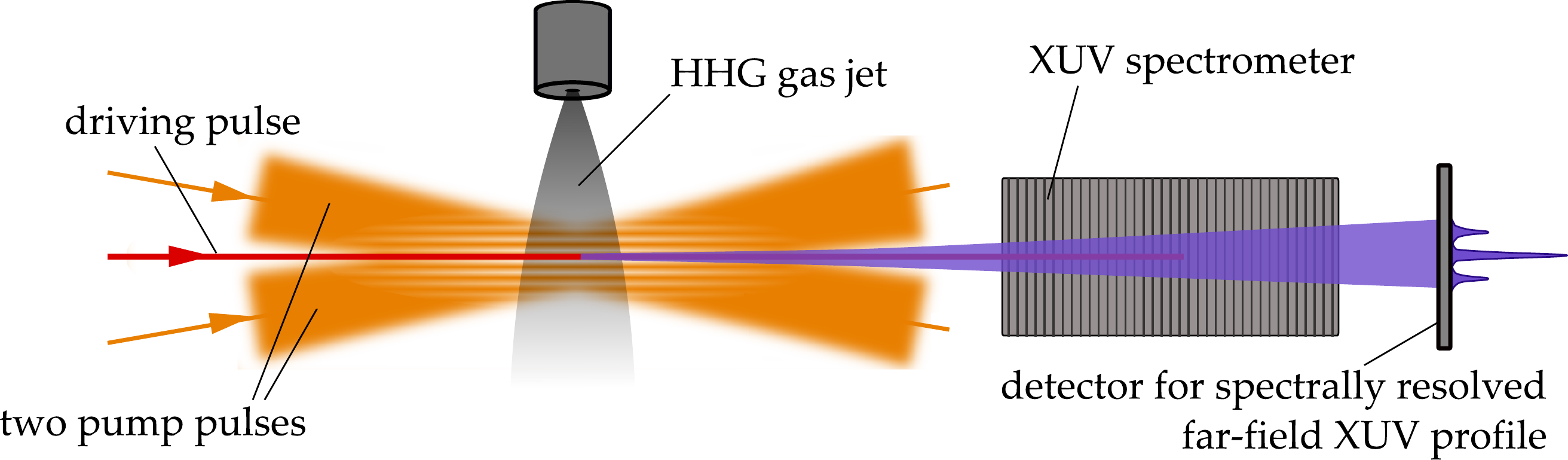}{Scheme for transient grating spectroscopy \cite{Mairesse2008Transient,Mairesse2010Phase} based on HHG in a medium with an ``excitation grating'' prepared by two pump beams with an angle. The XUV spectrometer disperses the spectrum in the direction perpendicular to the page and lets the beam diverge in the other dimension so that the diffration due to the transient grating in the HHG medium can be observed.}

Since the ``excitation'', in our case the alignment of the molecules, modifies in general both the intensity and the phase of the emitted harmonics, this constitutes an amplitude and phase mask. While the pump-intensity-grating is purely sinusoidal, the shape of this amplitude-and-phase mask is purely sinusoidal only if the XUV intensity and phase depend linearly on the excitation, which itself depends linearly on the pump intensity. Otherwise, the mask is inharmonic. In any case, the grating will lead to a $\pm$first-order diffraction at very small angles $\delta\approx\lambda_\mathrm{XUV}/\lambda_\mathrm{g}$, i.e. typically $\delta\sim1\:$mrad. Due to these small diffraction angles and the short length of gas jets ($\sim1\:$mm), the situation is always well described in the ``thin grating'' limit and the far-field interference pattern is given by the power-spectrum of the amplitude-and-phase mask. If the excitation grating is sinusoidal, there are only $\pm$first-order diffraction peaks and the analysis is simplified. From the measured diffraction efficiency, one can then extract the phase (and amplitude) modulation of the grating, and thus the phase difference of the emission from excited (aligned) and unexcited (unaligned) molecules. Refs \cite{Mairesse2010Phase} and \cite{Woerner2010} (in the supplementary information) describe two different ways of doing so.

\subsubsection{Gas mixing}
A different scheme for making the XUV emission from molecules interfere with a phase reference relies on using a gas mixture for the HHG medium \cite{Kanai2007Destructive,*Kanai2008,Wagner2007Extracting,McFarland2009n2phase}, pioneered bv Kanai et al.. If the molecules under study are mixed with, e.g., rare gas atoms in the HHG medium, the respective XUV emissions will interfere according to their relative phases. If the XUV spectra for the two components of the gas mixture are known, i.e. spectra have been measured with the pure gases under equal conditions, then the interference-term $\cos(\Delta\varphi)$ can be extracted from the spectrum measured with a medium of known mixing ratio. Here, $\Delta\varphi$ is the relative phase of the XUV emissions from the two components of the mixture. If the two components are a molecule and a corresponding ``reference atom'' with the same ionization potential, then this relative phase corresponds to the molecular recombination dipole phase up to possible sign changes contained in the molecular ionization amplitude, $\eta^\mathrm{mol}(\theta_\rmi)$ ; see section \ref{sec:measdip}. Varying the alignment/orientation of the molecules in the mixture, which will of course leave the reference atoms unaffected, then allows to measure $\partial\varphi/\partial\theta$.

Experimentally, this method requires the preparation of a gas mixture with precisely known partial pressures, as well as very similar phase-matching conditions between the gas mixture and the pure gases. Finally, a very reliable and precise measurement of XUV spectral intensities is needed.

\subsection{Selectively probing excited molecules} \label{sec:exp:highcontrast}

When an experiment aims at probing excited molecules to be prepared by a preceding pump pulse, the excited state is always populated with a certain probability only. The question of how to separate the background signal due to unexcited fraction of emitters from the signal due to the excited fraction is, of course, a classic one for pump-probe spectroscopy and methods for enhancing the detection contrast or even completely suppressing the background do exists. Two of these have already been mentioned in this tutorial at other instances: 
Transient grating spectroscopy \cite{Mairesse2008Transient,Woerner2010} (see section \ref{sec:phasemeas:spatial}), and polarization resolved detection \cite{Mairesse2008Polarizationresolved} (see section \ref{sec:exp:polariz}). 

\section{An example: Experimental Orbital Reconstructions} \label{sec:tomoexp}
	
As an illustration and application of many concepts introduced in the preceding sections, let us review our study published in \cite{Haessler2010tomo} on orbital tomography with the N$_2$ molecule. In the experiments, HHG was driven with a linearly polarized Ti:Sa-based laser, i.e. $\lambda_0=800\:$nm and typical electron excursion durations $\approx1.5\:$fs. Ref. \cite{Patchkovskii2009} assures us that the movement of the nuclei in the N$_2^+$ ion (in its ground state) upon ionization is negligible in this timespan: the nuclear overlap integral (\ref{eq:nucloverlap}) remains $\gtrsim0.95$. 

We treat the electronic structure of N$_2$ in the Hartree-Fock framework and use Koopmans' approximation, thus neglecting exchange contributions (cf. section \ref{sec:multielectron}), which is also supported by \cite{Sukiasyan2010}. In this framework, multi-channel contributions to HHG are described as multi-\emph{orbital} contributions: the channel-specific Dyson orbitals are simply the ionized Hartree-Fock orbitals. The small energy difference of $\Delta\varepsilon\approx1.4\:$eV between the first excited and the ground state of N$_2^+$, i.e. between the HOMO and the \mbox{HOMO-1,} as well as the fact that the \mbox{HOMO-1} is much larger in the y-direction than the HOMO (see figure \ref{fig:full_n2orbs}) make it likely that both orbitals contribute to HHG in N$_2$ \cite{McFarland2008}. For a symmetrical molecule like N$_2$, all orbitals have a defined parity: the HOMO is gerade and the \mbox{HOMO-1} is ungerade. We thus can factorize the complex XUV spectrum for each orbital without having to limit recollision to one side of the molecule (cf. section \ref{sec:factorize}). Since the laser is linearly polarized, ionization and recollision angles are the same.

From (\ref{eq:tomo:multichannelDME}), we know that the time-domain recombination DME for multi-channel HHG writes as a coherent sum over the individual channel contributions.  Its Fourier transform and consequently the emitted complex XUV spectrum, $\epsilon(\omega)$, thus have the same form.  In this study, we disregard any coupling of the orbital contributions, i.e. (de-)excitation of the N$_2^+$ ion during the electron excursion, such that the channel weights $b_{(j)}$ in (\ref{eq:fullionwp}) \emph{ff.} are time-independent complex numbers set by the tunnel ionization process, i.e. they depend on $\theta$. We can thus use (\ref{eq:factorizedsignal}) for each contributing channel with constant weight and sum to obtain (using the length form):
\begin{equation}
	\epsilon(\omega,\theta) = b_\mathrm{X}(\theta)\alpha_\mathrm{X}(\bfk) \braket{\psi_\mathrm{X} \vert \hat{\mathbf{r}} \vert \bfk}+  b_\mathrm{A}(\theta)\alpha_\mathrm{A}(\bfk) \braket{\psi_\mathrm{A} \vert \hat{\mathbf{r}} \vert \bfk},
\label{eq:example:eps}
\end{equation}
where the subscripts X and A denote the HOMO and HOMO-1 contribution, as in section \ref{sec:multiorbital}. This corresponds to equation 1 in ref. \cite{Haessler2010tomo}. We use $\bfk=\sqrt{2\omega}(\cos\theta,\sin\theta)$, i.e.the XUV frequency is associated with the same electron wavenumber for both channels. Although one may want to include a shift corresponding to $\Delta\varepsilon$, the uncertainty in the correct link between XUV frequency and electron wavevector is much larger than this issue anyhow. Note that the channel weights $ b_\mathrm{X/A}(\theta)$ in (\ref{eq:example:eps}) correspond to the ionization-angle-dependent scaling of the EWP amplitudes, expressed by $\eta^\mathrm{mol}$ in (\ref{eq:exp:moldip})---the relative ionization amplitudes fix the relative weights for the contributing channels.

From the experimental data, we will retrieve the total recombination DME as described in section \ref{sec:measdip}. The reference atom is argon since it has nearly the same ionization potential as the N$_2$ HOMO contribution ($\Ip^\mathrm{Ar}=15.76\:$eV $\approx\Ip^\mathrm{N_2}=15.58\:$eV). This calibrates for $\alpha_\mathrm{X}(\bfk)$, but $\alpha_\mathrm{A}(\bfk)$ is somewhat different due to the corresponding $\Ip$ being larger by $\Delta\varepsilon$. Of the three terms in (\ref{eq:factorizealpha}), we can neglect the variation of the first and the last: the slightly different EWP spreading for the two contributions will only negligibly affect the relative weight of both contributions; and the difference in ionization DMEs for both contributions is neglected based on the argument of the tunnel-ionization step acting as a strong spatial filter (i.e. its $k$-dependence will be very similar for both contributions; note, however, that their $\theta$-dependence is very different ). For the remaining phase factor, we have already found the variation with $\Ip$ in (\ref{eq:dSdIp}): due to the quasi-classical action along the trajectories being stationary, the only significant phase variation comes from the difference in phase acquired by the ions during the excursion duration. Thus, $\alpha_\mathrm{A}(\bfk) / \alpha_\mathrm{X}(\bfk)\approx\exp[-\rmi\Delta\varepsilon\tau]$. Consequently, the total recombination DME retrieved from our measurements is described by
\begin{align}
	\mathbi{d}_\mathrm{exp}(k,\theta) &= b_\mathrm{X}(\theta) \braket{\psi_\mathrm{X} \vert \hat{\mathbf{r}} \vert \bfk}+  b_\mathrm{A}(\theta)\exp[-\rmi\Delta\varepsilon\tau] \braket{\psi_\mathrm{A} \vert \hat{\mathbf{r}} \vert \bfk}, \label{eq:expdipole} \\
					&= \braket{ \psi_\mathrm{hole} \vert \hat{\mathbf{r}} \vert \bfk}, \nonumber
\end{align}
with
\begin{equation}
\psi_\mathrm{hole}=b_\mathrm{X} \psi_\mathrm{X}+   b_\mathrm{A}\exp[-\rmi\Delta\varepsilon\tau] \psi_\mathrm{A}.
\label{eq:example:psihole}
\end{equation}
We have arrived at an expression describing a recombination DME between a plane-wave and a time-dependent hole-wavefunction, given by the superposition of the two channel-specific Dyson orbitals beating with relative phase according to their energy-difference. This is the same result as in (\ref{eq:theo:holeDME}), and we can of course rephrase what we just did in the terms employed to derive (\ref{eq:theo:holeDME}): We have approximated the continuum EWP for both orbital contributions by the \emph{same} packet of plane waves with complex spectral amplitude, $\alpha_\mathrm{X}(\bfk)$. The attentive reader will have noticed that in (\ref{eq:example:psihole}), we have dropped the $\theta$-dependence of the channel weights. This is a simplification we are forced to make if we want to define a hole-wavefunction for our experiment. Actually, the created hole is different for each $\theta$ and the $b_\mathrm{X/A}$  in the expression for $\psi_\mathrm{hole}$ are thus to be understood as channel weights averaged over all angles.

Before starting with the tomographic reconstruction, we have to take a close look at what exactly has been measured. The experiments, described in \cite{Haessler2010tomo}, collected data by a series of RABBIT measurements of the HHG emission from N$_2$ aligned at angles $\theta=0^\circ, 10^\circ, 20^\circ, \ldots 90^\circ$ between the molecular axis and the linear polarization direction of the  driving laser (the angular distribution of the molecules is simulated in figure \ref{fig:align-N2-example}). These were normalized to the results of a  RABBIT measurement of the HHG emission from argon under the same experimental conditions. Two reflections off gold-coated mirrors made us preferentially detect the XUV polarization component parallel to the driving laser polarization, i.e. parallel to $\bfk$. A perpendicular component is neglected---as simulated in section \ref{sec:sampling}.

The experiment \emph{measured} the spectral intensities and the group delay $\partial\varphi / \partial\omega$ (cp. section \ref{sec:phasemeas:atto}) for the harmonic orders 17 to 31 and for each angle $\theta$. In order to obtain the spectral phase as function of both $\omega$ and $\theta$, we had to add the \emph{assumption} that $\partial\varphi / \partial\theta = 0$ for the lowest detected harmonic order (supported by HHG 2-source interferometry measurements done in Ottawa). The angular variation of the ionization probabilities, $[\eta^\mathrm{mol}(\theta)]^2$, are not corrected for, because \emph{(i)} we had no means of reliably calculating them and \emph{(ii)} because we actually included them in our definition of the hole wavefunction (\ref{eq:example:psihole}). Note, however, that we include the sign-changes of $\eta^\mathrm{mol}(\theta)$, as explained in the following.

The x- and y-components of the total DME, $\mathbi{d}_\mathrm{exp}(k,\theta)$, have been determined from measured data for angles $\theta=0^\circ \ldots 90^\circ$, i.e. points in the first quadrant of the Fourier-space representation of the objects $x \psi_\mathrm{hole}$ and $y \psi_\mathrm{hole}$. Due to the ``phase memory'' of the EWP, we could not simply continue the measurements for the remaining quadrants (see section \ref{sec:signchanges}), and instead have to complete $\mathbi{d}_\mathrm{exp}(k,\theta)$ according to the symmetry of the contributing orbitals, i.e. we have to include the sign-changes of $\eta^\mathrm{mol}(\theta)$. But how could we do so if two contributions with different symmetries interfere?

As mentioned in section \ref{sec:factorize}, the recombination DME for the gerade HOMO is purely imaginary-valued and that for the ungerade HOMO-1 is purely real-valued. This means that, if the excursion durations, $\tau$, for the used harmonic orders are close to a multiple of the half-period of the beating between both contributions, i.e. if $\Delta\varepsilon\approx n\pi,\: n\in\mathds{Z}$, then the two orbital contributions to the total recombination DME (\ref{eq:expdipole}) are in quadrature. For a finite spectral range, this situation can of course only be approximated, and as we show in \cite{Haessler2010tomo}, for our experimental conditions, $0.75\pi\lesssim\Delta\varepsilon\tau\lesssim1.2\pi$, so that HOMO and HOMO-1 contributions are indeed approximately separated in the imaginary and real part of the total recombination DME (\ref{eq:expdipole}), respectively. Note that since $\mathbi{d}_\mathrm{exp}$ is determined up to a global phase only, we can of course arbitrarily rotate it in the complex plane and this separation in imaginary and real part is only found at one particular complex rotation. We found the most consistent tomographic reconstructions, if the arbitrary global phase was set to zero.

The so-separated contributions to $\mathbi{d}_\mathrm{exp}(k,\theta)$ can now be completed according to the symmetry of the orbitals, i.e. such that: for $\mathrm{Im}[\mathbi{d}_\mathrm{exp}(k,\theta)]$, the x-component is even in $k_x$ and odd in $k_y$, and the y-component is odd $k_x$ and even in $k_y$; while for $\mathrm{Re}[\mathbi{d}_\mathrm{exp}(k,\theta)]$, the x-component is odd in both $k_x$ and $k_y$, and the y-component is even both in $k_x$ and $k_y$. This probably sounds confusing, but the reader will find that it is straightforward to derive from the symmetry properties of the Fourier transform that these are the symmetries of the DME components that immediately follow from those of the orbitals.

\myfigure{1}{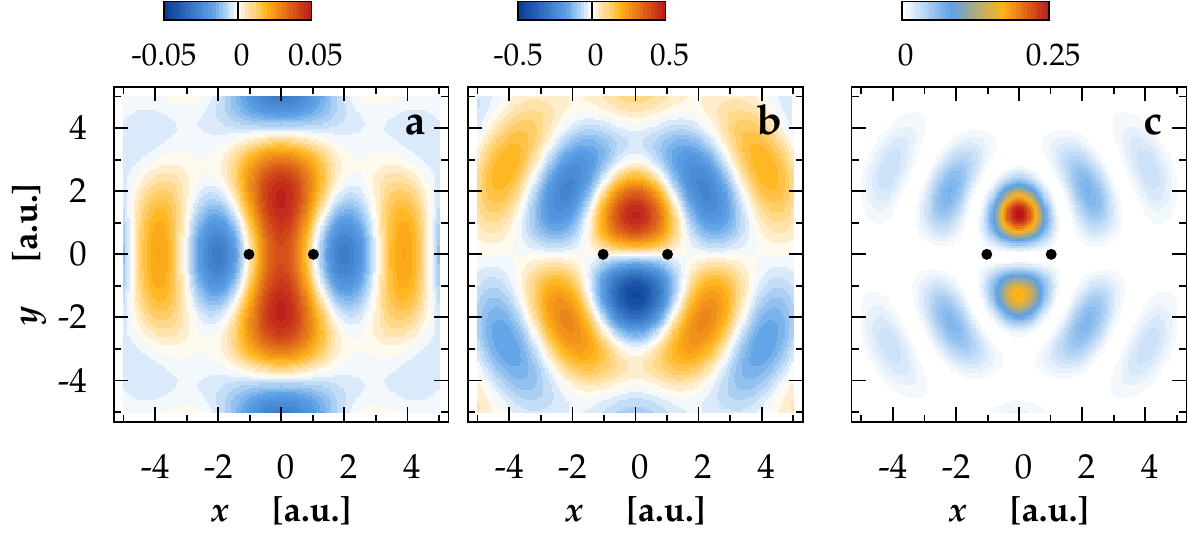}{Experimental orbital reconstructions for N$_2$. (a) Orbital obtained using only the imaginary part of  $\mathbi{d}_\mathrm{exp}(k,\theta)$, (b) using only the real part, (c) squared orbital obtained using the full  $\mathbi{d}_\mathrm{exp}(k,\theta)$, interpreted as the hole-density at the recollision instant. The black dots mark the positions of the nuclei.}
Now, we are ready to use $\mathbi{d}_\mathrm{exp}(k,\theta)$  as input to the tomography scheme described in section \ref{sec:tomo}, using (\ref{eq:tomo-halfsum}) and (\ref{eq:tomo:recon-int}). As we show in \cite{Haessler2010tomo}, whenplugging in only the imaginary part of $\mathbi{d}_\mathrm{exp}(k,\theta)$, we indeed reconstruct an experimental image of the N$_2$ HOMO (see figure \ref{fig:n2-tomo-rec}a) with distortions very similar to those obtained in the simulations (figure \ref{fig:krange2}c). When transforming only the real part, i.e. the HOMO-1 contribution, we prefered using the velocity form expression (\ref{eq:tomo:recon-int-velocity}) in order to avoid division by $x$/$y$ in real space, which leads to numerical problems with orbital node at $y=0$. The result is shown in figure \ref{fig:n2-tomo-rec}b.

While the reconstruction of the HOMO contains a clear spatial structure that is not simply imposed by the symmetry of the DME, this is not the case for the reconstruction of the HOMO-1. It is essentially a result of the considered spectral range and the imposed symmetry: when setting the DME amplitudes to unity and the phases to zero for all angles and frequencies, the obtained image is almost the same as that extracted from the experiment. Although this is expected for the HOMO-1, the DME of which does not have any particular structure in the considered spectral range, the absence of structure in both simulation and experimental result does not allow to claim a reconstruction of the HOMO-1. It is rather an indication of the consistency of our experimental observations and their interpretation.

While the limited spectral range (harmonics 17 to 31,  also used for the simulations shown in figures \ref{fig:sampling}b and \ref{fig:krange2}c) is certainly the most important limitation for a precise reconstruction in real space, there are a few others. The separation of both contributions in real and imaginary part of the DME is not perfect and there is an approximately linear variation of their relative phase over the spectral range. Furthermore, the relative weight of both contributions is not constant but varies both with frequency and angle, which introduces an additional filter-function in Fourier space and thus distortions in real space.

Finally, we can of course use the full complex-valued DME $\mathbi{d}_\mathrm{exp}(k,\theta)$ for a tomographic reconstruction to obtain $\psi_\mathrm{hole}$. The same result is obtained when simply summing the two purely real-valued reconstructions obtained from real and imaginary part of the DME. The square of this sum is shown in figure \ref{fig:n2-tomo-rec}c, in order to emphasize the asymmetry. This reconstruction corresponds to the hole density, $\psi_\mathrm{hole}^2$, in the N$_2^+$ ion at after the ``self-probing delay'', which is given by the mean excursion duration of the harmonics orders considered. This delay is $\tau=1.5\:$fs, and the ``exposure time'', i.e. the range of excursion durations covered by the considered harmonic range is $\Delta\tau=0.6\:$fs.

To conclude this example, we can ask how this result could be improved and whether it could have been obtained without separating the orbital contributions and imposing the symmetries. Yes, we could achieve an improved result without knowing in advance what we expect to reconstruct: if we limit recollision to one side of the molecule and control the EWP trajectories such that $\theta_\rmi$ remains approximately constant while measuring $\partial\varphi/\partial\theta$, we could in principle \emph{measure} the correct symmetries by beating the ``phase memory'' issue. Extending significantly the spectral range is possible via shaped waveforms \cite{Chipperfield2009} and/or by using longer wavelengths for the driving laser. Of course, a broader spectrum will in general also cover a longer range of excursion durations---the trade-off between spatial and temporal resolution inherent to ultrafast imaging with chirp-encoded recollision is pointed out in the comment by Smirnova and Ivanov on \cite{Haessler2010tomo}. However, conrol over the continuum electron trajectories will be able to handle this issue as well.

\myfigure{.95}{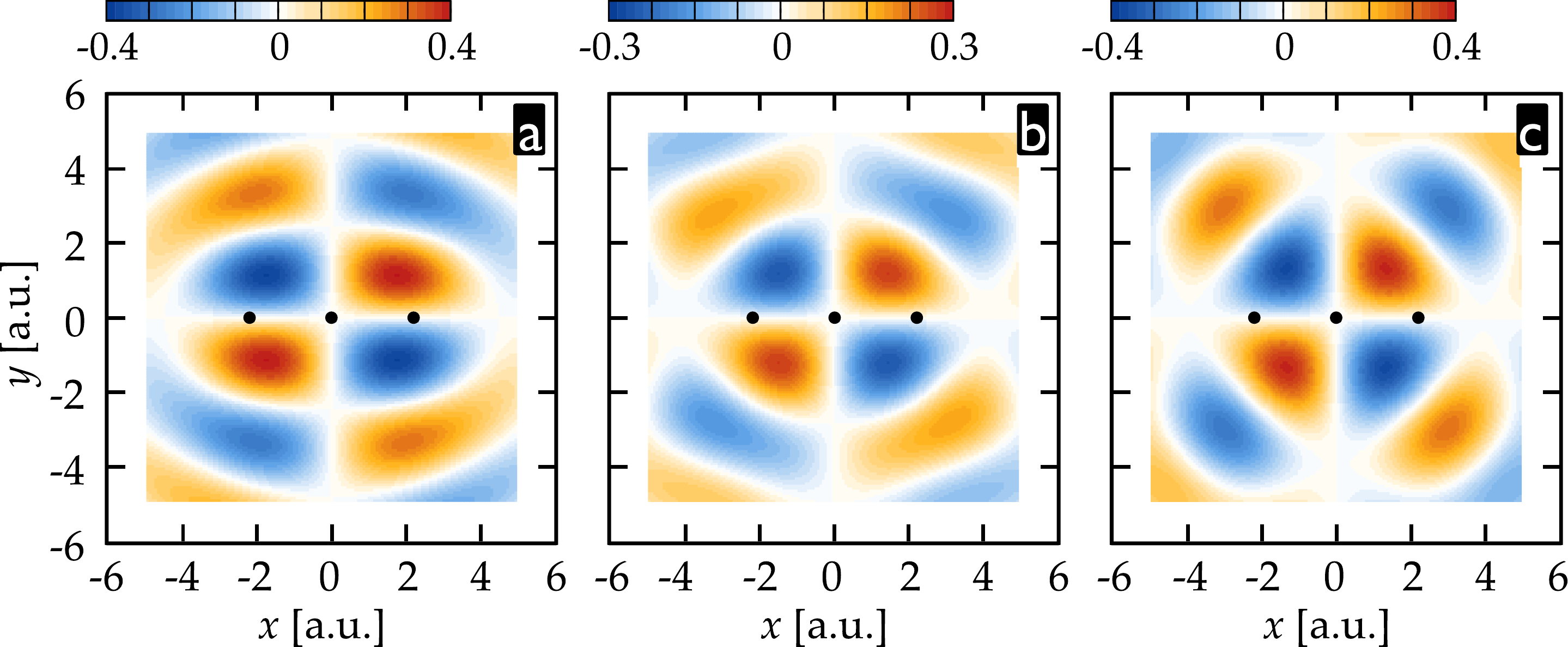}{Tomography of the CO$_2$ HOMO, based on the velocity form. (a) Simulation using the experimental sampling ($\Delta\theta=10^\circ$, harmonics 17--29). (b) Using the experimental DME, retrieved from the measured data reported in \cite{Boutu2008Coherent}. (c) Using a DME with $\pm$unity amplitude and $\pi/2$ phase, i.e. without any information content except the imposed symmetry and that the orbital is real-valued. The black dots mark the positions of the nuclei.}
As a second example, let us briefly consider the CO$_2$ HOMO. Our data shown in figure 3 of \cite{Boutu2008Coherent}, measured in the exact same way as just described for N$_2$, can be used to perform a tomographic reconstruction as well. Supposing that the X-channel (i.e. the HOMO) dominates HHG, we can impose the corresponding symmetry to the retrieved recombination DME (the x-component is even in $k_x$ and odd in $k_y$, and vice versa for the y-component), and finally obtain the result shown in figure \ref{fig:co2tomo}b. While they are not very different from the reconstruction assuming only the symmetry, shown in figure \ref{fig:co2tomo}c, one can notice that instead of the diamond shape of the latter, the core of the orbital presents an oval shape, closer to the simulation presented in figure \ref{fig:co2tomo}a. Nontheless, this is certainly an example for an inconclusive case due to a too narrow spectral range and corresponding too low precision in the real-space reconstruction.

\section{General Conclusions and Outlook}\label{sec:outlook}

We hope we could convince many newcomers that the self-probing paradigm holds great potential for measurements combining \AA ngstr\"om and attosecond resolution. Numerous works applying self-probing in numerical or real-world experiments, cited throughout this tutorial, have already accumulated overwhelming evidence that the molecular dipole in HHG does encode information on the structure and dynamics in the molecule at  these scales.

Many of the models upon which information retrieval is based are still under active development and continuous improvement can be expected for the years to come. As the most important construction sites let us mention the description of the continuum EWP as well as the inclusion of multi-electron effects.

The experimental tool-box for self-probing already contains a great many techniques to access the relevant observables serving as input to model-based information retrieval methods. Enabling new techniques will in the future complement this tool box; for instance laser-waveform shaping will provide ever finer control over the probe EWP. For experiments with non-symmetric molecules, progress in the \emph{orientation} of molecules in sufficiently dense gas samples will be crucial.

The example of experimental orbital tomography has shown that one has to take care in realizing which information is actually \emph{measured} in an experiment, and appreciate the amount of pre-knowledge that has to be added in order to extract information. On the other hand, it seems unnecessarily ambitious to aim at imaging methods that do not require any pre-knowledge about the system under study. For example, the static structure of molecules \emph{before} dynamics are launched is almost always known with good precision. In any case, orbital tomography is by no means a necessary goal for every useful self-probing experiment. It is simply one possible way of analyzing the fundamental quantity which encodes information about the molecule: the complex molecular dipole. More accurate models may not allow a similar direct information retrieval scheme, but there will always be ways to disentangle valuable information.

The most important merit of self-probing is clearly that it combines atomic scale spatial resolution with femtosecond or even attosecond \emph{temporal resolution}. While images retrieved from experiments on \emph{static} systems will probably never be precise enough to be a benchmark for advanced quantum-chemistry calculations, the attainable spatial resolution is sufficient to be able to follow intra-molecular dynamics. While self-probing cannot be taken as a scheme to ``directly film'' electrons in molecules---every frame is retrieved from data based on a model and has to be analyzed carefully---we are confident that ``molecular movies'' will come to a (lecture) theater near you soon.

\acknowledgements
We thank Willem Boutu, Bertrand Carr\'e, Pierre Breger, Patrick Monchicourt, Thierry Ruchon, Michel Perdrix, Olivier Gobert, Jean-Fran\c{c}ois Hergott, David Garzella, Fabien LePetit and Delphine Jourdain and \emph{LUCA} for invaluable help on the experiments, and acknowledge crucial theoretical contributions by C\'ecilia Giovanetti-Teixeira, Richard Ta\"ieb and Alfred Maquet to our joint work. We enjoyed many fruitful discussions with Jon Marangos, Manfred Lein, Elmar van der Zwan, Yann Mairesse, David Villeneuve, Misha Ivanov, Olga Smirnova and Chii-Dong Lin. 

S.H. acknowledges funding by the Lise Meitner fellowship M1260-N16 of the Austrian Science Fund (FWF). P.S. acknowledges financial support from the EU-FP7-ATTOFEL, and, together with J.C., from the ANR-09-BLAN-0031-01 Attowave program.

\section*{List of abbreviations}
	\newcommand{\abbritem}[2]{
	\noindent
	\begin{tabular}{p{1.2cm}p{7cm}}
	#1 & #2
	\end{tabular}
	\\
		}
	 \abbritem{BO}{Born-Oppenheimer}	 
	\abbritem{CEP}{Carrier-Envelope Phase}
	\abbritem{DME}{Dipole Matrix Element}	
	 \abbritem{EWP}{Electron Wave Packet}
	 \abbritem{HHG}{High Harmonic Generation}
	 \abbritem{OPA}{Optical-Parametric Amplifier}
	 \abbritem{SFA}{Strong-Field Approximation}
	 \abbritem{Ti:Sa}{Titatium:Sapphire}
	  \abbritem{TDSE}{Time-Dependent Schr\"odinger Equation}
	\abbritem{XUV}{eXtreme Ultra-Violet}

\bibliography{bib/stefan}

\end{document}